\documentclass[11pt]{article}
\usepackage{fancyhdr}
\usepackage{lipsum}

\fancyheadoffset{0cm}

\fancypagestyle{plain}{%
   \fancyhf{}%
   \fancyhead[R]{\thepage}%
}

\usepackage{dingbat}
\usepackage{subfig}
\usepackage{lscape}
\usepackage{pdflscape}
\usepackage{graphicx}
\usepackage[flushleft]{threeparttable}
\usepackage{multirow}
\usepackage{multicol}
\usepackage{diagbox}
\usepackage{comment}
\usepackage{hyperref}
\usepackage{bm}
\usepackage[margin=1in]{geometry}
\usepackage{booktabs}
\newcommand{\PreserveBackslash}[1]{\let\temp=\\#1\let\\=\temp}
\newcolumntype{C}[1]{>{\PreserveBackslash\centering}p{#1}}
\newcolumntype{R}[1]{>{\PreserveBackslash\raggedleft}p{#1}}
\newcolumntype{L}[1]{>{\PreserveBackslash\raggedright}p{#1}}
\usepackage{arydshln}
\usepackage{caption}
\usepackage{siunitx}
\usepackage[table]{xcolor}
\usepackage{dblfloatfix,float}
\usepackage{amsmath,amssymb}
\usepackage{mathabx}
\usepackage{bbm}
\usepackage[font=small,labelfont=bf,tableposition=top]{caption}

\usepackage{parskip} % gaps between paragraphs
\DeclareCaptionLabelFormat{andtable}{#1~#2  \&  \tablename~\thetable}

\usepackage{natbib}
 \bibpunct[, ]{(}{)}{,}{a}{}{,}%
\newcommand{\Cov}{\mathrm{Cov}}

\newcommand{\Var}{\mathrm{Var}}

\begin{document}

\title{Joint Models for Cause-of-Death Mortality in Multiple Populations}
  \author{Nhan Huynh\thanks{Department of Statistics and Applied Probability, University of California at Santa Barbara} and Mike Ludkovski\thanks{Department of Statistics and Applied Probability, University of California at Santa Barbara, \url{ludkovski@pstat.ucsb.edu}}}

  \maketitle \thispagestyle{empty}

\begin{abstract}
We investigate jointly modeling Age-specific rates of various causes of death in a multinational setting. We apply Multi-Output Gaussian Processes (MOGP), a spatial machine learning method, to smooth and extrapolate multiple cause-of-death mortality rates across several countries and both genders.
To maintain flexibility and scalability, we investigate MOGPs with Kronecker-structured kernels and latent factors. In particular, we develop a custom multi-level MOGP that leverages the gridded structure of mortality tables to efficiently capture heterogeneity and dependence across different factor inputs. Results are illustrated with datasets from the Human Cause-of-Death Database (HCD). We discuss a case study involving cancer variations in three European nations, and a US-based study that considers eight top-level causes and includes comparison to all-cause analysis. Our models provide insights into the commonality of cause-specific mortality trends and demonstrate the opportunities for respective data fusion.

%Our analysis considers smoothing and extrapolating raw cause-of-death mortality statistics  We propose the application of Multi-output Gaussian Process (MOGP), a spatial machine learning method, to jointly model multiple causes across several countries and both genders.
%
%leverage high-quality mortality datasets in the HCD while capturing the complex mortality dynamics between causes and across countries. Various models within MOGP are exploited to boost the flexibility and scalability  in GP. Particularly one model relaxes the assumption of a global spatial covariance kernel over Age-Year inputs to allow spatial heterogeneity across causes. We also employ a scalable model that makes use of the gridded structure in the mortality table to speed up GP computation, graciously handling datasets with cause-specific mortality segregated by multiple levels. Formulated within a Bayesian paradigm, MOGP provides rich quantification for both in-sample smoothing and out-of-sample forecasts. The close-form expression of the predictive distribution in MOGP allows us to generate the forecast distribution of the aggregate mortality rates. Through various illustrations, we demonstrate the opportunity for data fusion to maximize the predictive gains (more accurate results with higher credibility) over models incorporating causes of death separately. Finally, jointly modeling the mortality rates of multiple causes provides insights about the key drivers of the future aggregate mortality rates in a short and medium range.
\end{abstract}

\section{Background and motivation}

In-depth modeling of the evolution of human mortality necessitates analysis of the prevalent causes of death. This is doubly so for making mortality forecasts into the future across different age groups, populations and genders. In this article we develop a methodology for probabilistic forecasting of cause-specific mortality in a multi-population (primarily interpreted as a multi-national) context. Thus, we simultaneously fit multiple cause-specific longevity surfaces via a spatio-temporal model that accounts for the complex dependencies across causes and countries and across the Age-Year dimensions.

While there have been many works on modeling mortality across several populations \citep{Dong2020,andrew2017,guibert2017forecasting,hyndman2013coherent,KLEINOW2015,li2017coherent,Tsai2019}, as well as an active literature on cause-of-death mortality, there are very few that do both simultaneously. As we detail below, there are many natural reasons for building such a joint model, and this gap is arguably driven by the underlying ``Big Data" methodological challenge. Indeed, with dozens of mortality datasets that are indexed by countries, causes-of-death, genders, etc., developing a scalable approach is daunting. We demonstrate that this issue may be overcome by adapting machine learning approaches, specifically techniques from multi-task learning \citep{Bonilla2008,Caruana1997,letham2019bayesian,mogp_robot2009}. To this end, we employ multi-output Gaussian Processes (MOGP) combined with linear coregionalization. GPs are a kernel-based data-driven regression framework that translates mortality modeling into smoothing and extrapolating an input-output response surface based on noisy observations. It yields a full uncertainty quantification for mortality rates and mortality improvement factors. Coregionalization is a dimension reduction technique that enables efficiently handling many correlated outputs.

This work is a continuation of our series of articles  \cite{LUDKOVSKI2016GAUSSIAN}, \cite{HLZ20} and \cite{huynh2020} that discussed the application of GPs to model all-cause mortality in the single-population and multi-population contexts, respectively. Unlike all-cause mortality in different geographic regions, which tends to exhibit strong correlation and long-term coherence, different causes have less commonality, and thus require a more flexible structure for the respective cross-dependence. Moreover,
joint analysis of 10+ mortality surfaces brings computational scalability challenges of potentially hundreds of model parameters to calibrate. Thus, in order to carry out cause-specific mortality analysis, we make methodological innovations along two directions.  First, we compare two distinct versions of the MOGP that implement dimension reduction by fusing outputs through linear combinations of latent functions: the Semiparametric Latent Factor Model (SLFM) and the Intrinsic Coregionalization Model (ICM). ICM assumes a fixed spatial kernel in Age-Year, while SLFM does not. This distinction corresponds to different assumptions about the structural commonality in the modeled mortality surfaces. We conduct sensitivity analysis to assess these two choices for the tasks of in-sample fitting and of forecasting cause-specific mortality. Second, we implement a multi-level MOGP-ICM model that separates latent factors across the different types (countries, causes, genders, etc.) of categorical inputs describing the populations. The separability assumption in the joint covariance kernel yields a product-type cross-population correlation structure, providing insights about the relationships between the mortality improvement trends.

Our models are driven by the scalability issue which has been a critical obstacle to analyze \emph{in bulk} the large-volume mortality datasets that have become available recently. Thus, to cope with many mortality surfaces, we leverage  the twin pillars of multi-output models \citep{Teh2005SLF,letham2019bayesian,mogp_robot2009}  and the structured Kronecker covariance that mitigates the typical cubic computational complexity of GPs \citep{flaxman15, Gilboa2015, Saatci2011, zhe19a}.

Decomposing all-cause mortality rates  leads to reduced signal-to-noise ratio since less common causes intrinsically have limited death counts. Consequently, cause-specific analysis must contend with much noisier data. One motivation for the MOGP approach is to explicitly enable data fusion across populations,  sharing information to improve model fitting. We demonstrate significant de-noising of mortality experience that successfully captures cause-specific trends, including for causes with low data credibility. We also document the benefit of joint models to reduce model risk, i.e.~improved inference of model hyper-parameters. As another feature, our framework can handle non-rectangular datasets, e.g.~countries with different period coverages. In one of our case studies we exploit this to borrow the most recent data from other countries to update predicted domestic mortality rates.

%We look at several extensions of MOGP-ICM, one of which is the Semi-latent Factor Models (SLFM) \cite{Teh2005SLF}. The SLFM does not require the assumption of sharing the spatial kernel over Age-Year inputs among populations being modelled. We conduct sensitivity analysis to distinguish the two models for tasks of in-sample fitting and forecasting. Lastly, we investigated the multi-level MOGP-ICM to cope with multidimensional categorical inputs. Given GP suffers the cubic computational complexity, we take advantage of the structured covariance kernel via Kronecker methods to cope with the scalability issue. The application of the structured Kronecker covariance has been exploited for efficient GPs in various setting, see \cite{flaxman15, Gilboa2015, Saatci2011}.

\textbf{Literature Review}

One of the few works approaching cause-specific mortality modeling within a multi-population context is the recent study in \cite{Lyu2021}. The authors introduced a nested model in the spirit of \cite{LI2005} to jointly model major causes from three European countries by capturing the cross-cause and cross-country dependencies through common factors.

Since a given death is associated with a single cause-of-death, direct dependence among causes is not observable. As such, many researchers choose to model and forecast each cause in isolation. A variety of forecasting methods for individual causes are employed: univariate time-series, such as ARIMA methods in \cite{Caselli1996}, \cite{McNown1993}, \cite{McNown1992}, dynamic parametrization in \cite{Tabeau1999}, least squares methods and variations of the Lee-Carter model in \cite{Caselli2019}.
To capture dependence between causes, a common approach is via copulas within the framework of dependent competing risks. The main effort is to characterize the joint distribution of survival times in terms of un-observed cause-specific mortality rates, see \cite{Dimitrova2013}, \cite{Wilke2010} and \cite{Li2019}. Alternatively, \cite{Alai2018} utilized multinomial logistic regression. Outside the competing risks framework, \cite{Severine2013} proposed a multivariate Vector Error Correction  time-series model to examine the existing cointegration relationships.
% The cointegration was characterized by the linear combination of  different causes exhibiting stationary property in the long run, in turn making VEC a more efficient technique to model multiple time-series datasets that individually are non-stationary.
Another approach is to link multiple causes through a list of clinical factors, then applying a stochastic model to forecast these factors \citep{Foreman2018Forecasting}. %Such studies however require extensive resources from teams of experts in many research fields to large datasets that contain highly-detailed clinical information.

Several studies relied on compositional data analysis to achieve coherence in the sense that cause-specific forecasts sum to the all-cause forecast. This entails modeling the by-cause distribution of deaths, directly incorporating dependence between causes, see \cite{Bergeron2017}, \cite{Kjaergaard2019}. %, \cite{Oeppen2008}.
%
%the coherent forecasts in the sense that cause-specific deaths must sum to the total number of deaths. Rather than modeling and forecasting cause-specific mortality rates, the key innovation was to model cause-specific distribution of deaths as a way to directly incorporate dependence between causes, see \cite{Bergeron2017}, \cite{Kjaergaard2019}, \cite{Oeppen2008}.
%
We refer to \cite{Wilmoth1995,Caselli2019,Tabeau1999} for discussions on the usefulness and limitations in using cause-specific models to project all-cause mortality.

%the decomposition of aggregate mortality into causes of death for better mean forecasts has not been consistently observed. While \cite{Crimmins1981} found segregating causes of death achieve more accurate forecasts, \cite{Wilmoth1995} illustrated that the forecasted all-cause mortality using cause-specific models were more pessimistic than those based on all-cause models. Other limitations in using cause-specific models to project all-cause mortality can be found in \cite{Alho1991}, \cite{Booth2008}, and \cite{Caselli1996}.

The remainder of this paper is organized as follows. Section \ref{sec:hcd} introduces cause-of-death mortality datasets from the HCD. Section~\ref{sec:method} describes the MOGP-ICM framework and its extensions within multinational context. Section~\ref{sec:results} focuses on how MOGPs can maximize predictive gains over single-population models and provide insights about the projected trends of aggregate mortality. Section~\ref{sec:multi-national} compares the results from Multi- and Single-level ICM. Finally, Section~\ref{sec:summary_cod} concludes with main findings and directions for further analysis.

\section{The Human Cause-of-death Database}\label{sec:hcd}

The Human Cause-of-death Database (HCD) \citep{HCD} provides detailed cause-specific mortality data for more than a dozen developed countries. The HCD offers three levels for the classification of causes. The short list (with 16 broad categories) and the intermediate list (103 categories) are the same across countries, while the full list is country-specific. The data for each country is organized by calendar years, age groups, gender, and causes. Datasets for different countries do not line up, both due to historical availability and different timelines for updates, see Figure \ref{fig:hcd} in Appendix \ref{app:hcd_cover}. % displays the respective historical period for the 16 countries available in the HCD.

As part of their extensive and well-documented post-processing the HCD conducted a series of bridge-coding studies to reduce disruptions in mortality trends due to changes in the International Classification of Disease (ICD). The current 10th Revision of the ICD is far more detailed with the addition of 8,000 categories compared to the 9th Revision \citep{Anderson_icd2001}. As cause-of-death records switched to the 10th Revision, death counts were shifted among some categories. To minimize such jumps, the HCD reconstructed death assignments between the old and the new ICD Revisions; this reconciliation is already part of the HCD datasets we used.

In our first case study we analyze sub-categories of Cancer, available from the 103-category intermediate list. Being the leading cause of death worldwide \citep{cancer_who2021}, it is useful to understand the trends in cancer mortality for different age groups and explore the dependence between its common variations. Moreover, cancer types generally do not feature any substantial trend discontinuity due to ICD revisions \citep{Anderson_icd2001}, allowing a better assessment of model performance in terms of in-sample smoothing and out-of-sample forecasting.

\begin{figure}[!t]
    \centering
    \includegraphics[width=10cm]{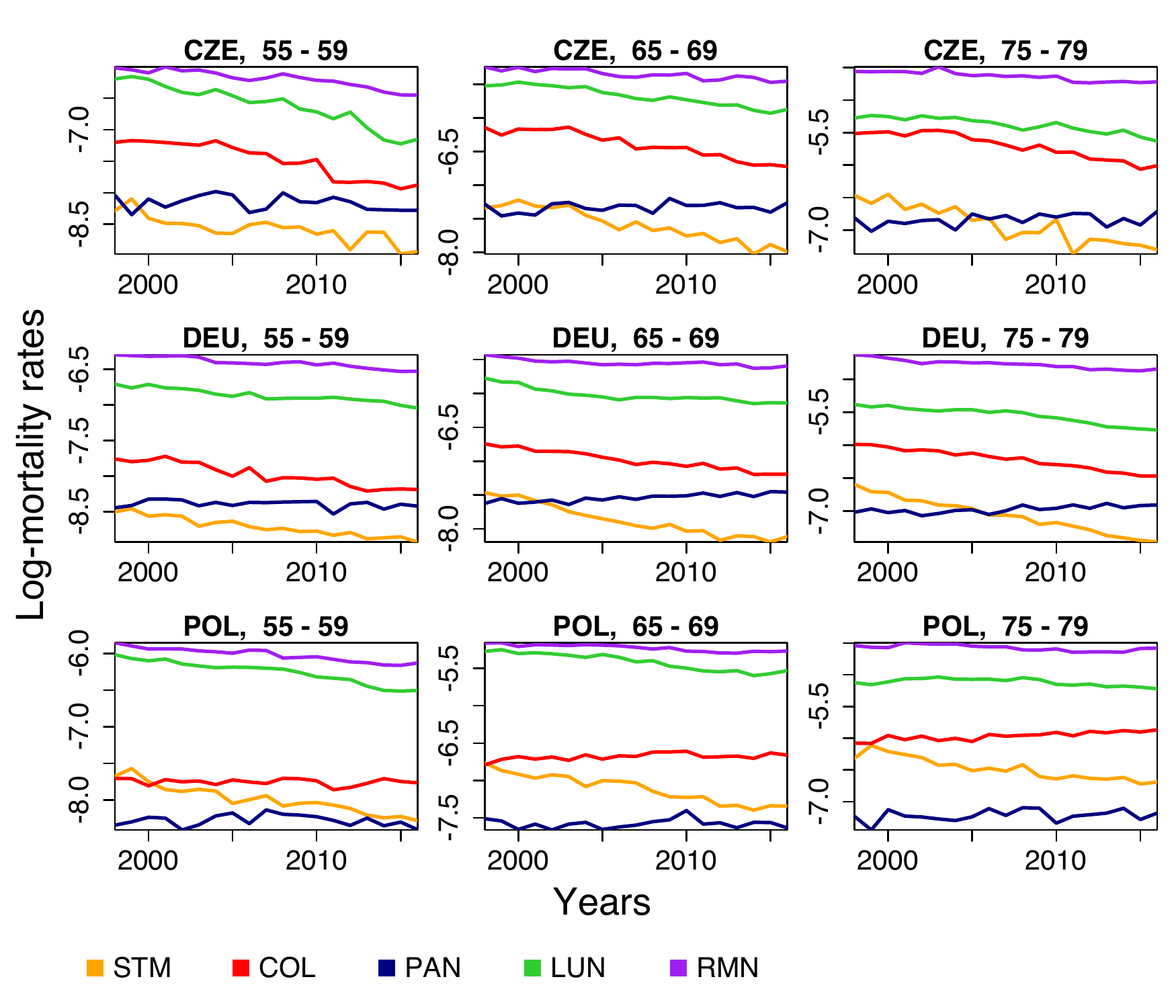}
    \caption{Log-mortality rates of selected Cancers by Country \& Age groups among Males. Note the different $y$-axes in each panel. Source: HCD. \label{fig:raw-logmort}}
\end{figure}

For our testbed we select 5 variations of cancer: lung and bronchus (LUN), colon and rectum (COL), pancreas (PAN), stomach (STM), and all other cancers (RMN) from 3 countries in the HCD: Czech Republic (CZE), Germany (DEU), and Poland (POL). The chosen Cancer causes-of-death are common in both Male and Female populations, enabling us to jointly model mortality rates across Cause, Country, and Gender factors. For example, we exclude breast cancer from the study as the respective male counts are very limited. Age in the HCD is formatted as discrete five-year age groups; we treat it as a continuous covariate encoded via the respective group average (52 for Age group 50--54, 57 for group 55--59, etc).  In Figure \ref{fig:raw-logmort}, we visualize the raw log-mortality rates across different cancer types in Male populations from Czech Rep., Germany, and Poland. Lung and Colorectal are the leading causes of cancer deaths, while Pancreatic and Stomach are less prevalent and exhibit more volatility. Figure~\ref{fig:age_quad} in the Appendix visualizes the respective patterns in Age; we observe a strong convexity in Age, especially for LUN.

%\subsection{CDC Wonder US Cause-of-Death Dataset}

Our second case study uses a subset of the HCD data for United States, based on the CDC's National Center for Health Statistics, where we decompose all-cause mortality into its major categories. This analysis is inspired by a similar study conducted by the SOA \citep{codUS2019} that employed a multivariate Lee-Carter model. The SOA study looked at 11 major causes, all Age groups between 0 and 95+ in 1999--2016 with data combined from the HCD and the Global Burden of Disease project. %Through cause-of-death analysis, their key objectives are to understand the interaction of causes of death and the key drivers of the future aggregate mortality dynamics in both short and medium term. Distribution of causes and various data visualization are provided in their report. For demonstration, we employed our MOGP framework
We restrict analysis to Ages 40--69 and years 1999--2018, and moreover reduce to the eight most common causes for the examined age groups:  Heart (HEA), Stroke (STK), Diabetes (DIA), Cancers not induced by smoking (CAN), (lung) Cancers induced by smoking (CANL), Respiratory (RES), Drug abuse (DRU), and all other remaining causes (RMN). The cause mapping for this case study closely follows the SOA guidance.

\subsection{Stacking Sub-Populations}

For joint modelling purposes, datasets are stacked together. Generically we have a total of $L$ populations, indexed by the subscript $l=1,2,\ldots, L$. When needed, to indicate country, cause, and gender factors, we re-index by $l=(c,s,g)$. Throughout, the two main independent variables are Age and Year, $(x^i_{ag},x^i_{yr})$,  and the observed mortality rate in the $l$-th population is denoted by:
\begin{equation}
    \mu^i_{l} := \dfrac{\text{Death counts at }(x^i_{ag},x^i_{yr})\text{ of type }l}{\text{Exposed-to-risk counts at }(x^i_{ag},x^i_{yr})}= \dfrac{D^i_{l}}{E^i_{l}},
\end{equation}
where $D^i_{l}$ is the number of deaths in the $l$th population for the $i$th observation with the corresponding Age \& Year inputs. The denominator $E^i_{l}$ refers to the corresponding exposed-to-risk counts. Note that different causes in the same country-gender combination with have the same $E^i_l$, but different $D^i_l$'s.
%The all-cause mortality rate for the $i$th observation in population $l$ is then given by:
%\begin{equation}
%    \mu^i_{l}:=\sum_{s} \mu^i_{s,l} = \dfrac{\sum_s D^i_{s,l}}{E^i_{l}}.
%\end{equation}
%
Our GP models work with log-mortality rates so that our data is $$y^i_{l} = \log \mu^i_{l}.$$
The overall dataset is then represented as $(x^i, y^i_l)$, $i=1,\ldots,N_l$, $l=1,\ldots, L$.
 %Let $L$ be the total number of populations and let $\mathcal{D}_l = \{(x^i,y^i)\}_{i=1}^N, l=1,\ldots, L$ be the dataset for the $l$th population, interpreted as a given  combination of factor inputs.

% \begin{figure}[!t]
%     \centering
%     \includegraphics[width=16.5cm, height = 5cm]{cancer-distribution.pdf}
%     \caption{2016 cancer statistics in Male/Female populations in 3 countries considered in this study.  \label{fig:cancer-dist}}
% \end{figure}

% \begin{figure}
%     \centering
%     % \captionsetup{justification=centering}
%     From multi-country ICM models:\\
%     $r_{CZE,GER} = 0.42$, $r_{CZE,POL} = 0.90, r_{GER, POL} = 0.76$
%     {{\includegraphics[width=15cm]{stomach-age-pattern.pdf}}} \\
%     $r_{CZE,GER} = 0.96$, $r_{CZE,POL} = 0.03, r_{GER, POL} = -0.22$
%     {{\includegraphics[width=15cm]{colorectal-age-pattern.pdf}}} \\
%     $r_{CZE,GER} = 0.84$, $r_{CZE,POL} = 0.89, r_{GER, POL} = 0.89$
%     {{\includegraphics[width=15cm]{lung-age-pattern.pdf}}} \\
%     $r_{CZE,GER} = 0.89$, $r_{CZE,POL} =  0.99, r_{GER, POL} = 0.85$
%     {{\includegraphics[width=15cm]{Prostate-age-pattern.pdf}}}
%     \caption{Comparison the log-mortality trend in 3 countries by cancer type and age group.}
% \end{figure}

\section{Methodology}
\label{sec:method}
\subsection{Gaussian Process Regression for Mortality Tables}

Consider the non-parametric additive regression model for the $L$ populations to be jointly analyzed:
\begin{equation}
\label{eqn:regression}
    y^i_l = f_l(x^i) + \epsilon^i_l, \qquad\qquad i=1,\ldots, N,
\end{equation}
where $x^i$ represents an individual entry in the mortality table (indexed by Age, Year), $y^i_l$ is the observed output in the $l$th population, and $f_l(\cdot)$, $l=1,\ldots,L$ is the underlying latent log-mortality surface,  obscured with the observation noise $\epsilon^i_l$.

We first summarize the \emph{spatial} structure that concerns the dependence of mortality rates  as a function of Age and Year. Momentarily focusing on a single-output Gaussian Process (SOGP) regression,  we put a GP prior on the latent function $f_l \sim \mathcal{GP}(m,C)$, meaning that any finite vector $f_l(\mathbf{x})=(f_l(x^1),\ldots,f_l(x^n))$ at $n$ inputs follows the multivariate Gaussian distribution
\begin{equation*}
    f_l(x^1),\ldots ,f_l(x^n) \sim \mathcal{N}\big(\mathbf{m_l(x)},\mathbf{C}_l(\mathbf{x,x})\big),
\end{equation*}
where ${m}_l(x)=\mathbb{E}[f_l(x)]$ is the mean vector of size $n$ and ${C_l(x,x')}=\mathbb{E}[(f_l({x})-{m_l(x)})(f_l({x'})-{m_l(x')})]$ is the $n$-by-$n$ covariance matrix. All properties of a GP are thus completely described by its mean and covariance functions.

The functions $m_l(\cdot), C_l(\cdot, \cdot)$ characterize our prior beliefs about the response surface $f_l$. \textit{The covariance kernel} of the GP defines a similarity between pairs of data points. It characterizes the smoothing process by determining the influence of observations on the distribution of the output. Data points that are close are expected to behave more similarly than data points that are farther away. In terms of spatial dependence on (Age, Year) we concentrate on a common family of covariance functions known as the Mat\'ern class, equipped with automatic relevance determination. Specifically, the Mat\'ern-5/2 kernel defines the covariance between two mortality table entries $x, x'$ as follows:
\begin{equation}
    \label{eqn:kernel}
        C_l(x,x') = \prod_{k \in \{ag,yr\}}
        \bigg(1+\dfrac{\sqrt{5}}{\theta_{k,l}}|x_k-x_k'|+\dfrac{5}{3\theta_{k,l}^2}|x_k-x_k'|^2\bigg)
        \exp\bigg(-\dfrac{\sqrt{5}}{\theta_{k,l}}|x_k-x_k'|\bigg).
\end{equation}
This kernel is parametrized by the Age lengthscale $\theta_{ag,l}$ and the Year lengthscale $\theta_{yr,l}$.

\textit{The mean function} $m_l(x)$ describes the relevant trends in log-mortality rates. We fit a parametric prior mean to capture the long-term longevity features: $m_l(x)=\beta_{0,l}+\sum_{j=1}^p \beta_{j,l} h_j(x),$ where $h_j$'s are given basis functions and the $\beta_{j,l}$'s are unknown coefficients.
For example, to reflect a quadratic trend in Age (Figure \ref{fig:age_quad} in Appendix \ref{app:raw-age}) and linear trend in Year dimension we use:
\begin{equation}
\label{eq:mean_sogp}
    m_l(x^i) = \beta_{0,l} + \beta^{ag}_{1,l}x^i_{ag} + \beta^{ag}_{2,l}(x^i_{ag})^2 + \beta_{yr,l}x^i_{yr}.
\end{equation}

%\textbf{Mean function:} We make $m(x)$ to be population-specific in order to maximize model flexibility in describing the mortality trend of each population. Specifically, we incorporate the interaction terms between Population(s) and both Age and Year covariates in \eqref{eq:mean_sogp}:
%\begin{equation}
%     m(x^i) = \beta_0 + \sum_{l=2}^L\beta^{ag}_{1,l}x^i_{ag}x^i_{pop,l} + \sum_{l=2}^L\beta^{ag}_{2,l}(x^i_{ag})^2x^i_{pop,l} + \sum_{l=1}^L\beta_{yr,l}x^i_{yr}x^i_{pop,l}.
%\end{equation}

We are interested in the posterior distribution for $\mathbf{f_*} \equiv f_l(\mathbf{x_*})$ at specified inputs $\mathbf{x_*}$ given the dataset $\mathbf{y}_l = (y^{1}_l, \ldots, y^N_l)$, $p(\mathbf{f_*|y}_l)$, in other words the likelihood of the true response surface being $\mathbf{f_*}$ given what we have observed. Using $\Cov(y^i_l,y^j_l)=\Cov(f_l(x^i),f_l(x^j))+\sigma^2_l\delta(x^i,x^j)$ where $\delta(x^i,x^j)$ is the Kronecker delta, we have the Gaussian observation likelihood $\mathbf{y}_l \sim \mathcal{N}(\mathbf{m_l(x)},\mathbf{C_l(x,x)}+\mathbf{\Sigma}_l)$ where the error terms are assumed to be independent and Gaussian-distributed, $(\epsilon^1_l,\ldots,\epsilon^N_l) \sim \mathcal{N}(\mathbf{0}, \mathbf{\Sigma}_l = \text{diag}(\sigma_l^2))$. Incorporating the evidence from observations, as reflected in the likelihood function and the prior, we obtain the Gaussian posterior
\begin{equation*}
    p(\mathbf{f_*|y}_l) \sim \mathcal{N}(\mathbf{m_*(x_*,l),C_{*,l}(x_*,x_*)}).
\end{equation*}
The Universal Kriging equations below provide not only the posterior mean $m_*(\cdot,l)$ and posterior variance $s^2_*(\cdot,l)$ but also the estimated coefficients $\bm{\beta}_l=(\beta_{1,l},\ldots,\beta_{p,l})^T$. Let $\mathbf{h}(x)=\big(h_1(x), \ldots,h_p(x)\big)$, $\mathbf{H}=\big(\mathbf{h}(x^1),\ldots,\mathbf{h}(x^N)\big)$, and $\mathbf{D=(C+\Sigma)^{-1}H}$ where $\mathbf{C}$ is the covariance matrix $C_l(x^i,x^j)_{i,j=1}^N$. The posterior mean of $\bm{\beta}_l$ along with the predicted posterior mean $m_*(x_*,l)$ and respective variance $s_*^2(x_*,l)=C_{*,l}(x_*,x_*)$ for any input $x_*$ are:
\begin{align}
    \bm{\hat{\beta}}_l &\ =(\mathbf{H}^T\mathbf{D})^{-1}\mathbf{H}^T\mathbf{(C+\Sigma)}^{-1}\mathbf{y}; \\ \label{eq:GP-mean}
    m_*(x_*,l) &\ = \mathbf{h}(x_*)^T\bm{\hat{\beta}}+\mathbf{c}(x_*)^T\mathbf{(C+\Sigma)}^{-1}(\mathbf{y-H}\bm{\hat{\beta}}); \\ \label{eq:GP-var}
    s^2_*(x_*,l) &\ =  (\mathbf{h}(x_*)^T -\mathbf{c}(x_*)^T\mathbf{D})^T(\mathbf{H}^T\mathbf{D})^{-1}
    (\mathbf{h}(x_*)^T-\mathbf{c}(x_*)^T\mathbf{D})
\end{align}
where $\mathbf{c}(x_*)=(C_l(x^1,x_*), \ldots, C_l(x^N, x_*))$ is the vector of covariances between inputs in the training set and desired test input $x_*$. % $\mathbf{C}_{**}$ is the covariance matrix of $f(\bx_*)$, $\mathbf{\Sigma_{**}}=\text{diag}(\sigma^2)$ is the noise variance matrix of the test inputs $\bx_*$.
Note that the predictive distribution of observation ${y_l}$ at $x_*$ is similarly obtained as ${y_l} \sim \mathcal{N}(m_{*}(x_*,l),s^2_*(x_*,l)+\sigma^2_l)$.

Below we use $m_*(x_*,l)$ as the model prediction for the respective (log)-mortality rate of the $l$th population in cell $x_*$, and $s_*(x_*, l)$ as the corresponding posterior uncertainty which is used to obtain predictive quantiles around the former prediction.

\subsection{Semi-parametric Latent Factor Model}

The vector-valued latent response variable over the Age-Year input space is defined as $\mathbf{f(x)} = (f_1(\mathbf{x}),\ldots,f_L(\mathbf{x}))$, where the functions $\{f_l({x})\}_{l=1}^L$ are the log-mortality surfaces for the corresponding $l$th population. Similar to single-population GP above, we place a GP prior over the latent function $\mathbf{f}$ such that:
$$\mathbf{f} \sim \mathcal{GP}(\mathbf{m,C}),$$
where $\mathbf{m}$ is the mean vector function whose elements $\{m_l({x})\}_{l=1}^L$ are the mean functions of each output, and $\mathbf{C}$ is the fused covariance matrix. 
This implies that we separate the ``spatial" dependence, encoded in $x$ from the inter-task dependence encoded in $l$. While $x$ has a natural Euclidean metric that is used in \eqref{eqn:kernel}, the population indices $l$ are generally of factor-type. Hence,  to describe the dependence across $L$ outputs we need $L(L-1)/2$ hyperparameters $C^{(f)}_{lk}, 1 \le l,k \le L$ which becomes inefficient and unstable beyond 3-4 populations. Thus, with an eye towards reducing the number of hyperparameters, additional structure is needed for $C^{(f)}$.

In the Semiparametric Latent Factor Model (SLFM) dating back to~\cite{Teh2005SLF}, each output $f_l({x})$ is assumed to be a linear combination of $Q$ latent functions:
\begin{equation}
    f_l({x}) = \sum_{q=1}^Q a_{l,q} u_q({x}),
    \label{eqn:sfm-linear}
\end{equation}
where $u_q({x})$'s are independent realizations from GP priors with distinct covariances $C^{(u)}_q({x,x'})$, and $a_{l,q} \in \mathbb{R}$'s are the factor loadings ($q=1,\ldots,Q$), considered part of the kernel hyperparameter space $\Theta$. Thus, the semiparametric name of the model comes from the combination of a nonparametric component (several GPs) and a parametric linear mixing of the functions $u_q({x})$.

The role of $Q \le L$ is to achieve dimension reduction for the correlation structure across the $f_l$'s. Let $\mathbf{a}_q = (a_{1,q}, \ldots, a_{L,q})^T$ be the vector representing the collection of coefficients associated with the $q$th latent function across the $L$ outputs. Then the covariance of the vector-valued function $\mathbf{f(x)} = \sum_{q=1}^Q\mathbf{a}_qu_q(\mathbf{x})$ is:
\begin{align}
    \notag
    \Cov(\mathbf{f}(x),\mathbf{f}(x')) & = \sum_{q=1}^Q (\mathbf{a}_q\mathbf{a}_q^T) \otimes C^{(u)}_q({x,x'}) \\
    & = \sum_{q=1}^Q (A_qA_q^T) \otimes C^{(u)}_q({x,x'}) = \sum_{q=1}^Q B_q \otimes C^{(u)}_q({x,x'}) \label{eqn:slfm-def}
\end{align}
where $\otimes$ symbolizes the Kronecker product, $A_q = \mathbf{a}_q = (a_{1,q},\ldots, a_{L,q})^T$ and each $B_q$ has rank one. 

\subsection{Intrinsic Coregionalization Model}

Similar to SLFM, ICM assumes each output function $f_l({x})$ is generated from a common pool of $Q$ latent functions, cf.~\eqref{eqn:sfm-linear}. However, the latent $u_q({x})$ all share the \emph{same} GP prior with the covariance kernel $C^{(u)}({x,x'})$. Then, the covariance for $\mathbf{f}(x)$ is:
\begin{align}
    \Cov(\mathbf{f}(x),\mathbf{f}(x')) & =  \bigg(\sum_{q=1}^Q \mathbf{a}_q\mathbf{a}_q^T\bigg) \otimes \Cov(u_q({x}),u_q({x'}))
    %= AA^T \otimes C^{(u)}(\mathbf{x,x')} \\
     = B \otimes C^{(u)}({x,x'}), \label{eqn:icm-def}
\end{align}
where $B = AA^T \in \mathbb{R}^{L \times L}$ has rank $Q$. In other words, the cross-output correlation is of rank $Q \le L$, while the spatial covariance of each output is the same. The $l$th element in the diagonal of the cross-covariance matrix  $B$ (or $\sum_{q=1}^Q B_q$ in SLFM) represents the process variance of $f_l(\cdot)$. %We can infer the correlation matrix $R$ and extract $r_{l_1,l_2}$'s ($1 \leq l_1,l_2 \leq L$) accordingly, see Appendix \ref{app:cross_corr}.
Since $B   = \sum_{q=1}^Q \mathbf{a}_q\mathbf{a}_q^T$, the individual entries are $B_{l,k} = \sum_{q=1}^Q a_{l,q}a_{k,q}$ and the diagonals are $B_{l,l} = \eta^2_{l} = \sum_{q=1}^Q a^2_{l,q}$, $1 \leq l \leq L$.
%We can expand the cross-covariance kernel $B$ in Eq \eqref{eqn:icm-def} such as:
%\begin{align*}
%    B = AA^T & = \sum_{q=1}^Q \mathbf{a}_q\mathbf{a}_q^T \\
%             & = \begin{bmatrix}
%             \sum_{1=q}^Q a^2_{1,q} & \sum_{1=q}^Q a_{1,q}a_{2,q} & \dots & \sum_{1=q}^Q a_{1,q}a_{L,q} \\
%             \sum_{1=q}^Q a_{2,q}a_{1,q} &  \sum_{1=q}^Q a^2_{2,q} & \dots &
%             \sum_{1=q}^Q a_{2,q}a_{L,q} \\
%             \vdots & \vdots & \ddots & \vdots \\
%             \sum_{1=q}^Q a_{L,q}a_{1,q} & \sum_{1=q}^Q a_{L,q}a_{2,q} & \dots &
%             \sum_{1=q}^Q a^2_{L,q} \\
%             \end{bmatrix} \\
%\end{align*}
%The process variance of the l$th$ function output is the l$th$ element of the diagonal in $B$, or  $\eta^2_{l} = \sum_{q=1}^Q a^2_{l,q}$, $1 \leq l \leq L$.
We can similarly infer the correlation matrix $R=(r_{l_1,l_2})$ between population $l_1$ and $l_2$ ($1 \leq l_1,l_2 \leq L$); for ICM it is
\begin{align}\label{eq:R}
r_{l_1,l_2}:=\frac{\Cov(f_{l_1}({x}),f_{l_2}({x}))}{\sqrt{\Var(f_{l_1}({x}))\times\Var(f_{l_2}({x}))}} =  \frac{B_{l_1,l_2}}{\sqrt{ B_{l_1,l_1} B_{l_2, l_2}}} = \frac{\sum_{q=1}^Q a_{l_1,q}a_{l_2,q}}{\sqrt
{\big(\sum_{q=1}^Qa^2_{l_1,q}\big)\big(\sum_{q=1}^Qa^2_{l_2,q}\big)}}.
\end{align}

% Thus, the number of hyperparameters required to estimate the cross-covariance kernel $B$ is $Q \times L$.

In both SLFM and ICM, the fused covariance kernel belongs to the class of separable kernels \citep{alvarez2011kernels} and decouples using the Kronecker product into: (1) the coregionalization matrices $B_q$ that measure the interaction between different outputs and (2) the spatial covariance over Age-Year dimensions $C^{(u)}_q({x,x'})$, cf.~Eqs \eqref{eqn:slfm-def} \& \eqref{eqn:icm-def}.  %Note that the fused covariance kernel is positive semi-definite when the coregionalization matrices $B_q$ are positive semi-definite and $C^{(u)}_q({x,x'})$ is a valid covariance function. %Both models guarantee that the coregionalization matrices $B_q$ are positive semi-definite by using the Cholesky decomposition $B_q = A_qA_q^T$, with $A_q \in \mathbb{R}^{L \times Q}$.
In ICM, all $L$ populations share the same spatial covariance kernel $C^{(u)}({x,x'})$. Such assumption of a common spatial covariance over Age-Year inputs links to the concept of commonality in the mortality structure (but not levels) across populations. Compared to ICM, SLFM offers more flexibility at the cost of adding more hyperparameters: the total number of kernel hyperparameters in the fused covariance matrix $\mathbf{C}$ of SLFM is $QL+2Q$ vis-a-vis $QL+2$ hyperparameters in the ICM.

% \textbf{Non-Rectangular Data Sets.} We have discussed the use of ICM for isotropic data. The ICM framework can be extended to deal with {partially heterotropic data} where only a portion of $L$ inputs are available and which arises in HCD due to missing data especially at the most recent years. Let $M'$ be the number of distinct inputs across $L$ populations and $M = N_1+\ldots+N_L$ be the number of observations in training data. We consider the setting that $M' < ML$ so that for some inputs not all $L$ outputs are observed. Define the vector-valued ``complete data" function $\mathbf{f(x)}$, with $\mathbf{f(x)} \in \mathbbm{R}^{LM' \times 1}$. We further introduce $\mathbf{f}^o\mathbf{(x)}$ as the vector-valued function corresponding to the observed outputs, $\mathbf{f}^o\mathbf{(x)} \in \mathbbm{R}^{M \times 1}$. The relation between $\mathbf{f(x)}$ and $\mathbf{f}^o\mathbf{(x)}$ is formulated through the ``communication" matrix $S$,
% $\mathbf{f}^o\mathbf{(x)} = S^T\mathbf{f(x)}$,
% where $S \in \mathbbm{R}^{LM' \times M}$. The column vectors in $S$ are orthonormal with values of 0 and 1 to eliminate the unobserved outputs, see \cite{Skolidis2011heterotopic}. Applying linear transformation to a MVN vector, we can then identify the distribution of $\mathbf{\mathbf{f}^o\mathbf{(x)}}$ as a GP with covariance: $$\Cov(\mathbf{f}^o\mathbf{(x)},\mathbf{f}^o\mathbf{(x')})=S^T \Cov(\mathbf{f(x)},\mathbf{f(x')})S = S^T(B \otimes K)S,$$ recovering the Kronecker structure.

\textbf{Selecting rank $Q$}. As $Q$ is not one of the hyperparameters to be optimized, ad hoc ways are needed to pick it.  We use the Bayesian Information Criterion (BIC) to select rank $Q$ that produces the most parsimonious model, see \cite{mogp_robot2009} and \cite{huynh2020}. As discussed in \cite{Bonilla2008}, taking $Q<L$ in ICM corresponds
to finding a rank-$Q$ approximation (based on an incomplete Cholesky decomposition) to the full-rank $C^{(f)}$. A similar interpretation holds for SLFM and the respective $B_q$'s.  While attractive for dimension reduction and computational speed-up, low $Q$ may not be adequate to describe the overall dependence structure and hence clashes with the original goal of capturing the variability present in the fused mortality dataset. In particular, we observe that BIC tends to select $Q \in \{2,3\}$ which may be too small for $L \ge 5$. Based on our case studies, we recommend $Q \in \{3, 4, 5\}$ for maximizing predictive performance.

%Effectively, both approaches impose low-rank approximations to $C^{(f)}$.

%This approach often yields low rank options and in such scenarios, MOGP-ICM/SLFM can be thought of as an attractive dimension reduction approach, since it reduces from $\mathcal{O}(L^2)$ to $\mathcal{O}(LQ)$ parameters. As suggested in \cite{Bonilla2008}, a low rank $Q$ implies we estimate the coregionalzation matrix $B_q$'s using the incomplete Cholesky decomposition to speed up GP computations. However, the original goal of the latent $u_q$'s is to summarize  When the total number of populations is very large, small rank values 

\subsection{Multi-Level ICM for Scalable GPs}

In the situation when we have multi-dimensional factor inputs (e.g.~Cause and Gender together), one approach is to combine all factor inputs into a single covariate with $L$ distinct outputs prior to applying ICM or SLFM. When $L$ grows large, ICM becomes less feasible due to its time complexity $\mathcal{O}(N^3 L Q^2)$ \cite{Bonilla2008}. In this section, we develop the structured Kronecker product kernel (multi-level ICM in \cite{liu2019gaussian, zhe19a}) to mitigate this scalability issue in GP. The structured covariance kernel exploits the fact that mortality tables are gridded along each factor dimension.

%Let $L$ be the total number of outputs we want to jointly model.
We  express the total number of outputs $L$ as the product across $P$ types of categorical inputs, $L =\prod_{p=1}^P L_p$,
where $L_p$ is the number of levels within the $p$th categorical input. We then decompose  the cross-population covariance $\Tilde{B}$ as the Kronecker product:
\begin{align}
    \label{eqn:hier_icm_kern}
    \Tilde{B} = \bigotimes_{p=1}^P \Tilde{B}_p
    % \Tilde{B}_1 \otimes \ldots \otimes \Tilde{B}_P
\end{align}
where  $\Tilde{B}_p,~1\leq p \leq P$ refers to the cross-covariance matrix between sub-populations within the $p$th categorical input, taken to have rank $Q_p \le L_p$.
 Directly marginalizing the cross-covariance matrices yields a convenient interpretation of the correlation between sub-populations within a factor input, and moreover allows for separate estimation of each cross-covariance sub-matrix $\Tilde{B}_p$, $1\leq p \leq P$.

The multi-level ICM setup implies that each output $f_l({x})$ is the weighted combination of $Q_1 \times \ldots \times Q_P$ independent latent functions, all with the spatial covariance kernel $C^{(u)}({x,x'})$:
\begin{equation}
    f_l({x}) = \sum_{j=1}^{Q_1\times \ldots \times Q_P} a_{l,j} u_j({x}).
\end{equation}
The improvement in scalability of the multi-level ICM can be analyzed via the ranks $Q_p$ of the cross-covariance sub-matrices. Thanks to the Kronecker product's property, $\text{rank}(\Tilde{B}) = \prod_{p=1}^P \text{rank}(\Tilde{B}_p) = \prod_{p=1}^P Q_p$
and using Cholesky decomposition, Eq.~\eqref{eqn:hier_icm_kern} can be rewritten as:
\begin{align}
    \Tilde{B} = \Tilde{A}\Tilde{A}^T = \bigotimes_{p=1}^P \Tilde{B}_p
     = \bigotimes_{p=1}^P \big(\Tilde{A}_p \Tilde{A}_p^T \big)
    =  \bigg(\bigotimes_{p=1}^P \Tilde{A}_p \bigg)\bigg(\bigotimes_{p=1}^P \Tilde{A}_p \bigg)^T,
    \label{eq:multi-icm}
\end{align}
where $\Tilde{A}_p=(\mathbf{a}_1^p,\ldots,\mathbf{a}_{Q_p}^p)$, $1 \leq p \leq P$, and each vector $\mathbf{a}_k^p=(a_{1,k}^p,\ldots,a_{L_p,k}^p)^T$, ($1 \leq k \leq Q_p$) represents the collection of scalar coefficients associated with the $k$th latent function across $L_p$ sub-populations in the $p$th categorical input. Thus, the number of hyperparameters required to estimate the cross-covariance $\Tilde{B}$ is $\sum_{p=1}^P Q_p L_p$, which can be much lower than for single-level ICM when $L$ is large. Note that when $\text{rank}(B) < \text{rank}(\Tilde{B})$, the multi-level ICM utilises more latent functions to generate the model outputs, compensating for the imposed structure in $\Tilde{B}$ in \eqref{eq:multi-icm}. In terms of overall complexity, multi-level ICM requires $\mathcal{O}(N^3 (\sum_p L_p Q_p^2) )$ time compared to $\mathcal{O}( N^3 L Q^2)$ for single-level ICM.

%\cite{Saatci2011} show that the typical ICM approach requires $\mathcal{O}(N^3)$ time while multi-level ICM only requires $\mathcal{O}(PN^{(P+1)/P})$ operations.

\textit{Remark:} %\begin{rem}
In the typical situation, $L_p$'s are small and so it is feasible to consider full-rank multi-level ICM, i.e.~$Q_p = L_p$. Otherwise, \eqref{eq:multi-icm} allows to exploit simultaneously the Kronecker product structure, as well as the low-rank approximation. See Table~\ref{tbl:mogp-rank} for results on the impact of $Q_p$ values in multi-level ICM.
%
%\edit{In Multi-level ICM, we can employ the low-rank approximation approach to estimate the cross-covariance matrices $B_p$'s in \eqref{eq:multi-icm} by choosing $Q_p = \text{rank}(B_p) < L_p$. }
%\end{rem}

%The structured Kronecker product in the multi-level ICM presents substantial scalability. Indeed, single-level ICM can have up to $\binom{L}{2}$ hyperparameters, which is much larger  than  $\sum_{p=1}^P \binom{L_p}{2}$ hyperparameters that the multi-level ICM at most has. In the low-rank setups, it is trickier to compare the two methods. When

\subsection{MOGP Hyperparameters}

To implement a GP model requires specifying its hyperparameters. Note that actual inference reduces to linear-algebraic formulas in \eqref{eq:GP-mean}-\eqref{eq:GP-var}, and the modeling task is to learn the spatial covariance, namely the mean and kernel functions.

\textbf{Mean function:} We make the prior $m_l(x)$ to be population-specific in order to maximize model flexibility in describing the mortality trend of each population.
Thus, we have $3 L+1$ coefficients $\bm{\beta}=(\beta_{0},\beta^{ag}_{1,l}, \beta^{ag}_{2,l}, \beta_{yr,l}: l=1,\ldots, L)$, cf.~\eqref{eq:mean_sogp}.
%\begin{equation}
%     m(x^i, l) := \beta_0 + \beta^{ag}_{1,l}x^i_{ag} + \beta^{ag}_{2,l}(x^i_{ag})^2 + \beta_{yr,l}x^i_{yr}.
%\end{equation}

\textbf{Observation Likelihood:} We assume a constant observation noise within each population $\sigma_l = \mathrm{StDev}(\epsilon^i_l)$. This accounts for heterogeneous characteristics when observations from multiple populations are combined, in particular $\sigma_l$ is smaller for larger populations and for more prevalent causes \citep{HLZ20}. The $\sigma_l$'s are estimated via Maximum Likelihood along with all other hyperparameters. More advanced GP models that either employ Poisson likelihood or infer input-dependent non-parametric $\sigma_l(x^i)$ are possible but require additional coding and are beyond the scope of this work.

%While assuming homogeneity of noise variance in terms of Age and Year is not entirely realistic, based on the discussion in \cite{LUDKOVSKI2016GAUSSIAN} the impact of more complex observation models is minimal. A common alternative is to assume a Poisson likelihood; however it is well known that mortality data are overdispersed, so a Poisson parametrization is also mis-specified.

\textbf{Estimating Hyperparameters:}
For (multi-level) ICM the set of hyper-parameters is  $\Theta = \left(\theta_{ag}, \theta_{yr}, (a_{l,q}), (\sigma^2_l), \bm{\beta} \right)$; for SLFM it is $\Theta = \left((\theta_{ag,q}), (\theta_{yr,q}), (a_{l,q}), (\sigma^2_l), \bm{\beta} \right)$. We use the \texttt{R} package \texttt{kergp} \citep{Oliver19kergp} to carry out the respective Maximum Likelihood Estimation via Kronecker decompositions.  Alternatively, to account for model risk and offer more robust results, one could employ a fully Bayesian hyperparameter inference. This could be done with the  \texttt{Stan} software \citep{CARPENTER2017} but is beyond the scope of this work. %We provide an example of the resulting factor loadings in SLFM model fitted on log-mortality rates over 3 Countries and 5 Cancer types in Appendix \ref{app:a_slfm}.

\subsection{Performance Metrics}

Given a test set of observed $y_*(x_*,l)$'s, we evaluate the effectiveness of different models using two metrics. First, we employ the mean absolute percentage error (APE) to examine the discrepancy between the observed and predicted outputs:
\begin{equation}
    \text{APE}(y_*, m_*) := \bigg|\dfrac{y_*-m_*(x_*)}{y_*}\bigg|
\end{equation}
where $y_*(x_*,l)$ is the observed value at test input $x_*$ in the $l$th population, and $m_*(x_*,l)$ is the predicted log-mortality rate. Note that APE is scale-independent, enabling us to compare model performance across populations with different exposures.

We also use the Continuous Ranked Probability Score (CRPS) metric to assess the quality of the probabilistic forecasts produced by a MOGP. Indeed, one of the major benefits of GP-based mortality models is a full distribution for future observations $y_*(x_*,l)$ which allows a more detailed uncertainty quantification beyond just looking at the predictive mean $m_*(x_*,l)$. CRPS assesses the closeness of the realized outcome $y_*(x_*,l)$ relative to its predictive distribution $F_*(\cdot; x_*)$ which in the GP contest is Gaussian and leads to
\begin{align}
    \notag \text{CRPS}( F_*, y_*) & := \int_{\mathbb{R}}\left[F_*(z)-\mathbbm{1}_{\{z \geq y_* \}} \right]^2 \, dz \\  %\\ & =
 %   \text{CRPS}\big(F_{m_*(x_*),\nu_*(x_*)},y_*(x_*)\big) & := \nu_*(x_*) \text{CRPS}\bigg(F, \dfrac{y_*(x_*)-m_*(x_*)}{\nu_*(x_*)}\bigg) \\ \notag
%    & = \nu_*(x_*) \text{CRPS}(F,z) \\
 %   & = \nu_*(x_*)\bigg[z(2F(z)-1) +2f(z) - \frac{1}{\sqrt{\pi}}\bigg], \\
    & = \sqrt{s_*^2(x_*)+\sigma^2_l}\big[ \tilde{y}_* (2\Phi(\tilde{y}_*)-1) + 2 \phi(\tilde{y}_*) - \frac{1}{\sqrt{\pi}}\big], \quad \tilde{y}_* := \frac{y_*-m_*(x_*)}{s_*^2(x_*)+\sigma^2_l}
    %\int_{\mathbb{R}}\left[F\left(\dfrac{z-m_*(x_*)}{\sqrt{s_*^2(x_*)+\sigma^2}}\right)-\mathbbm{1}_{\{z \geq y_* \}} \right]^2 \, dz,
\end{align}
where $\phi(\cdot), \Phi(\cdot)$ are the standard Gaussian density and cdf. Observe how CRPS penalizes both bias ($2\Phi(\tilde{y}^*)-1$) and excessive predictive variance.

% $\nu_*(x_*) = \sqrt{s_*^2(x_*)+\sigma^2}$, $F(z) = \displaystyle \int_{-\infty}^z f(v)dv = \displaystyle \int_{-\infty}^z \dfrac{1}{\sqrt{2\pi}}e^{-v^2/2}dv$.

We average both APE and CRPS across a test set of Age-Year-Population inputs. The resulting mean APE is interpreted as a normalized relative predictive error, and mean CRPS as the squared difference between the forecasted and the empirical cumulative distribution functions. Models with lower mean APE/CRPS are judged to have a better fit.

\textbf{Mortality Improvement Factors.}
A common way to interpret a mortality surface is via the (annual) mortality \emph{improvement factors} which measure longevity changes year-over-year.
The raw annual percentage mortality improvement is
%\begin{equation}
    %MI_{back}^{obs}(x):=
    $MI^{obs}_l(x):= 1-\frac{\exp\big(y_l(x_{ag};x_{yr})\big)}{\exp\big(y_l(x_{ag};x_{yr}-1)\big)}$.
%\end{equation}
The smoothed improvement factor is obtained by substituting in the GP posterior means $m_{*}$'s: %and integrating over the posterior distributions:
\begin{equation}\label{eq:mi}
    MI^{GP}_l(x):=\Bigg[1-\frac{\exp(m_{*}(x_{ag};x_{yr},l))}{\exp(m_{*}(x_{ag};x_{yr}-1,l))}\Bigg].
\end{equation}

\section{Modeling Multiple Causes of Death}
\label{sec:results}

%\subsection{Joint Models for Common Cancer Types in European Countries}
%\label{subsec:kernel_learning}

To understand the behavior of Age-specific mortality across different causes of death, we begin by generating MOGP models for Cancer variants. Using the HCD database we fit both ICM and SLFM and assess their performance in 3 European countries. We test the resulting predictive performance by computing APE and CRPS for one-year-ahead mortality forecasts in three separate test sets, using SOGP as the baseline. All models, including SOGP models, are trained on the same Ages from 50--84 (7 age groups) and three overlapping periods: 1998--2013 for the 2014 prediction, 1999--2014 for 2015 prediction, and lastly 2000--2015 for 2016 prediction. We report APE and CRPS for all-Cancer log-mortality rates since some of the cancer variations, such as Stomach and Pancreatic have relatively few (and therefore more noisy) recorded deaths. The differences in APE and CRPS between MOGP and SOGP models are expressed as the 3-year median percentage improvement over SOGP models. Positive improvement means joint models have smaller mean APE/ CRPS values. %Lastly, it is straightforward to compute APE for all-cancer log-mortality.
To compute all-Cancer CRPS, we leverage the closed-form expression of the MOGP multivariate predictive distribution in \eqref{eq:GP-mean}-\eqref{eq:GP-var}
 to simulate the forecast distribution of all-Cancer log-mortality for each Age group in the data. To do so, we first draw ($5 \times 10^5$) stochastic samples of the joint $\mathbf{f_*}(x_*)$ across all the Cancer types. After exponentiating and summing, we then obtain corresponding samples of  (unlogged) all-Cancer mortality rates.

\begin{table}[!t]
\centering
\caption{Comparison between MOGP ICM/SLFM with different ranks $Q$, reported as the relative improvement in APE and CRPS of MOGP vis-a-vis SOGP. The reported values are medians of one-year-out aggregated all-Cancer forecasts for age groups 50--84 based on 3 training periods: 1998--2013 (predict  2014), 1999--2014 (predict 2015), and 2000--2015 (predict 2016)\label{tbl:mogp-cause}. Both ICM and SLFM are fitted on 5 Cancer types, Male populations in each selected country.}
\begin{tabular}{llcrrrrrr}  \toprule
\multicolumn{2}{l}{\multirow{2}{*}{Cause ($L=5$)}} & \multirow{2}{*}{\begin{tabular}[c]{@{}c@{}}\# Kernel\\ Hyperparameters\end{tabular}}    & \multicolumn{2}{c}{Czech Rep.} & \multicolumn{2}{c}{Germany} & \multicolumn{2}{c}{Poland} \\ \cline{4-9}
\multicolumn{2}{l}{} &   &  APE        & CRPS        & APE         & CRPS         & APE         & CRPS        \\  \midrule
\multirow{3}{*}{ICM}   & $Q=2$    & 12  & 19.16      & 30.18       & 3.86        & 15.05        & 12.85       & 17.72      \\
& $Q=3$   &  17   & 17.69       & 19.08     &  6.53         & 15.16        & 13.46      & 18.64      \\
& $Q=4$   &  22   & 22.31       & 21.43       &  6.30      & 14.58        & 17.02      & 11.01        \\
\midrule
\multirow{3}{*}{SLFM}   & $Q=2$    & 14  & 18.04       & 29.49       & 5.09        & 13.51        & 10.82       & 16.05      \\
& $Q=3$   &  21   & 22.87       &  23.72      &  2.78       & 13.55        & 4.11       & 10.65     \\
& $Q=4$   &  28   & 16.97       & 18.24       & 8.75       & 16.73        &   5.13     &  3.81        \\ \bottomrule
\end{tabular}
\end{table}

Table \ref{tbl:mogp-cause} shows the 3-year median improvement in MAPE and CRPS for Multi-cause ICM and SLFM over SOGP models. Overall, joint models produce more accurate mean forecasts (positive improvement in APE) with higher credibility (positive improvement in CRPS). We observe the opportunity to borrow information across different cancers to better estimate the kernel hyperparameters. As a result, joint models can describe important trends for individual cancers, leading to the reduction in disparity between the predicted values and the observed all-Cancer mortality in the test sets.

\begin{figure}[!t]
    \centering
    \includegraphics[width=12cm]{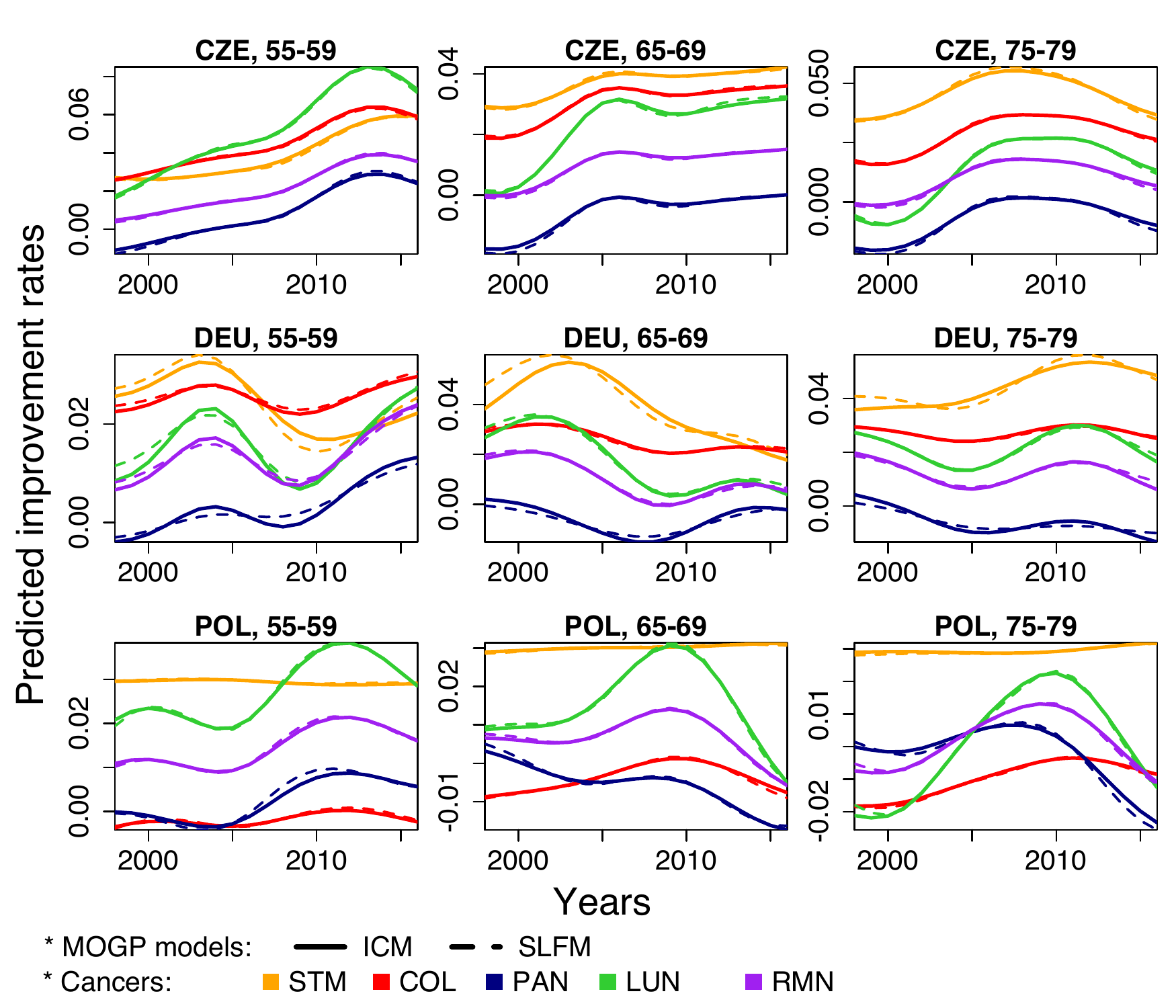}
    \caption{The predicted YoY improvement rates derived from multi-cause GP models by Country and Age groups among Male observations. In each country, MOGP-ICM ($Q=2$) and SLFM ($Q=2$) are fitted on Ages 50--84 (7 groups), Years 1998--2016 over 5 Cancer variations: Stomach, Colorectal, Pancreatic, Lung and Remaining types.
    \label{fig:impv-multicause}}
\end{figure}

\subsection{Commonalities in Cause-Specific Mortality Surfaces}

The main difference between ICM and SLFM is the underlying assumption about the latent factors $u_q(\cdot)$. ICM assumes that all factors have the same lengthscales $\theta_{ag},\theta_{yr}$, and is therefore appropriate for modeling homogenous mortality surfaces. SLFM is more general and fits distinct $\theta_{ag,q}, \theta_{yr,q}$; it is expected to perform better when the different surfaces exhibit heterogeneity (for example different degree of correlation across Age).
We examine this commonality assumption in Figure~\ref{fig:impv-multicause} by displaying side-by-side the mortality improvement factors of the various cancers derived from multi-cause ICM and SLFM. The BIC selection criterion yields $Q=2$ for both ICM and SLFM. In this case study, the results from SLFM are almost identical to ICM predictions. Therefore, the assumption of sharing the spatial kernel over Age-Year inputs across the considered cancer types is plausible. This conclusion is reinforced by Table~\ref{tbl:lengthscales} in Appendix A which shows that the inferred lengthscales $\theta_{ag,q},\theta_{yr,q}$ for SLFM are very similar for $q=1,2$ across all three countries. In other words, both of the latent factors learned by SLFM have similar Age-Year spatial dependence, and so there is little loss of fidelity in a priori forcing them to be equal, as is done in ICM. Indeed, the lengthscales in SLFM are close to the ICM ones.

%\subsection{Inferring insights from joint GP models}

%An important application of cause-of-death modeling is to gain understanding about the underlying trends across causes.
We can also inspect Figure \ref{fig:impv-multicause} for insights about the relative mortality improvement trends of different cancers. Stomach cancer has the largest improvement rates for most age groups in all three countries. Decline in stomach cancer incidence tends to be associated with economic improvements resulting in healthier diet, better food preservation, clean water supply, etc. We also observe the increasing improvement trend of lung cancer among age groups below 60 in Czech Rep.~and Poland,
reflecting lower smoking rates in birth cohorts after WWII.  Czech Males experienced large increase in the improvement rates in most cancers. In Germany, the improvement rates have been rising among age groups below 60 but slowing down among older age groups. In Polish Males, except stomach cancer, the improvement trends increased until early 2010s and then significantly declined, displaying the impact of an aging population and an increase in lifestyle exposure to risk factors for cancers \citep{Piotr2014}. The incidence of Stomach cancer has been flat over time, consistent with the finding in \cite{Etemadi2020}. %\todo{Move all this to earlier?}

\begin{figure}[!ht]
    \centering
    % \captionsetup{justification=centering}
    \subfloat[Multi-cause ICM in Czech R.]{{\includegraphics[trim={0cm 1.25cm 0cm 1.5cm},clip=true, width=5cm]{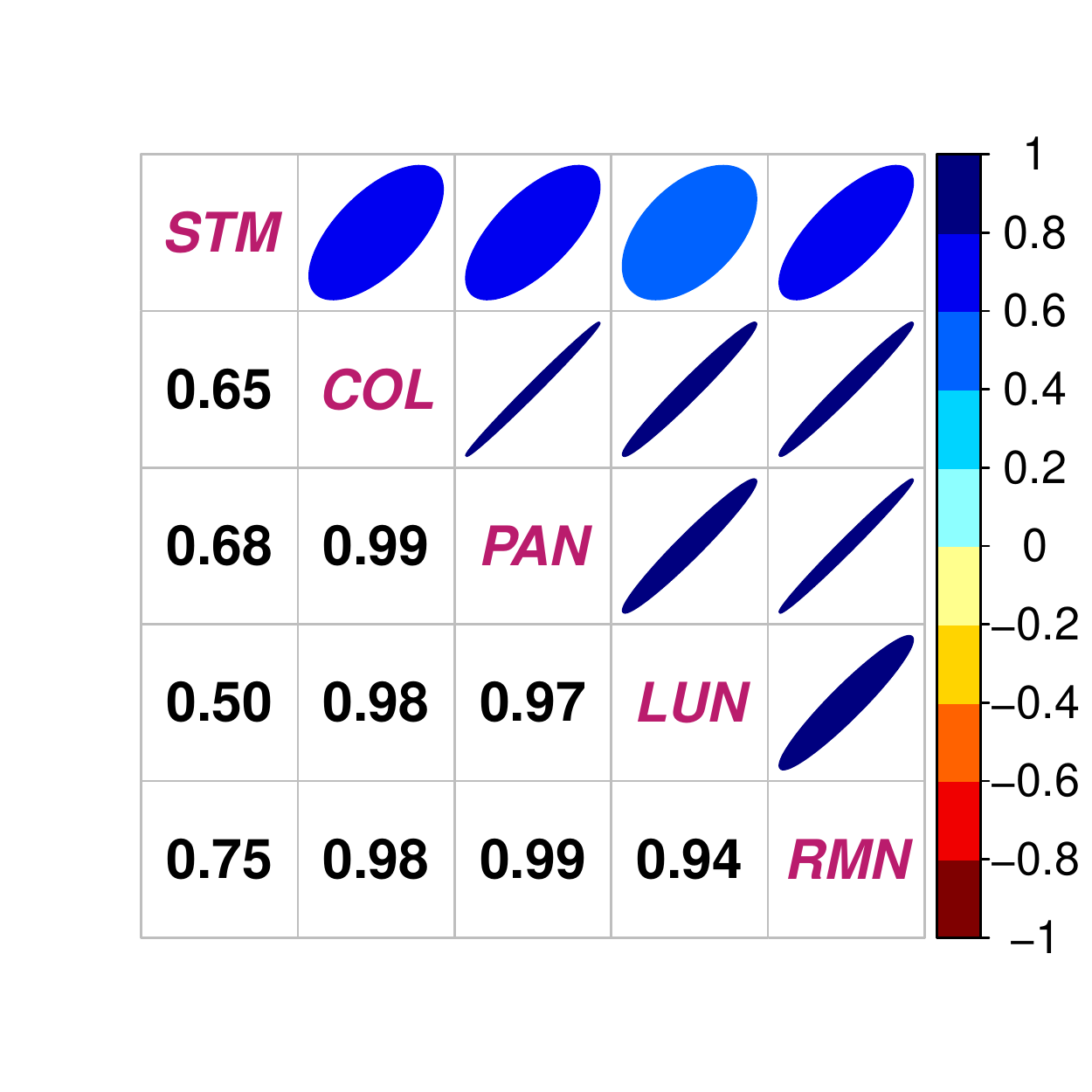}}}
    \subfloat[Multi-cause ICM in Germany]{{\includegraphics[trim={0cm 1.25cm 0cm 1.5cm},clip=true,width=5cm]{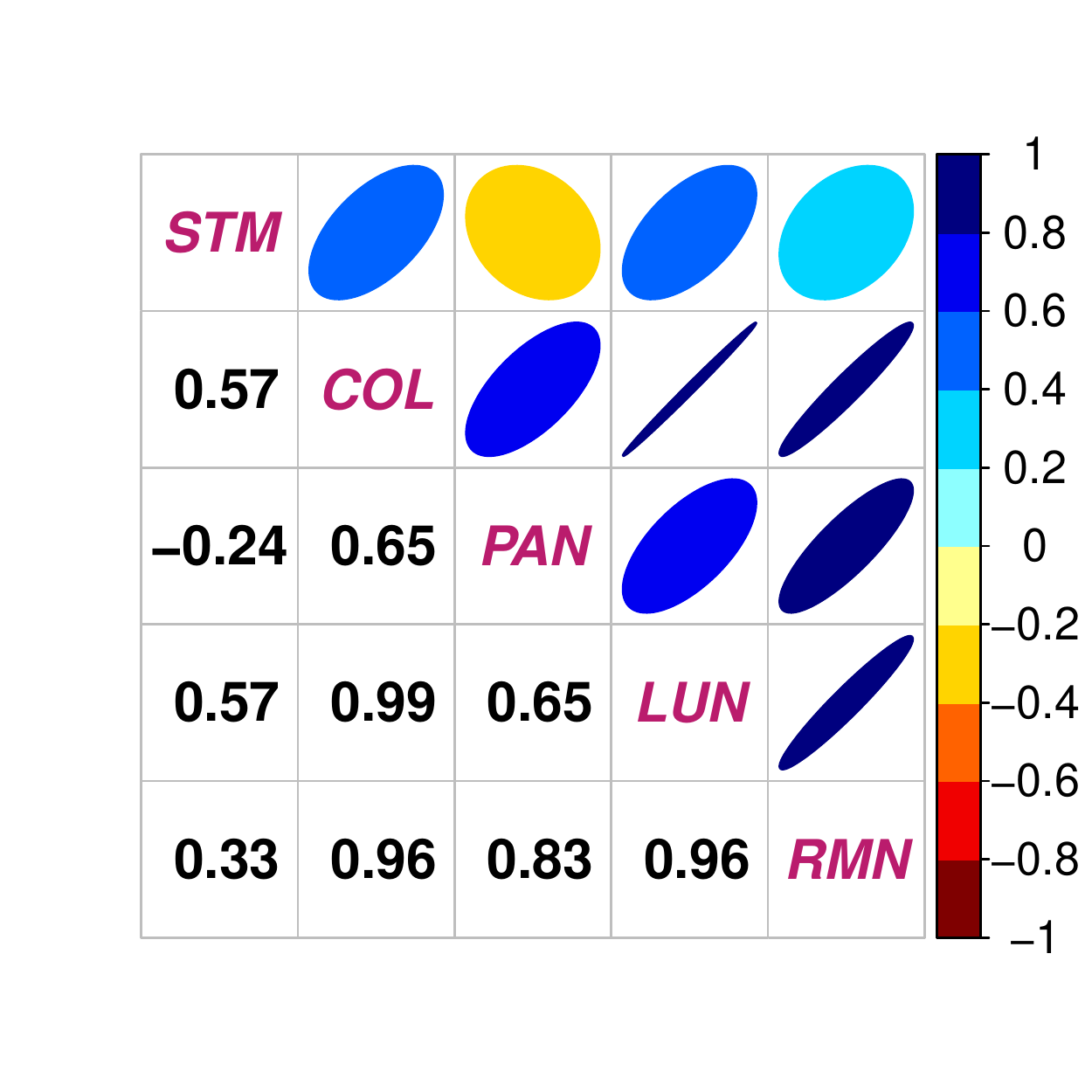}}}
    \subfloat[Multi-cause ICM in Poland]{{\includegraphics[trim={0cm 1.25cm 0cm 1.5cm},clip=true,width=5cm]{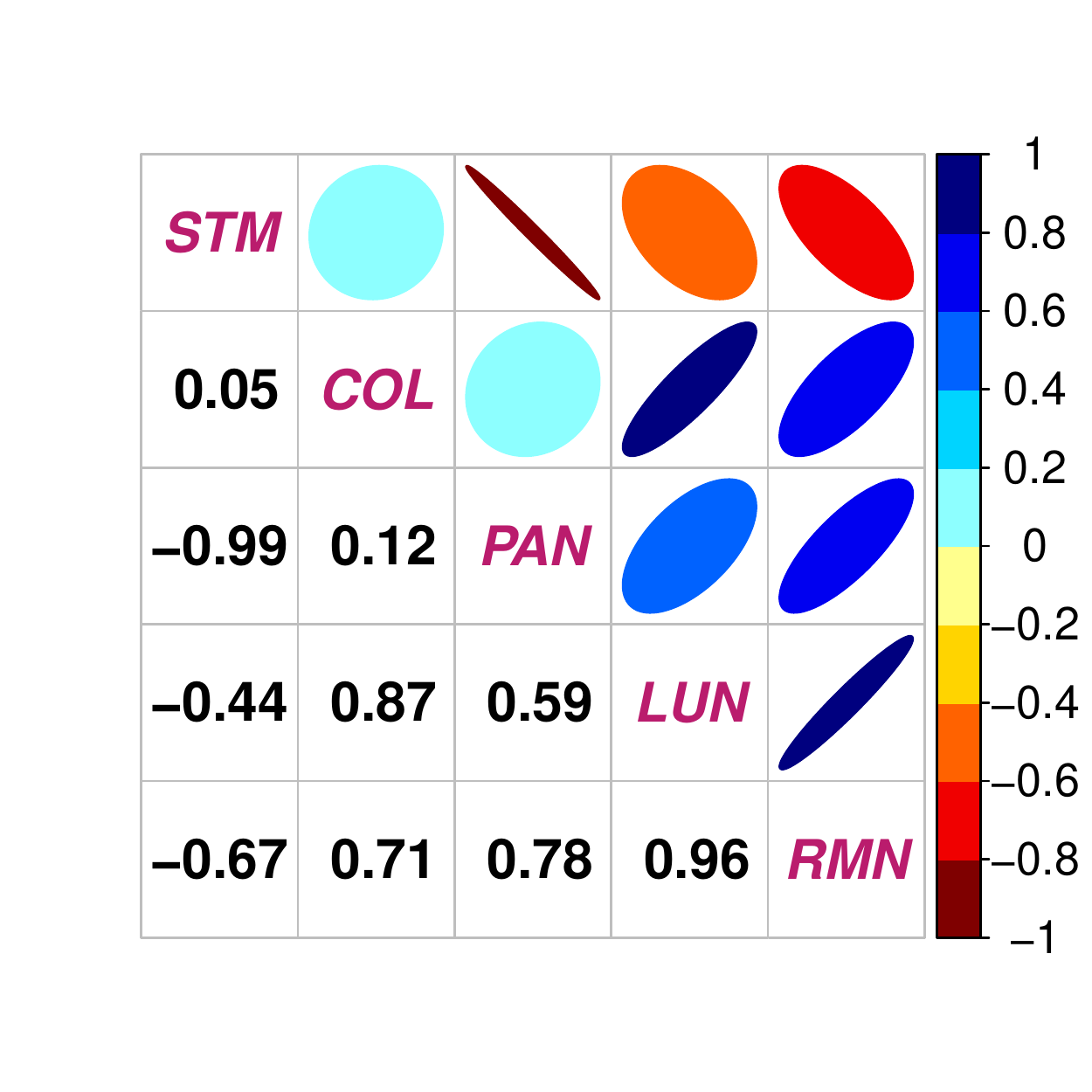}}} \\
    \subfloat[Multi-cause SLFM in Czech R.]{{\includegraphics[trim={0cm 1.25cm 0cm 1.5cm},clip=true,width=5cm]{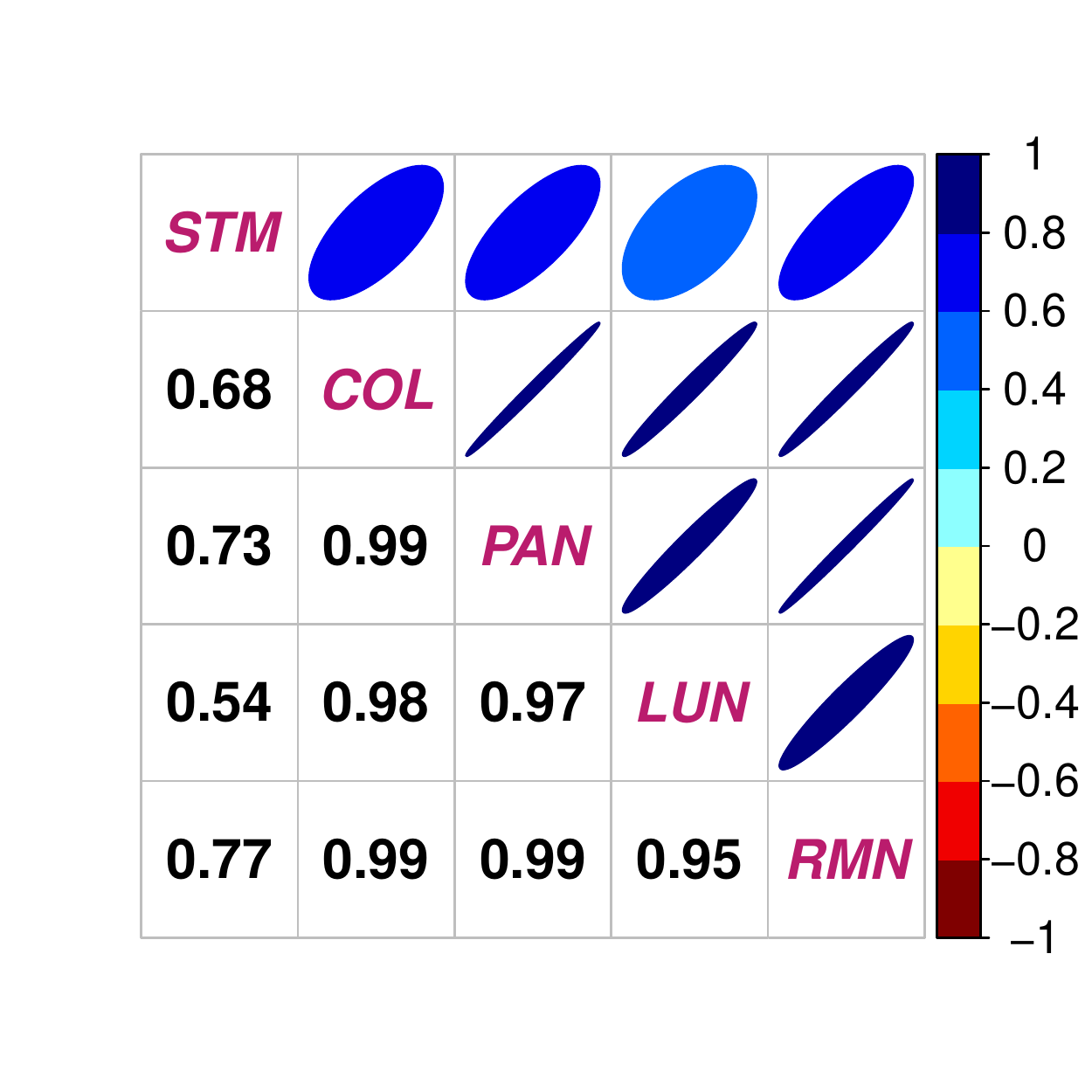}}}
    \subfloat[Multi-cause SLFM in Germany]{{\includegraphics[trim={0cm 1.25cm 0cm 1.5cm},clip=true,width=5cm]{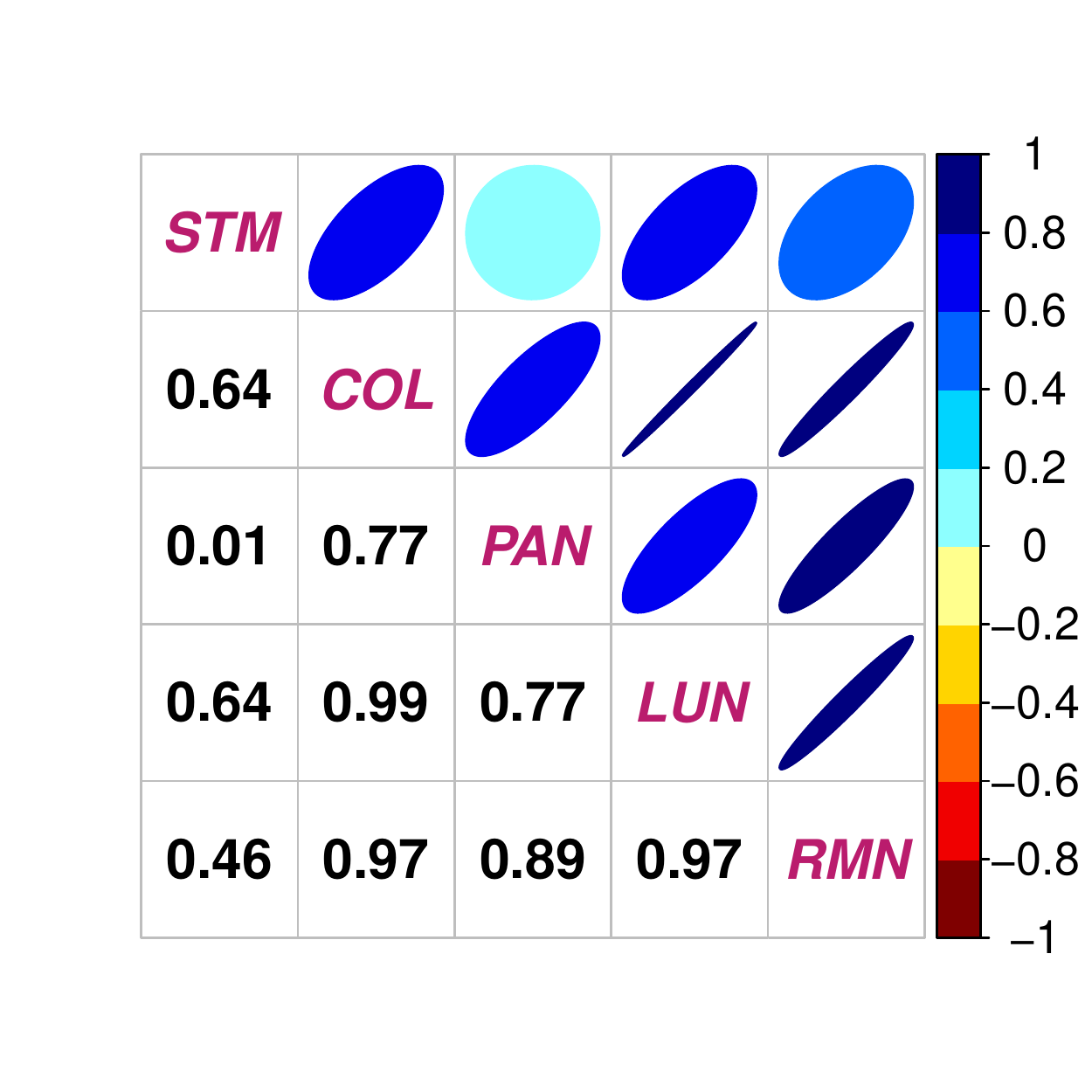}}}
    \subfloat[Multi-cause SLFM in  Poland]{{\includegraphics[trim={0cm 1.25cm 0cm 1.5cm},clip=true,width=5cm]{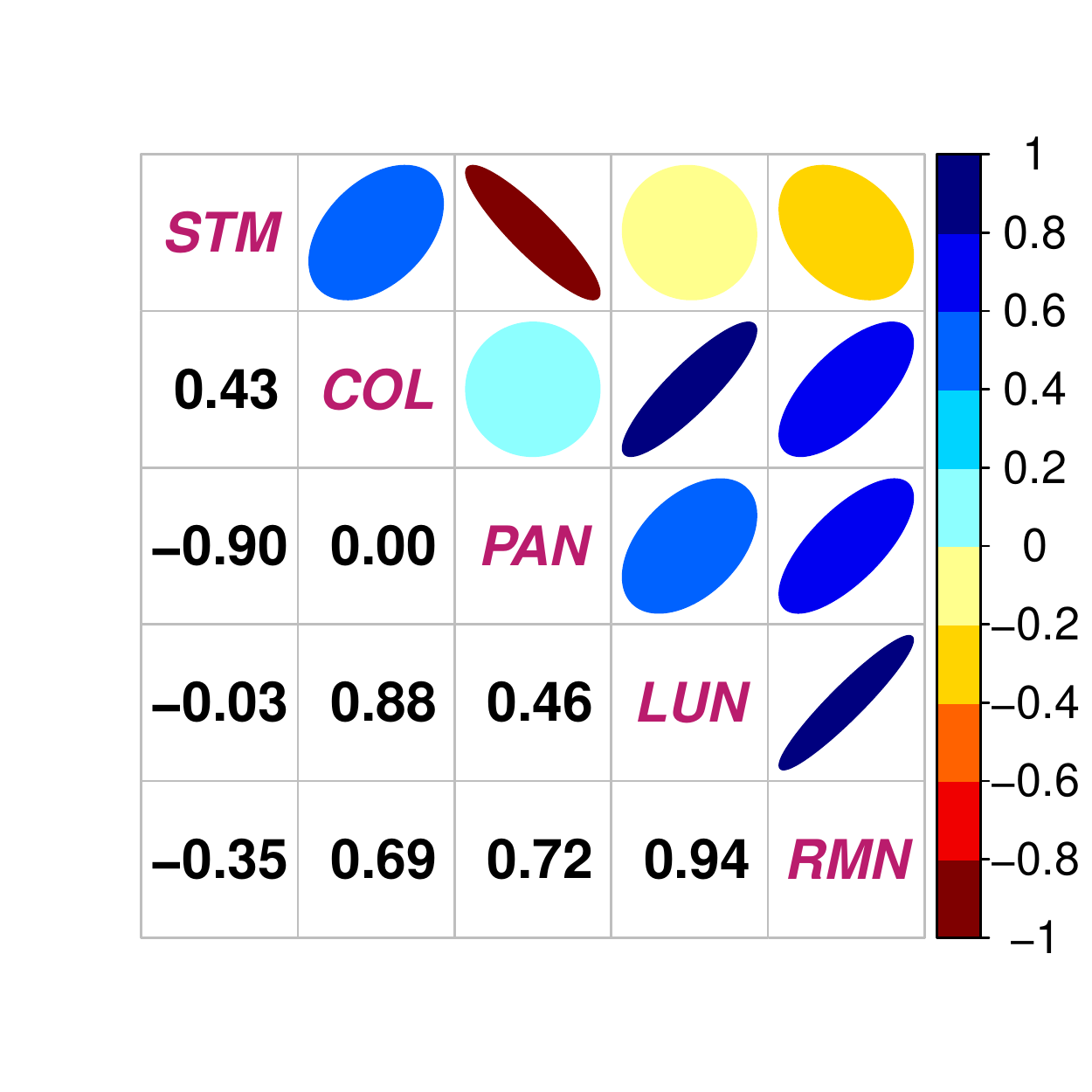}}} \\
    \caption{Correlation matrices $R$ derived from multi-cause GP models; rank $Q=2$ is chosen for both ICM and SLFM. In each country, the MOGP model is fitted on Males aged 50--84 (7 groups) during Years 1998--2016, over 5 Cancer variations: Stomach, Colorectal, Pancreatic, Lung and Remaining types.
    \label{fig:corr-multicause}}
\end{figure}

Figure \ref{fig:corr-multicause} visualizes the inferred cross-cause correlation matrices $(r_{l_1,l_2} : 1 \le l_1,l_2 \le 5)$. Both ICM and SLFM document a global positive association among mortality rates from these cancers in each country. It is consistent with strong resemblance in the mortality improvement trends between cancers in Figure \ref{fig:impv-multicause}. Since cancers within a given population share common risk factors, innovations in early detection and advancements in treatment for one cancer are likely to have positive effects on other cancer variations.  Negative correlation $r_{l_1,l_2} < 0$ reflects opposite trends, e.g.~Stomach and Pancreas cancers in Poland; this may be due to a competing risks context.

%\subsection{Top-Level Causes of Death}

%Through multiple examples, we demonstrated the application of cause-specific models via the MOGP framework to gain insights into the dynamics of all-cancer trends among 3 European neighboring countries.

For a different take on cause-of-death commonality, we applied the multi-cause GP ICM model to jointly model $L=8$ top causes in the HCD US dataset, separately for each gender. Using BIC as the criterion, models with rank $Q=6$ (for 8 populations) yield the largest BICs for both Males and Females. This indicates a higher degree of heterogeneity in these larger cause groupings, compared to $Q=2$ across five cancer causes in the previous study. Thus, the models employ more latent functions $u_q(\cdot)$ to adequately capture the total variability in the joint datasets. The inferred cross-cause correlation matrices are displayed in Figure~\ref{fig:corr-us} in Appendix B. Overall, our results are in agreement with the SOA study \citep{codUS2019}, for example confirming that there are moderate positive associations between most causes. The strongest correlations are found between Heart disease and Stroke, and between Heart and Drug overdose. As might be expected, there is little correlation between Remaining Causes (RMN) and most other categories. Some of the correlations vary between genders, possibly due to  strong observation noise in less common causes.

A complementary way to examine commonalities across causes is to inspect the inferred factor loadings $a_{l,q}$. Populations that have similar loadings will be highly correlated. Figure~\ref{fig:a-loadings} in Appendix A displays the factor loadings in a Country-Cause SLFM with $Q=3$ latent factors $u_q$. We observe that primary clustering is by Causes rather than by Countries. Some outliers, such as STM in Czech R., can also be seen and suggest idiosyncratic behavior of the respective mortality surface.

\subsection{Aggregating by-Cause models}

\begin{figure}
    \centering
    \includegraphics[width=15.5cm]{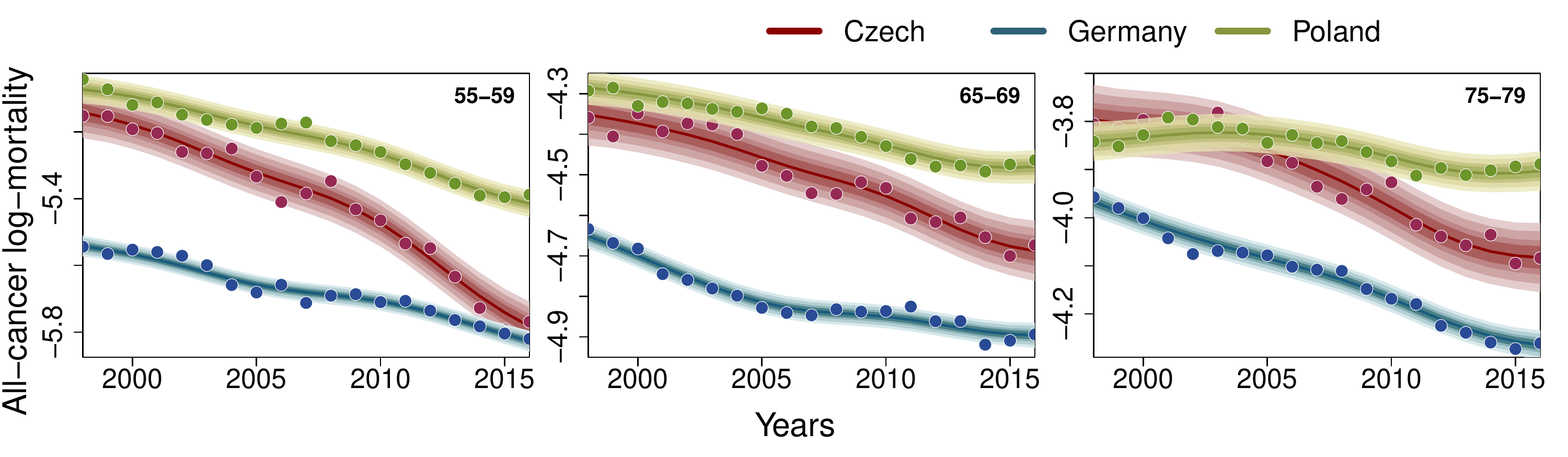}
    \caption{Predictive distribution of all-Cancer log-mortality rates for different age groups in 3 countries via multi-level Country-Cause ICM. The joint model is fitted on Males, Age groups between 50--84, Years 1998--2016 across 3 Countries ($Q_{ctry} = 3$) and 5 Cancers ($Q_{caus}=5$). Shading indicates the 60\%, 80\%, 95\% and 99\% predictive quantile bands.}
    \label{fig:all_cancer}
\end{figure}

An important motivation for our work was to use by-cause analysis to make more precise conclusions about all-cause mortality. For example, the mortality trends of individual Cancers give insights into the respective all-Cancer mortality trends. Figure~\ref{fig:all_cancer} shows the results from a multi-level Country-Cause GP ICM. It
visualizes the aggregated predictive distribution of all-Cancer log-mortality observations, $y_l(x_*)$,  for Male populations by Country and Age group, using shading to denote predictive quantiles. Note that since the model did not use all-Cancer mortality during training, the fact that the predictive in-sample bands closely match the historical movement of all-Cancer mortality data is a validation of a successful by-cause analysis.

The effort in fighting cancer has transpired in all three countries, but the improvement is not uniform. Despite Czech Republic and Poland being socio-economically similar, Czech Rep.~has a faster improvement pace than Poland. Although Germany continues to have the lowest log-mortality rates across all Age groups, the Czech Rep.~has drastically closed this gap in the last decade, see especially the left panel of Figure~\ref{fig:all_cancer} (Age group 55-59). The main driver  is the rapid improvements in all common Cancers in Czech Rep., for example the mortality improvement factor for LUN being recently more than double compared to Germany, cf.~Figure~\ref{fig:impv-multicause}.
%the most notable improvement occurring in Lung cancer that has been at least twice as large as other countries during recent years.

\subsection{Trends in US Top-Level Causes}

\begin{figure}[!t]
    \centering
    \textbf{AGE 40-44, US MALES} \\
    {{\includegraphics[width=5.40cm]{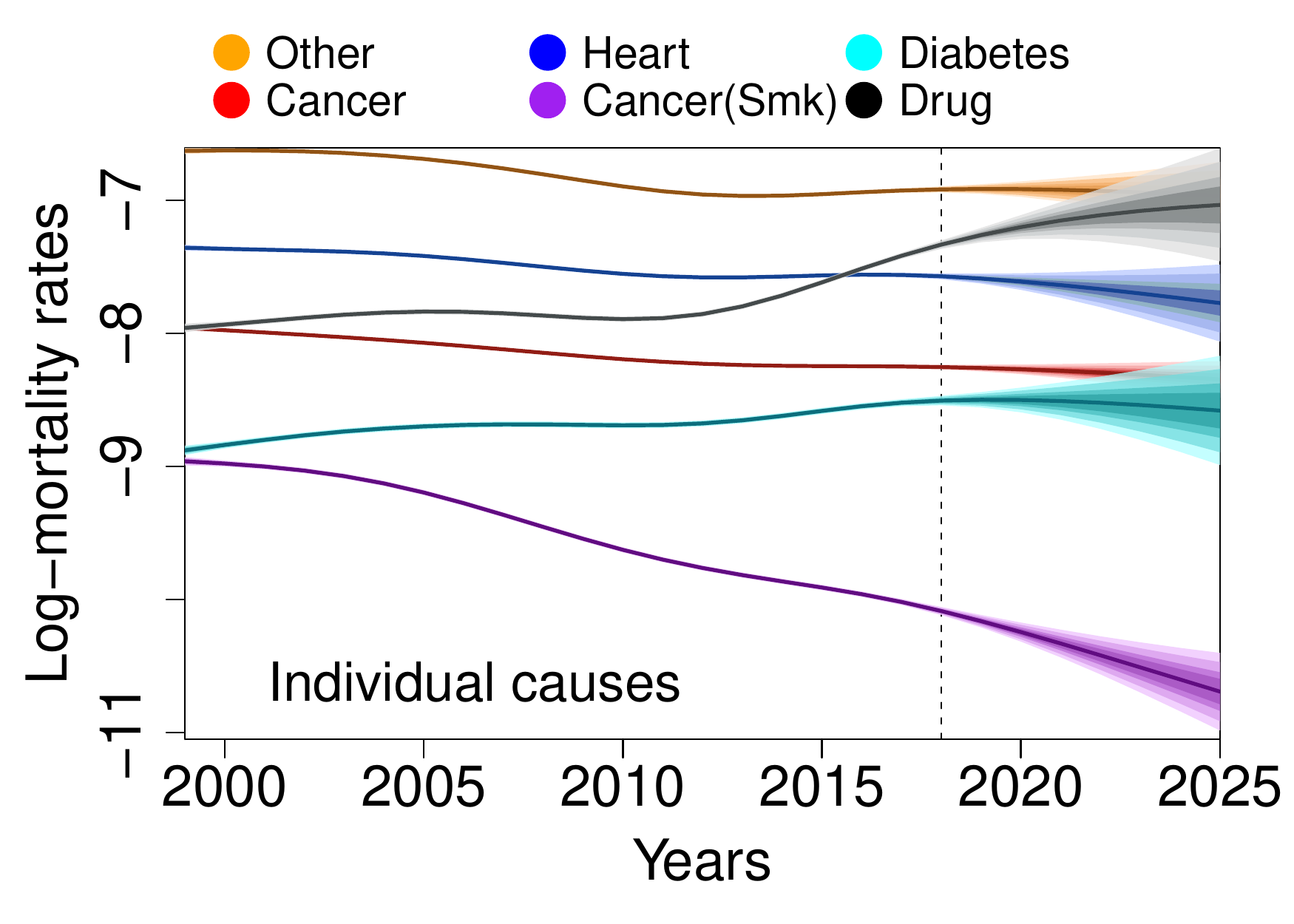}}}
    {{\includegraphics[width=5.40cm]{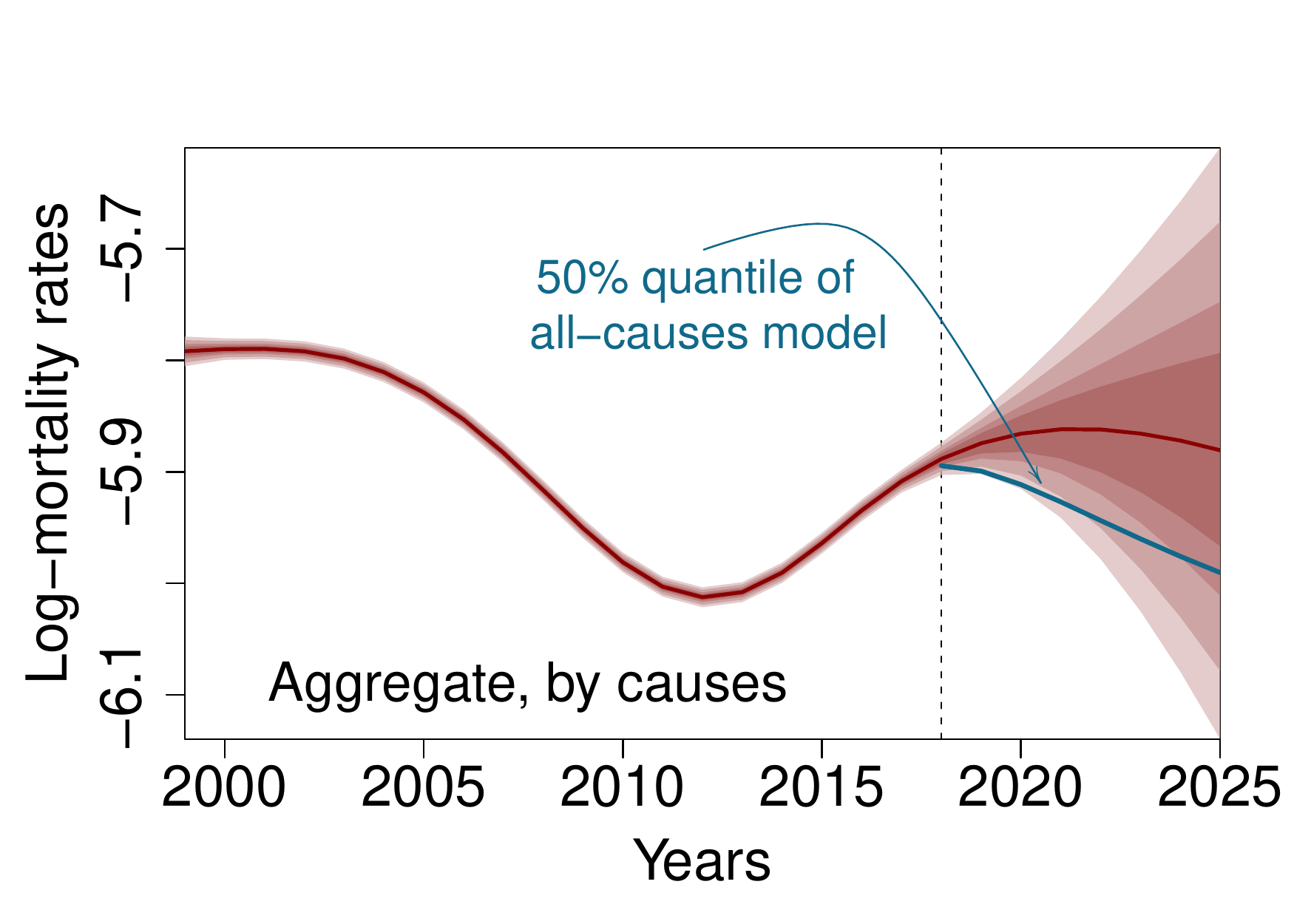}}}
    {{\includegraphics[width=5.40cm]{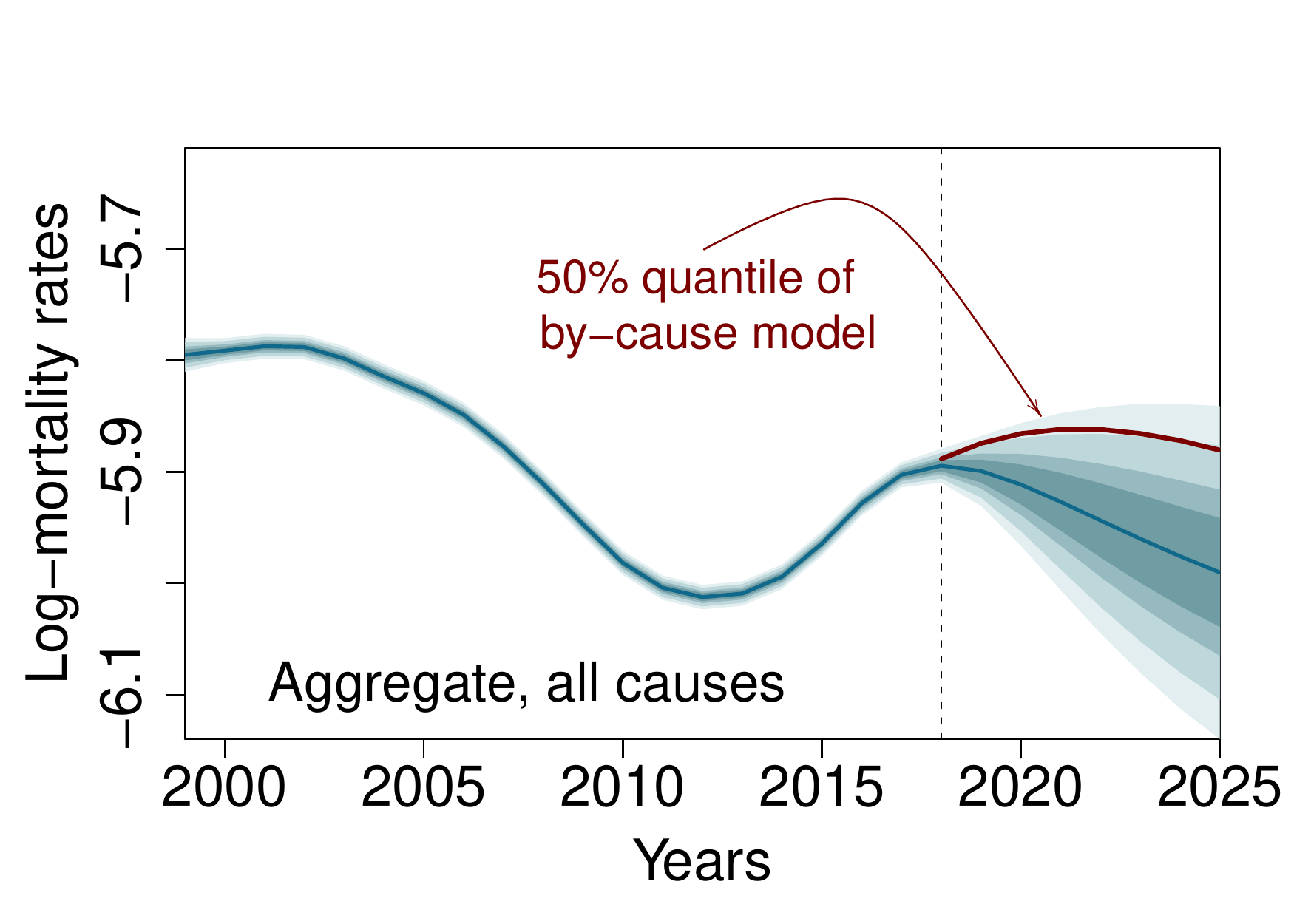}}} \\
    \vspace{2.5mm}
     \textbf{AGE 40--44, US FEMALES} \\
    {{\includegraphics[width=5.40cm]{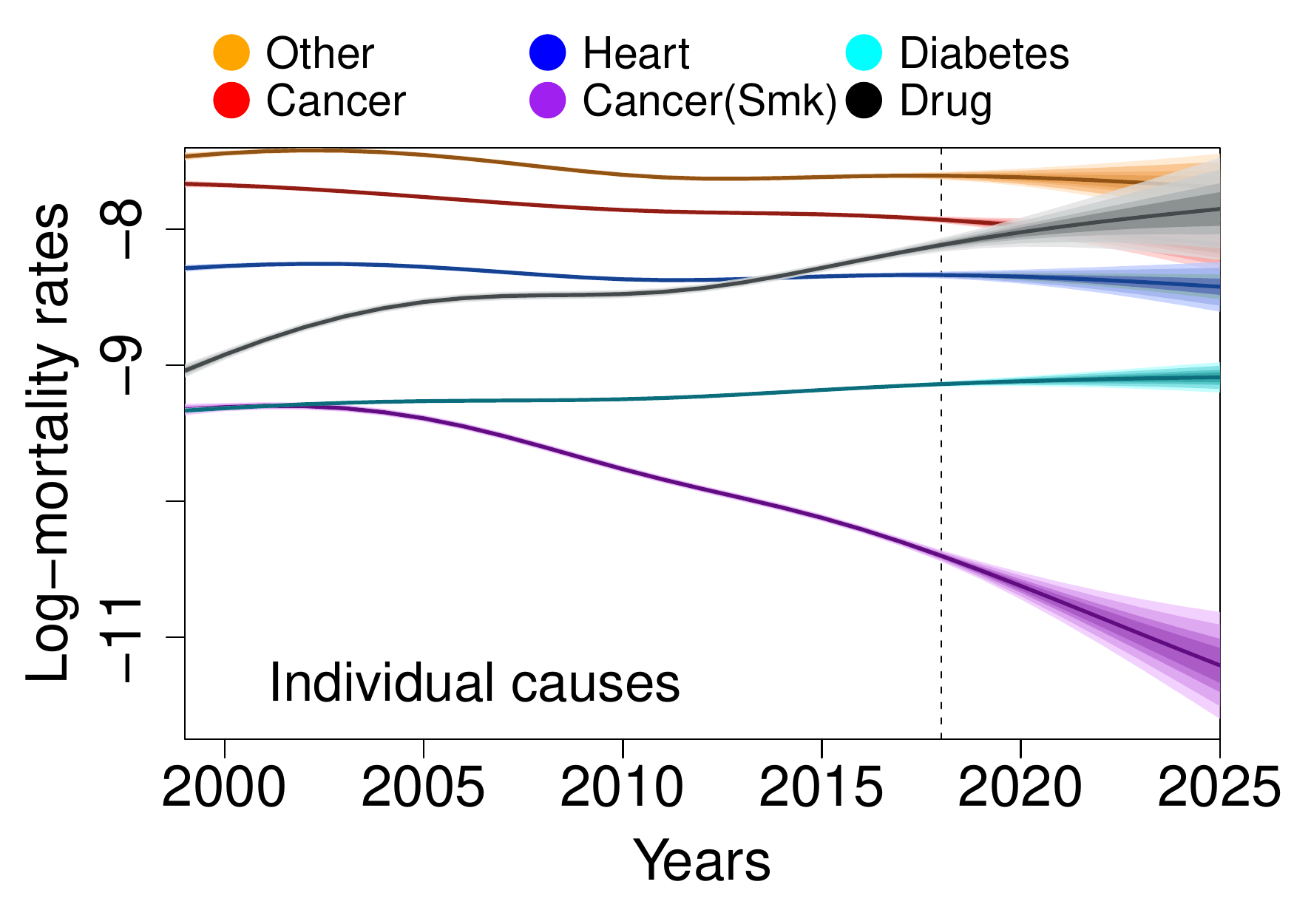}}}
    {{\includegraphics[width=5.40cm]{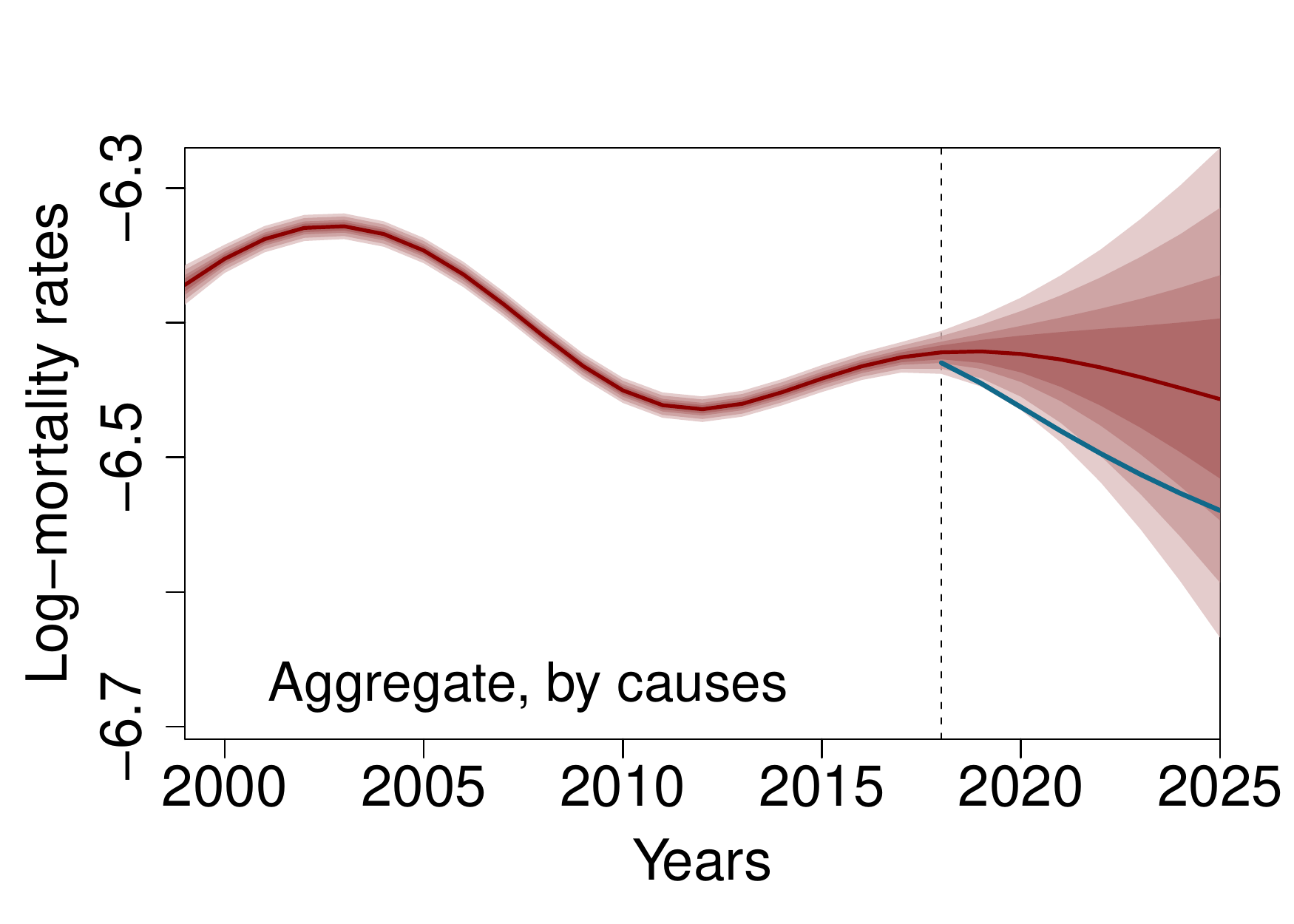}}}
    {{\includegraphics[width=5.40cm]{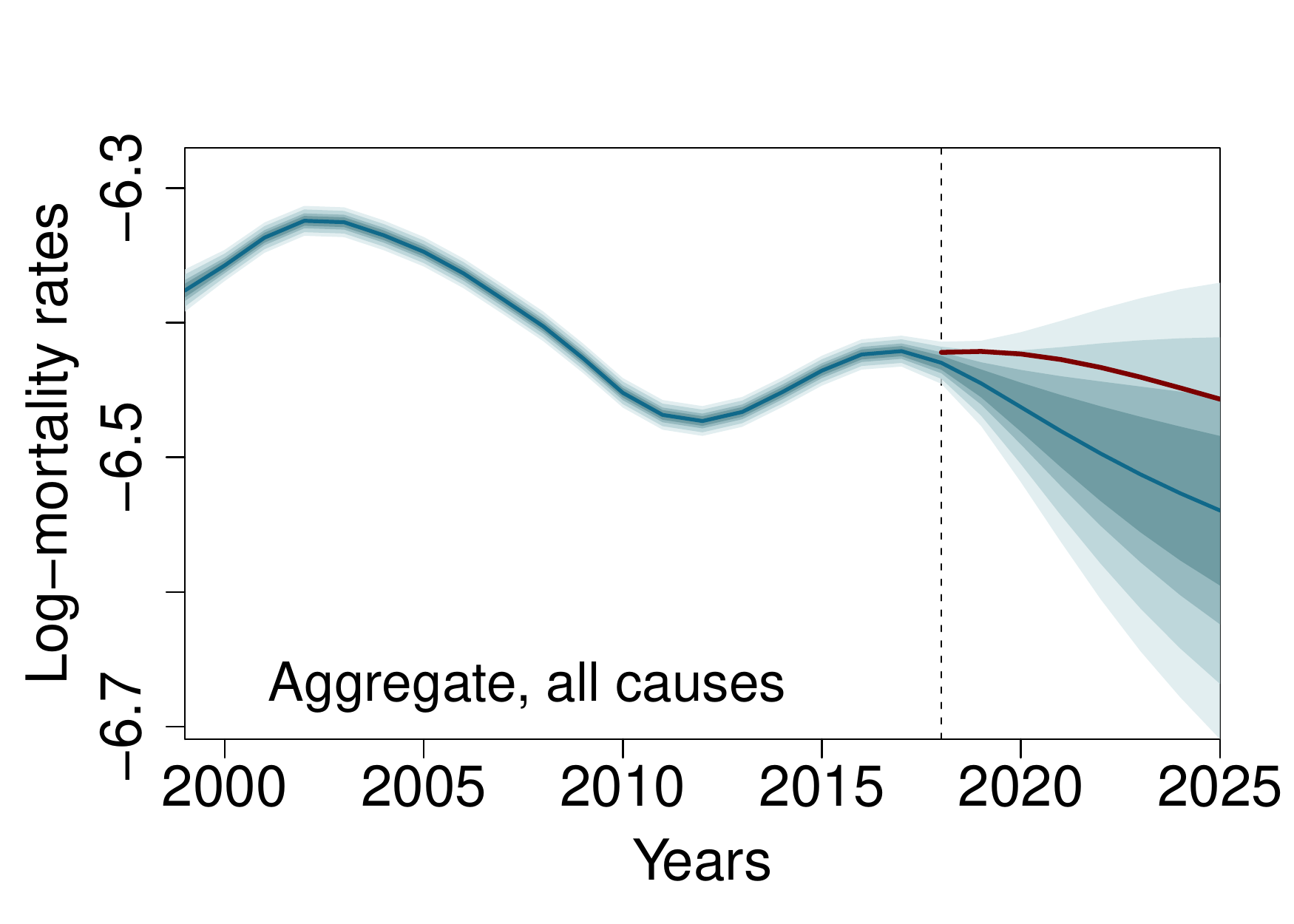}}} \\
    \vspace{2.5mm}
    \textbf{AGE 65--69, US MALES} \\
    {{\includegraphics[width=5.40cm]{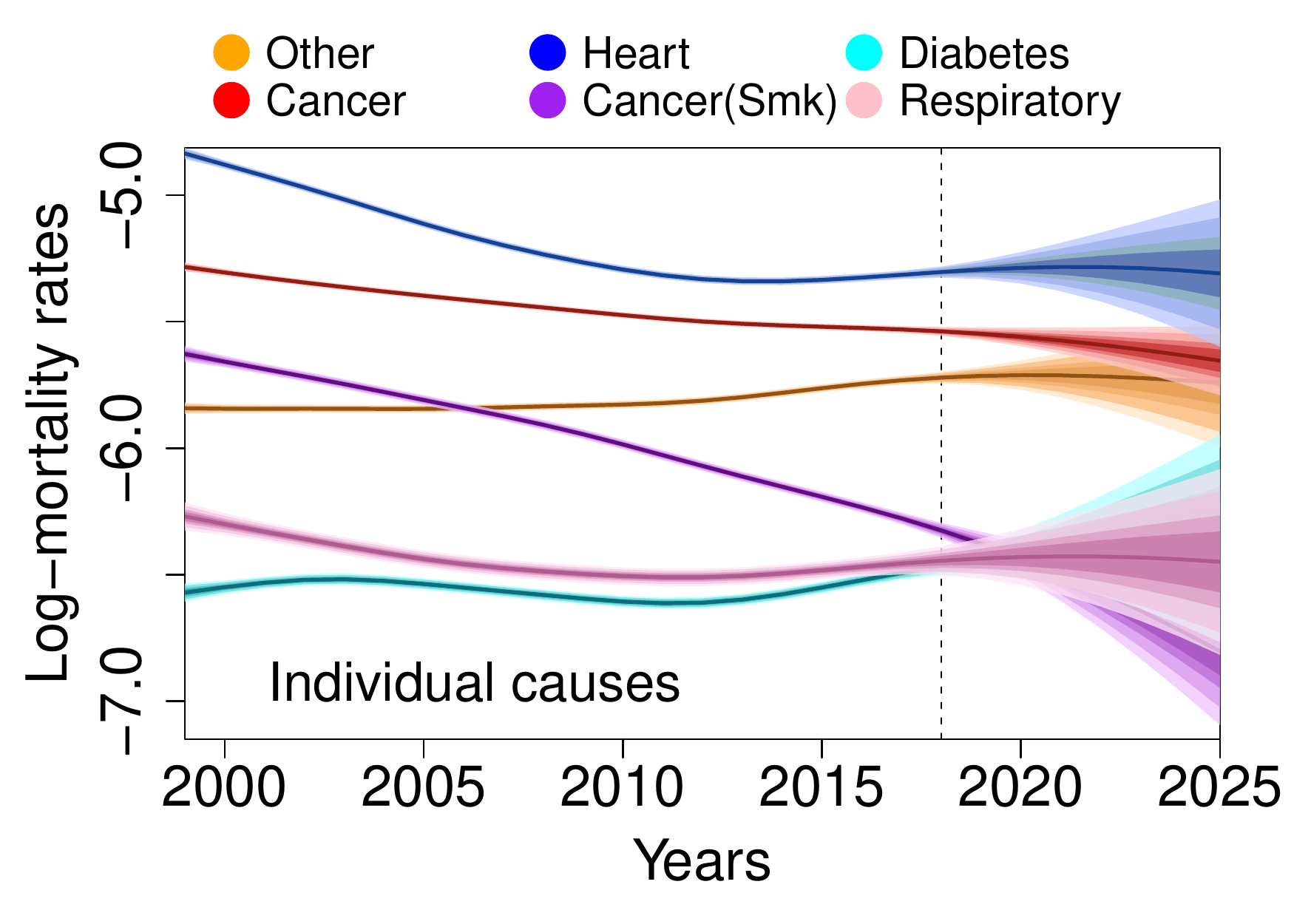}}}
    {{\includegraphics[width=5.40cm]{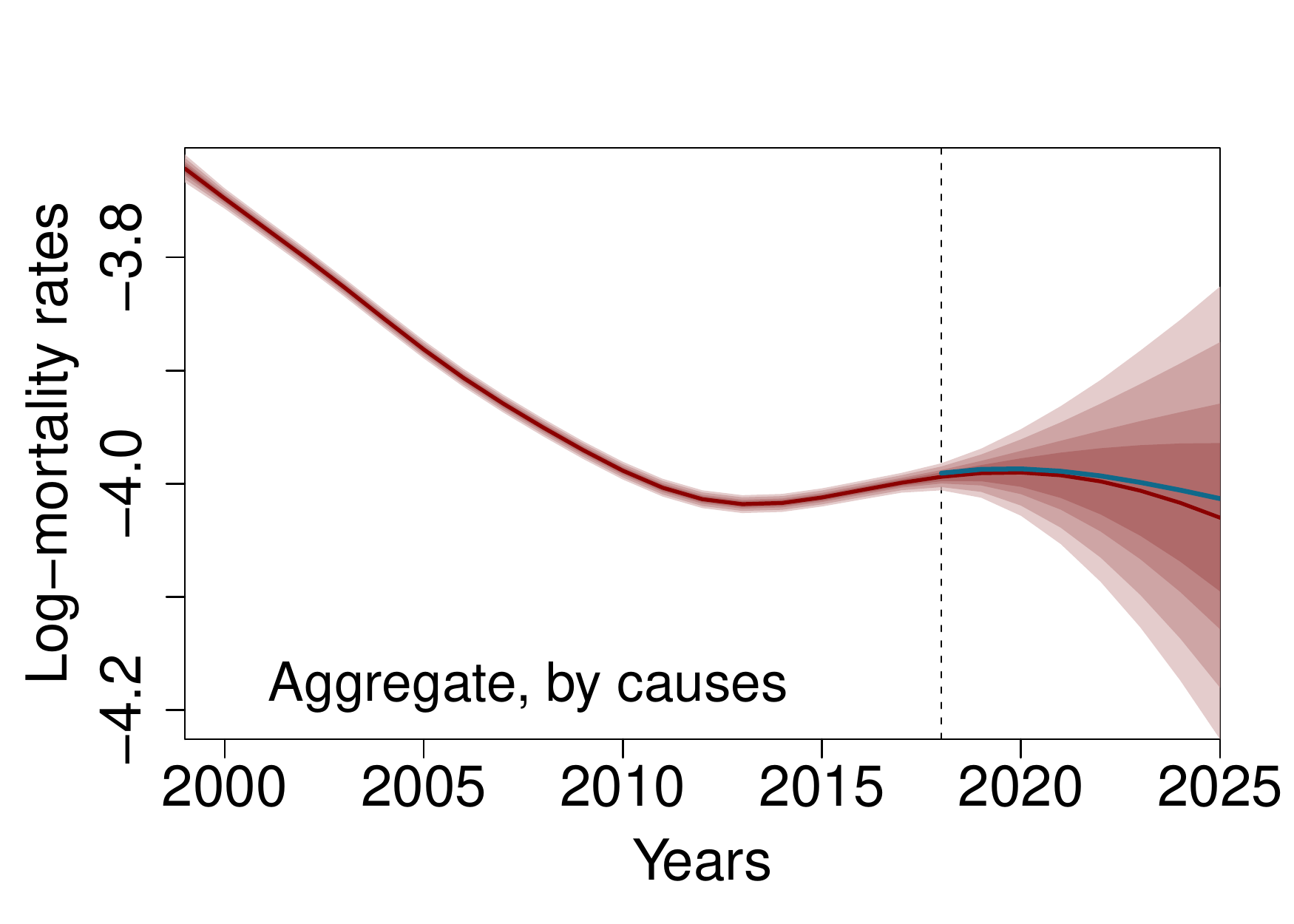}}}
    {{\includegraphics[width=5.40cm]{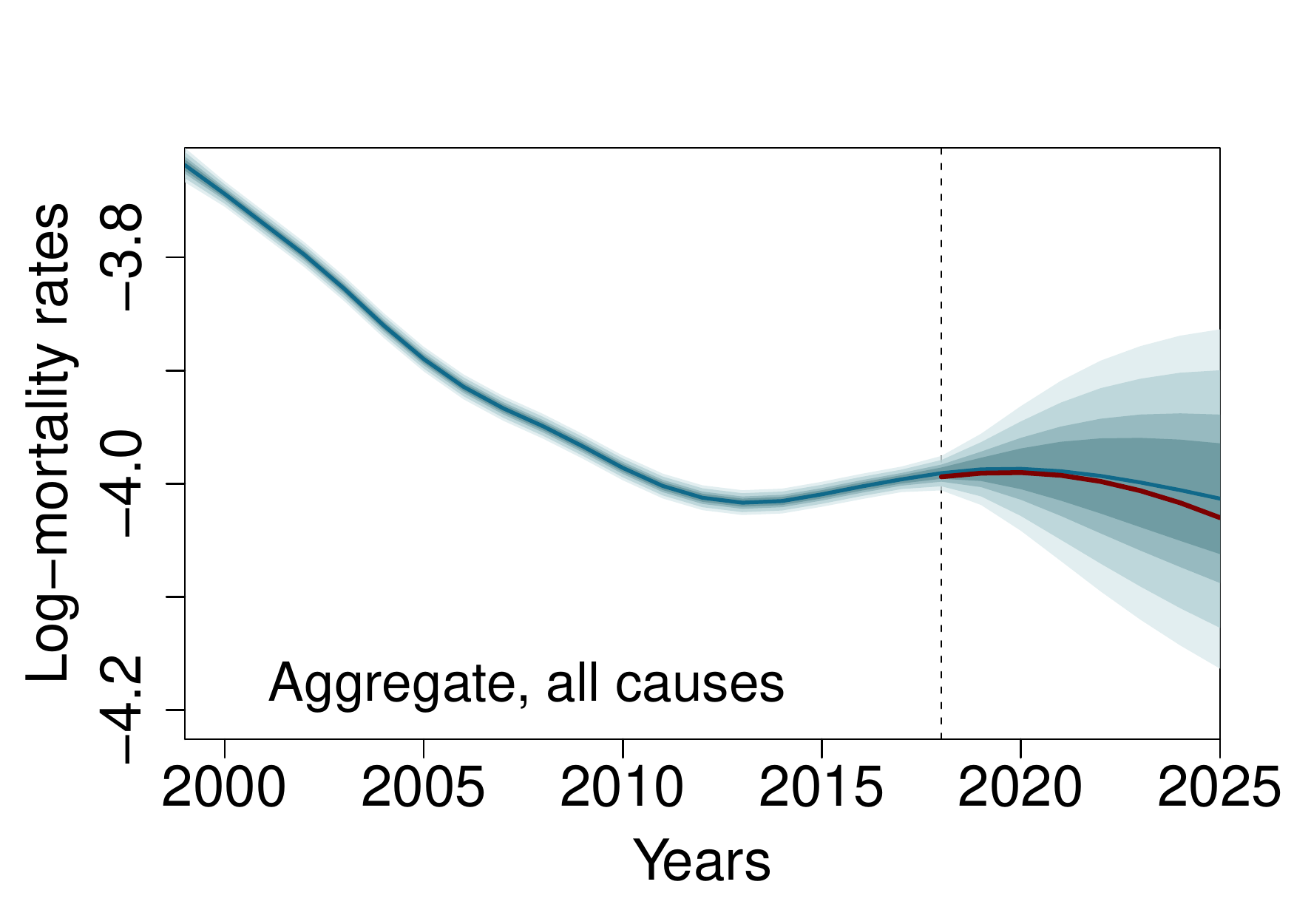}}} \\
    \vspace{2.5mm}
    \textbf{AGE 65--69, US FEMALES} \\
    {{\includegraphics[width=5.35cm]{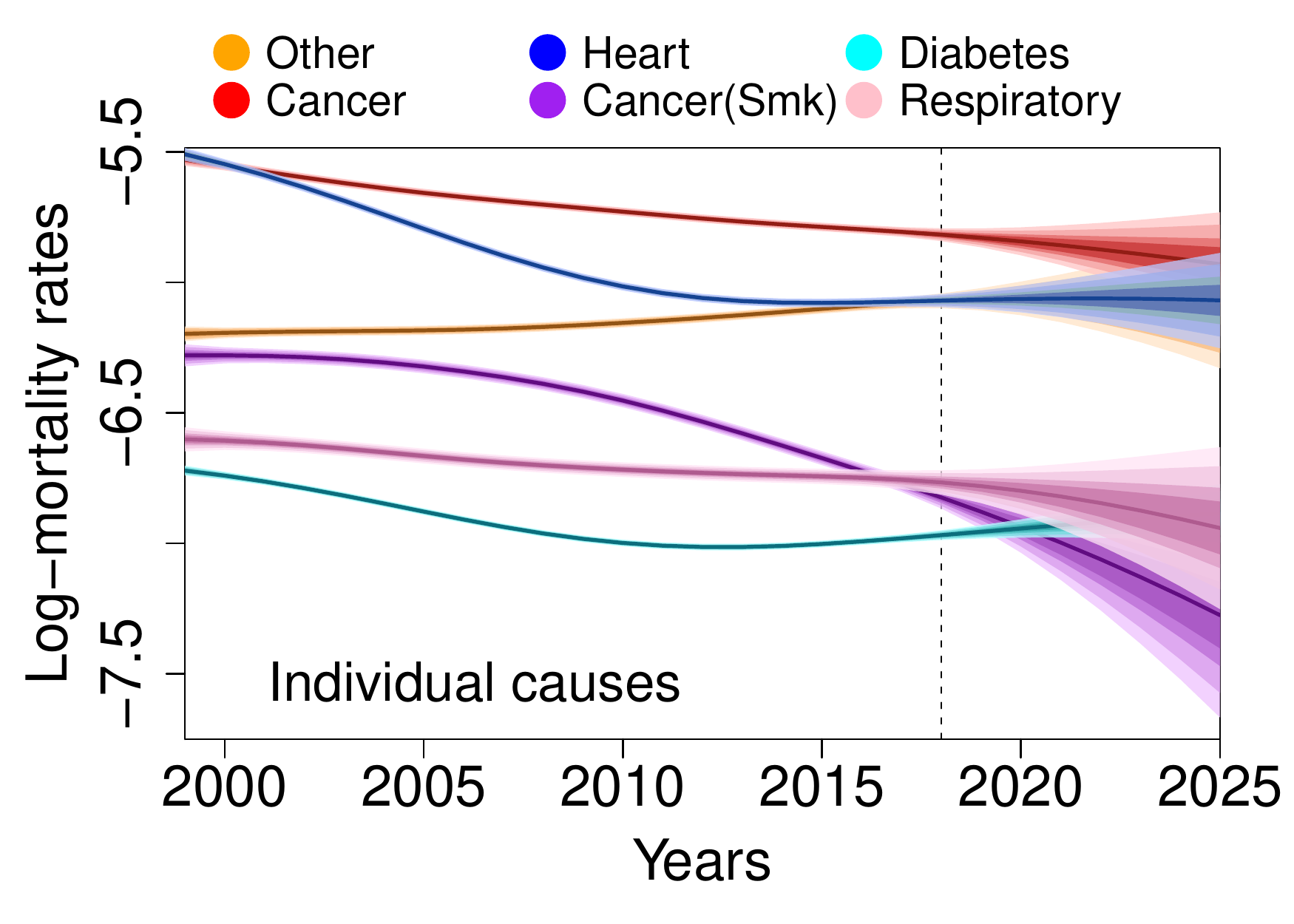}}}
    {{\includegraphics[width=5.35cm]{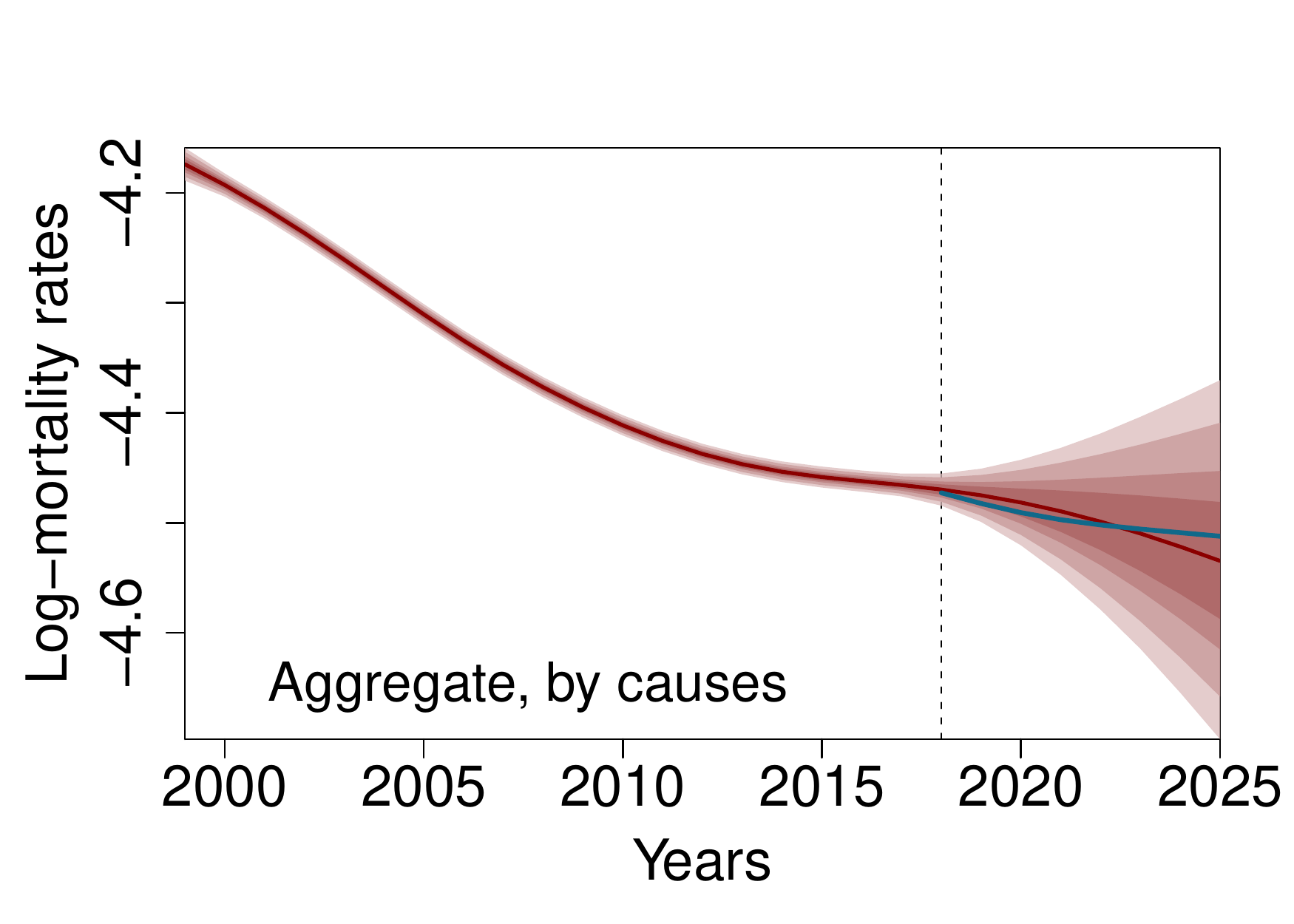}}}
    {{\includegraphics[width=5.35cm]{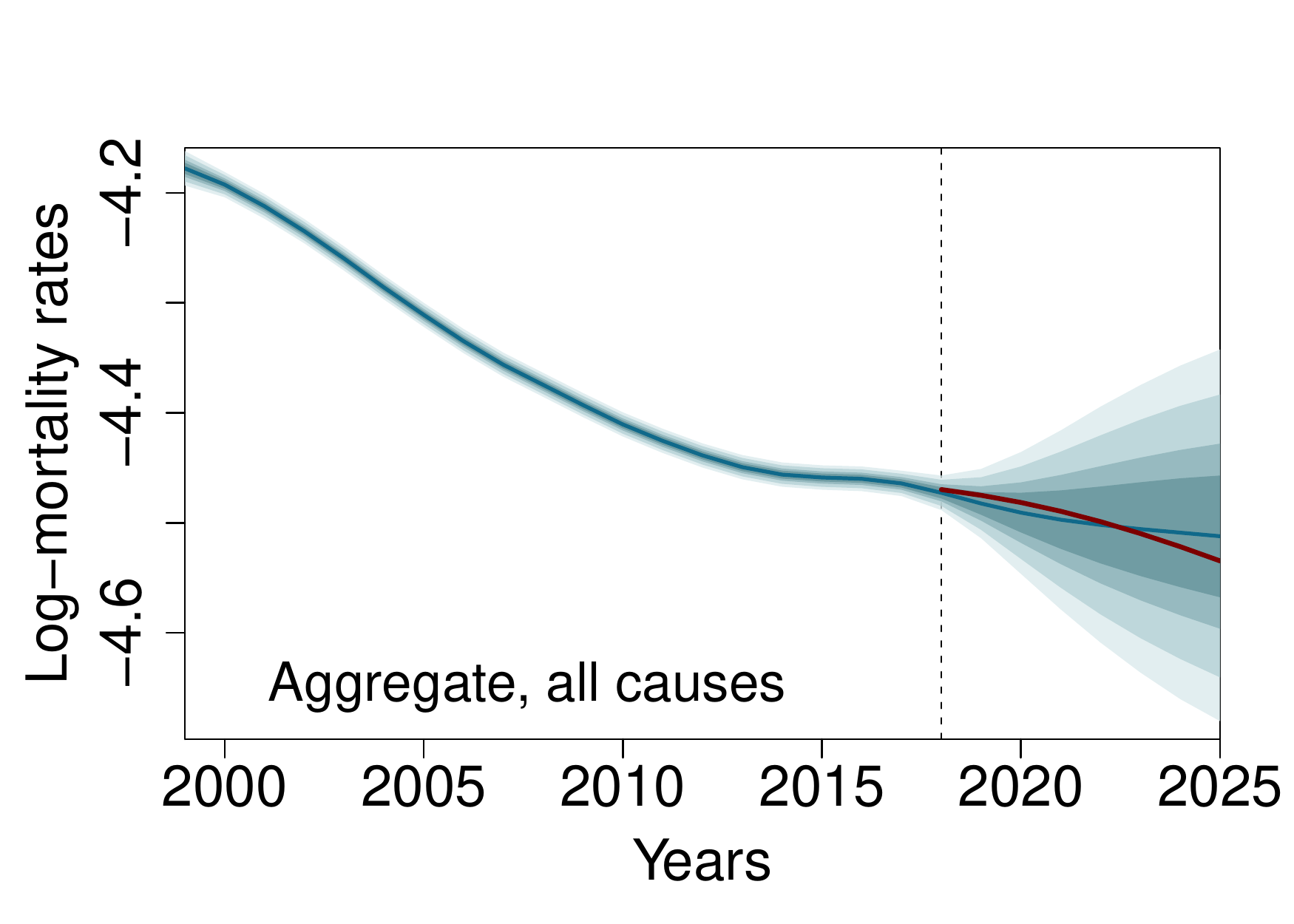}}} \\

    \caption{Posterior distribution of true log-mortality rates for US Males and Females for the 40-44 and 65-69 Age groups. In each row: left panel shows the posterior quantiles for top 6 individual causes (via multi-cause GP); middle panel shows aggregate log-mortality trends via by-cause model (multi-cause GP); right panel shows aggregate log-mortality trends via all-cause model (single-output GP). All models are fitted on Ages 40--69 and Years 1999--2018. The vertical lines indicate the boundary between in-sample (1999--2018) and out-of-sample forecast (2019--2025). Shading indicates the 60\%, 80\%, 95\% and 99\% predictive quantile bands. We further overlay the 50\%-quantile (predictive median) of by-cause and all-cause models for the out-of-sample period for convenient comparison.
    \label{fig:us-projections}}
\end{figure}

Figure \ref{fig:us-projections} presents the in-sample posterior distribution of log-mortality rates $f_l(x_*)$ during 1999--2018 along with the projected trends up to 2025 for two Age groups, in both Male and Female US populations. The left panels display the predictive trends for the top 6 causes in each age-group. We observe improvements in most causes and age groups; the largest improvement being in CANL. In contrast, Drug abuse deaths are rising rapidly among  young age groups (e.g: Age 40--44), see the 2 upper-left panels of  Figure~\ref{fig:us-projections}. In the older age groups, mortality from Heart disease experienced large declines in early 2000s, but essentially flattened out after 2015. \cite{codUS2019} emphasized the need for break-point detection to improve forecasts of such causes whose trends change over time. In the MOGP framework, the forecasts are driven by the most recent data and are thus automatically adjusted if trends shift. So for example, we do not need to do any break-point analysis to achieve the slow pace of future HEA improvement shown in Figure~\ref{fig:us-projections}.

%Heart disease has decreased over time but achieved relatively larger improvements in the older age groups (e.g., Age 65--69). However, the improvement has slowed down in recent years. In their report, the SOA emphasized incorporating break-point detection in the historical trends in different causes given trends are changing. By calibrating breakpoint detection within their proposed framework, the forecasted trends of heart disease will slow down, rather than extending the fast pace of decreasing in the early 2000s. In our study, we did not incorporate the breakpoint detection to `correct' the forecasted trends of heart disease. The forecasting mechanism in GP puts larger weights on more recent training data points. As a result, the projected trends via GP will be heavily influenced by the recent trends observed in the data, naturally resulting in the slow pace of the forecasted improvement trends of heart disease in the future for the next 5-10 years.

The middle and right columns in Figure~\ref{fig:us-projections} compare the aggregate all-cause projections for the US population. We witness the pessimism of the aggregate projected trends based on the by-cause models compared to an all-cause SOGP (right column), especially in the  younger ages. This discrepancy, driven by the growing importance of causes with increasing trends, such as Drug overdose, highlights the additional insights from by-cause modeling. The pessimism of  by-cause analysis was first mentioned in~\cite{Wilmoth1995} and re-iterated in~\cite{codUS2019}. For older age groups the underlying dynamics among common causes are more stable and all-cause and by-cause forecasts are broadly similar.

%Note that via simulation \todo{??}, the probabilistic forecasts of the aggregate trends depend highly on the dynamics between in-sample trends of the mortality improvement among individual causes. For Age 40--44 in both genders, we observe wider spread of the predictive aggregate trends (compared to older age groups) when going beyond 2020. This is driven by the consistently large improvement in cancer induced by smoking that leaves a significant gap with remaining causes by end of the training period.

Note that compared to Figure~\ref{fig:all_cancer}, the GP posterior uncertainty bands widen dramatically as we go from in-sample (up to Year 2018) to extrapolating for Years 2019--2025. This reflects the data-driven nature of GP forecasts which intrinsically leads to low uncertainty for in-sample smoothing and widening uncertainty as predictions are made further into the future. This phenomenon gets amplified as we add up the by-cause forecasts to obtain all-cause predictions, witness the wider band in the middle panels of Figure~\ref{fig:all_cancer} relative to the right panel based on a SOGP.

As discussed in \cite{huynh2020}, MOGPs are well-suited to generate expert-based projections. This is a useful feature to have given that by default projections are driven by the historical trends that might not continue in the future. For instance, the increasing trend of Drug abuse is largely fueled by the opioid epidemic. Assuming this crisis is addressed in the future, the projected DRU mortality should be adjusted downward. In MOGP this can be achieved by modifying the Year trend in $m(\cdot)$. Figure~
\ref{fig:expert-based} in the Appendix displays an illustration where the trend of Drug abuse is reduced by one third (through lowering the Year effect $\beta_{yr}$ by one third) of the original pace for both the Male and Female populations. The resulting adjusted forecast for aggregate mortality gets closer to that from the all-cause model, reducing the level of pessimism we have observed earlier among the US young population. Another potential adjustment could be for smoking-induced cancer, where the MOGP models extrapolate the historical trend of rapid longevity gains. However, the SOA report \citep{codUS2019} suggests that this pace might not take place in the intermediate term, lowering aggregate mortality gains.

 %Details on the mechanism of how MOGP framework combines expert opinions for mortality projections can be found in \cite{huynh2020}.

% \todo{Cross-reference the relevant equation and emphasize what it means to sample a 'trajectory'}

%\begin{figure}[!t]
%    \centering
%    \textbf{AGE GROUP 55-59, MALES} \\
%    \vspace{1.5mm}
%    \includegraphics[width=15cm]{all-cancer-sim-m57.pdf} \\
%    \textbf{AGE GROUP 65-69, MALES} \\
%    \vspace{1.5mm}
%    \includegraphics[width=15cm]{all-cancer-sim-m67.pdf} \\
%    \textbf{AGE GROUP 75-79, MALES} \\
%    \vspace{1.5mm}
%    \includegraphics[width=15cm]{all-cancer-sim-m77.pdf}
%    \caption{Predictive quantiles of all-cancer log-mortality rates by Country and Age group. The results are from  a full-rank multi-level ICM, fitted on Ages 50--84 and Years 1998-2013 and 3 categorical covariates of Country (3 levels), Cause (5 levels), and Gender (2 levels). The vertical dash lines in each plot separate in-sample and out-of-sample periods.}
%    \label{fig:predictive_all_cancer}
%\end{figure}
%

\section{Cause-of-Death Joint Modeling in a Multinational Context \label{sec:multi-national}}

We proceed to consider simultaneously 30 populations in the Cancer variations case study, arranged by the three factor inputs of Cause $(L_{caus}=5$), Country ($L_{ctry}=3$) and Gender ($L_{gndr}=2$).

\subsection{Multi-level vs Single-level Correlation Structure}

\begin{figure}[!t]
    \centering
    \subfloat[Multi-level ICM ($Q_{ctry}=3$, $Q_{caus}=5$, $Q_{gndr} =2$)  with 40 hyperparameters]{{\includegraphics[trim={0.7cm 0cm 1.15cm 0.35cm},clip=true,  width=7.5cm]{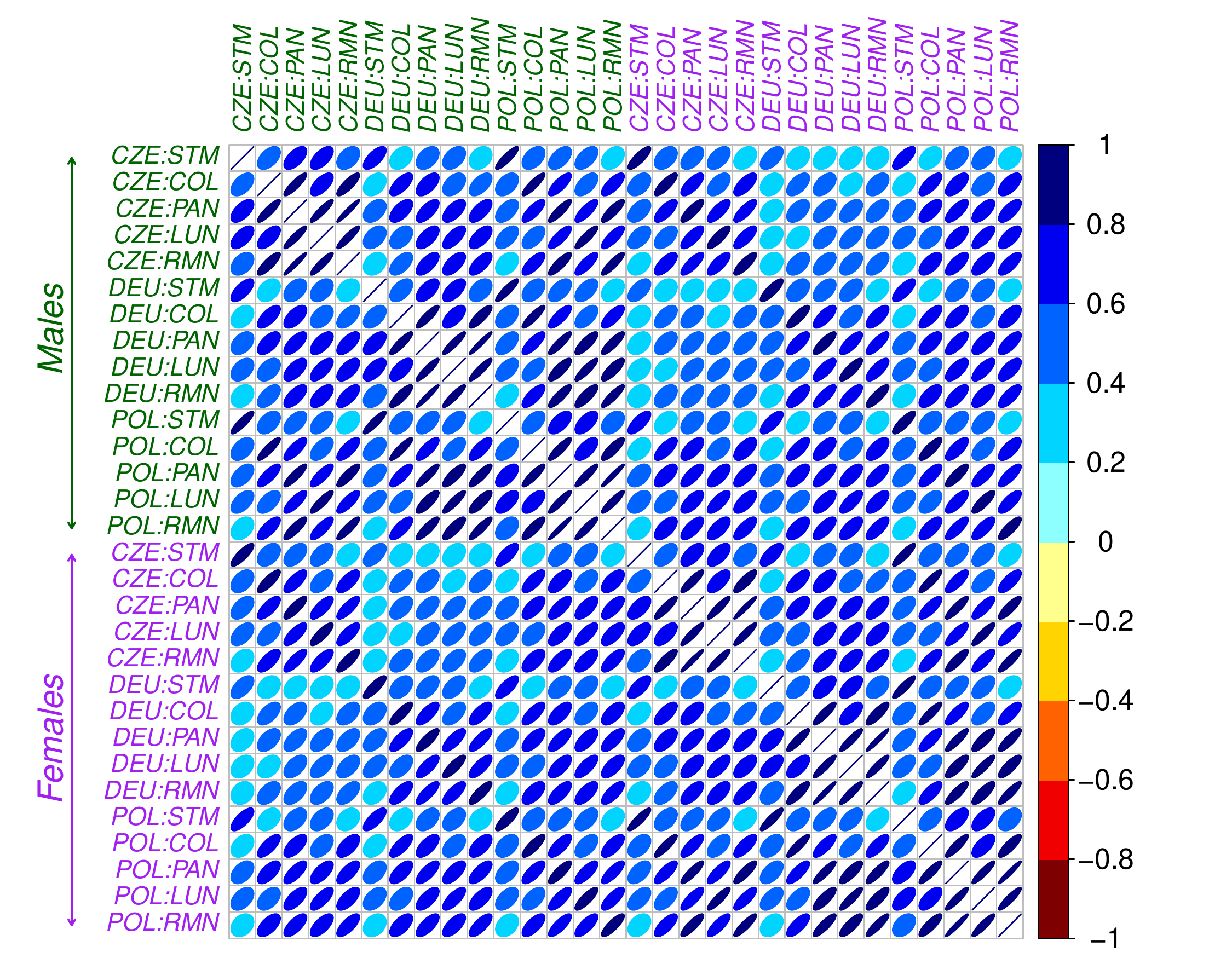}}} \qquad
    \subfloat[Single-level ICM ($Q=2$) with 62 hyperparameters ]{{\includegraphics[trim={0.7cm 0cm 1.15cm 0.35cm},clip=true, width=7.5cm]{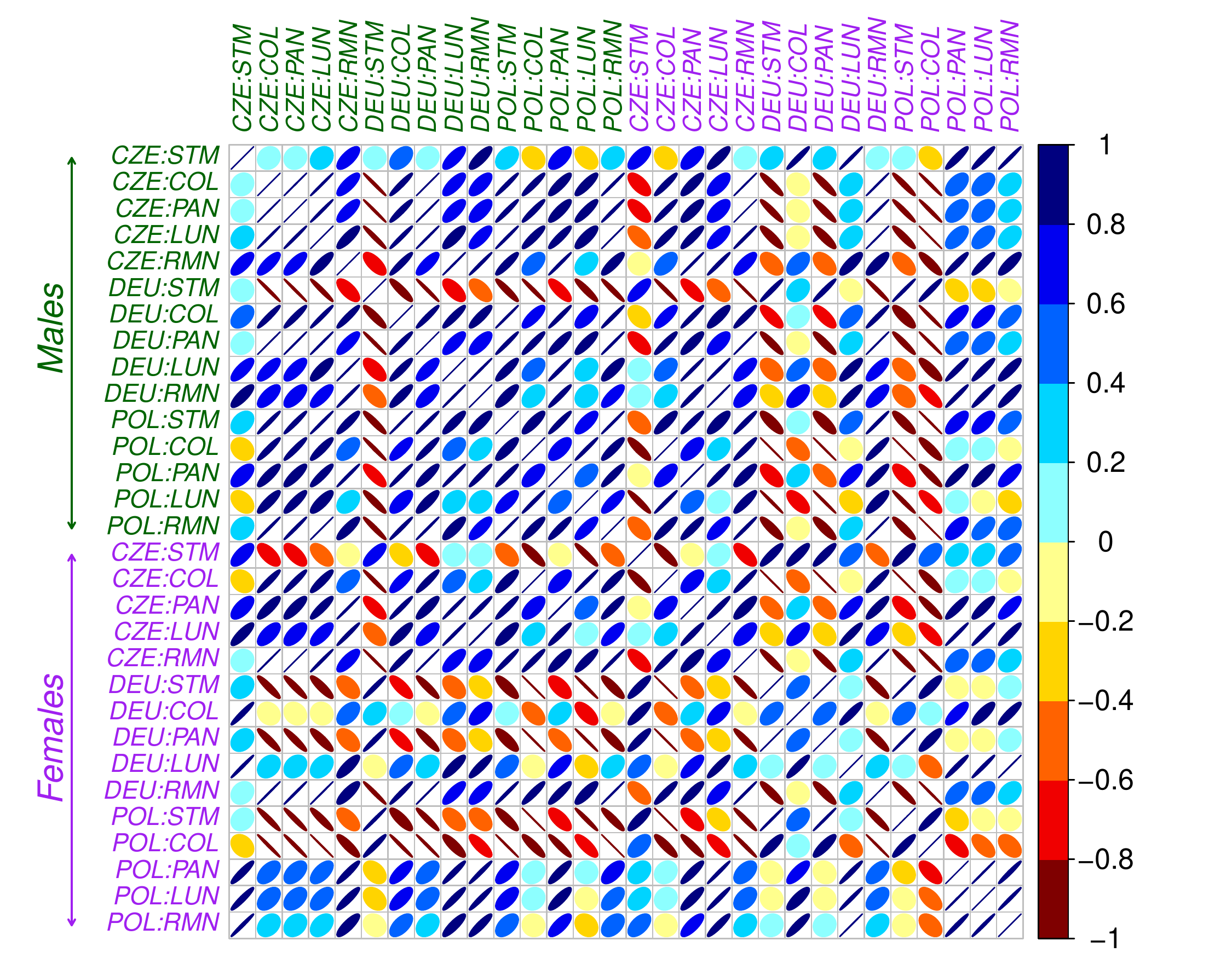}}} \\
    \caption{Cross-correlation matrices of MOGP models that incorporate Country, Cause, and Gender as categorical inputs. Thirty total populations.
    \label{fig:compare-hicm-icm}}
\end{figure}

\begin{figure}
    \centering
    \subfloat[Cross-country correlation ($R_C$)]{{\includegraphics[trim={0.1cm 1.5cm 0.1cm 1.75cm},clip=true, width=5.5cm]{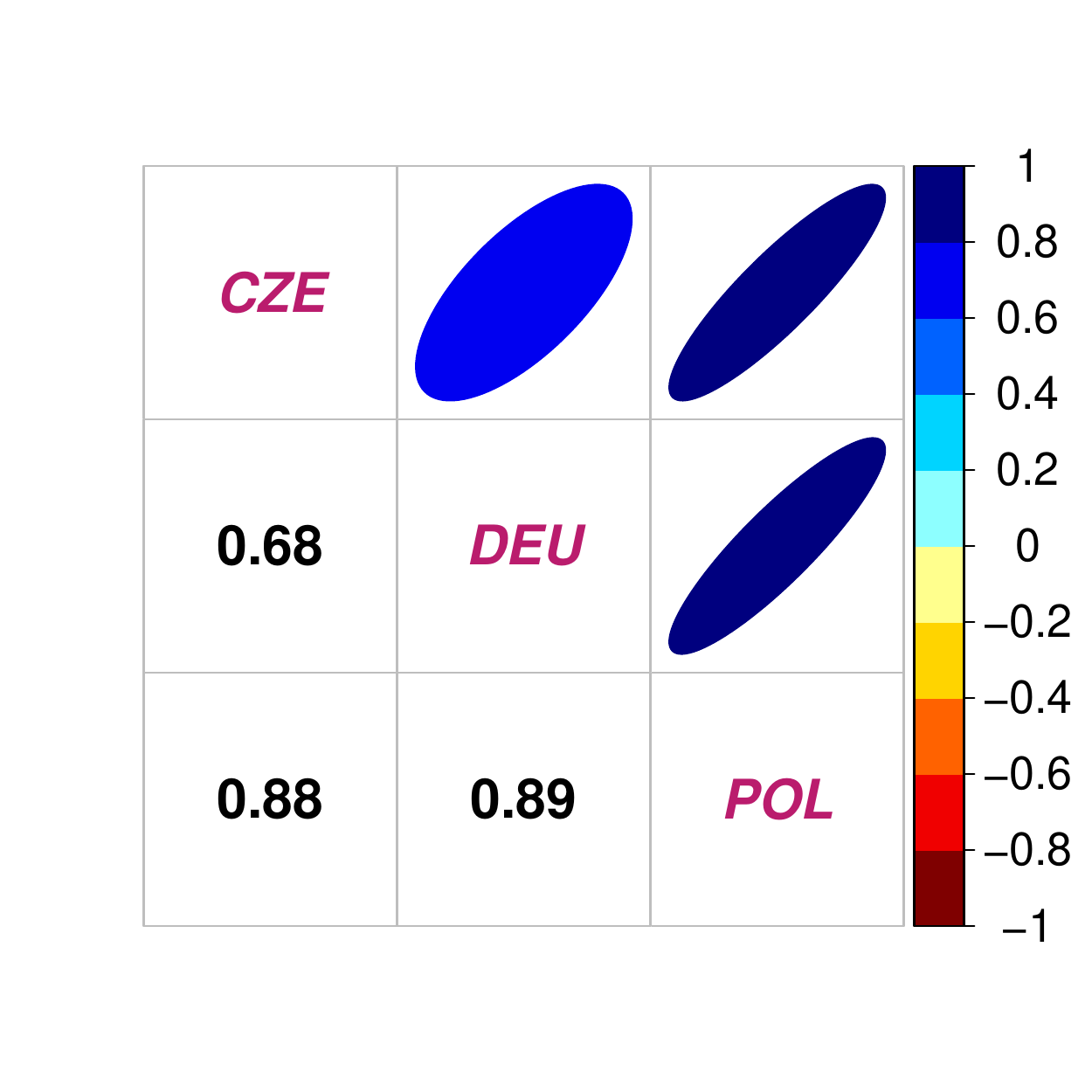}}}
    \subfloat[Cross-cause correlation ($R_S$)]{{\includegraphics[trim={0.4cm 1.7cm 0.3cm 1.75cm},clip=true, width=5.5cm]{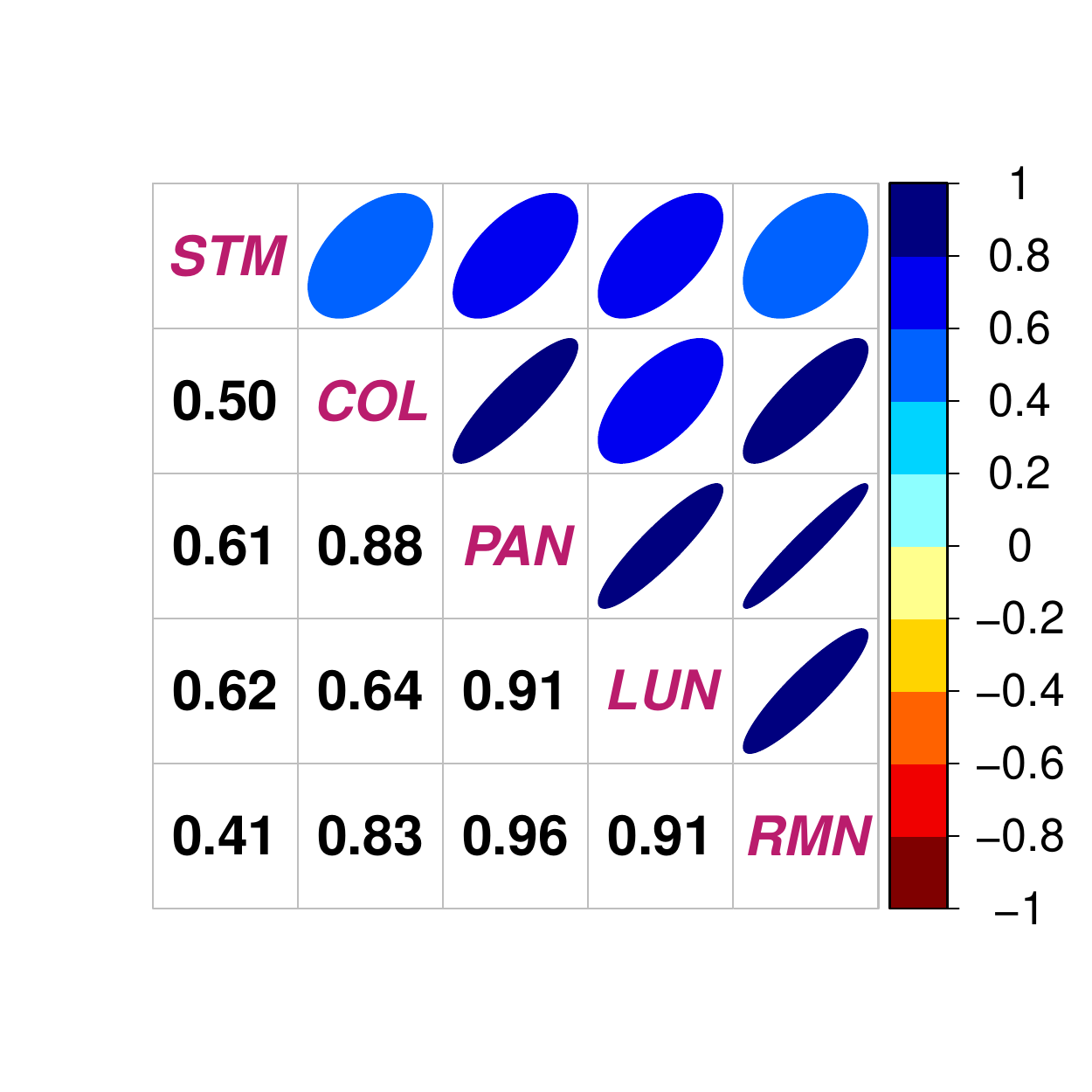}}}
    \subfloat[Cross-gender correlation ($R_G$)]{{\includegraphics[trim={0cm 1.5cm 0cm 1.6cm},clip=true, width=5.5cm]{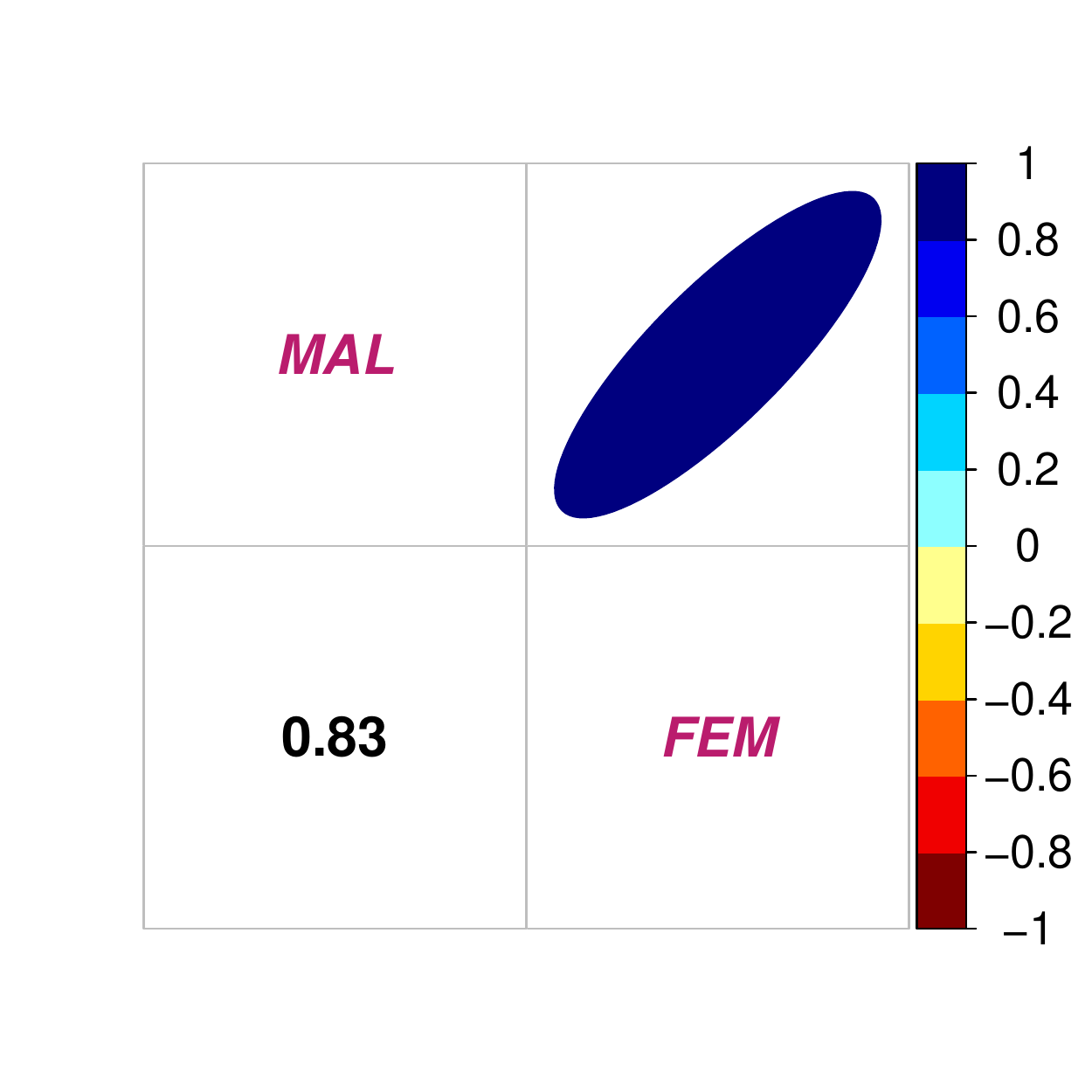}}}
    \caption{Cross-correlation matrices derived from a three-level Country-Cause-Gender multi-level ICM model ($Q_{ctry}=3,~Q_{caus}=5,~Q_{gndr}=2$), fitted on Age groups 50--84 and Years 1998--2016.}
    \label{fig:corr_hier_icm}
\end{figure}

Figure \ref{fig:compare-hicm-icm} displays the inferred correlation structure $r_{l_1,l_2} : 1 \le l_1,l_2 \le 30$ across the above 30 populations. On the left we fit a multi-level ICM ($Q_{ctry}=3,~Q_{caus}=5,~Q_{gndr}=2$)
 and on the right a single-level ICM ($Q=2$). Both models are fitted on Age groups 50--84, Years 1998--2016 for 3 Countries, 5 Cancers, and 2 Genders.  Observe that the multi-level model has $\sum_p Q_p L_p + 2 = 3 \times 3 + 5 \times 5 + 2 \times 2 + 2 = 38$ hyperparameters compared to $Q L + 2 = 62$ in the single-level model.
%
% Male and Female population across all countries and cancer types are highly correlated with $r_{Mal,Fem} \approx 0.80$.
 The right panel does not display any recognisable structure in the inferred correlations; because the model is not aware of the different factor dimensions, after dimension reduction the marginal associations between sub-populations within the original factor inputs are no longer accessible. In contrast, the multi-level model enforces a block structure in the correlation matrix $R$, see Figure~\ref{fig:compare-hicm-icm} (a). Recall that the correlation sub-matrices for each factor input are estimated separately and the Kronecker product structure implies that we can read off the correlation among any combination of factors. Figure~\ref{fig:corr_hier_icm} displays the derived sub cross-correlation matrices from the above three-level Country-Cause-Gender GP model. One can then multiply these factor-based correlations to get the total correlation $r_{(c_1,s_1,g_1),(c_2,s_2,g_2)} = \prod_{p \in \{c,s,g\}} r_{p_1,p_2}$ in  Figure~\ref{fig:compare-hicm-icm}(a). For example, the correlation of Cancer mortality between Czech-Lung-Male and Polish-Pancreas-Male is $r_{CZE,~POL} \times r_{LUN,~PAN} \times r_{MAL,~MAL} = 0.88 \times 0.91 \times 1 \approx 0.80$. This is in fact also the correlation between Czech-Lung-Female and Polish-Pancreas-Female ($r_{CZE,~POL} \times r_{LUN,~PAN} \times r_{FEM,~FEM}$) mortality.
The limited number of deaths in some causes explains differences in the correlation matrices between single- and multi-level ICM. For example, the correlations between Polish-Stomach-Female and other cancer types are mostly negatively correlated in ICM but (mildly) positively correlated in the multi-level ICM.

% We test the predictive performance of MOGP models by computing the 3-year median MAPE and CRPS from one-year-ahead mortality forecasts in three separate test sets. All models, including SOGP models, are trained on the same Ages from 50--84 and three overlapping periods: 1998--2013 for 2014 prediction, 1999-2014 for 2015 prediction, and lastly 2000-2015 for 2016 prediction. The differences in MAPE and CRPS between MOGP and SOGP models are expressed as the percentage improvement over SOGP models. Positive improvement means joint models have smaller MAPE/ CRPS values. We compute MAPE and CRPS for all-cancer log-mortality rates instead of individual cancers. The hold-out samples for cancer variations, such as Stomach and Pancreas may not be highly reliable due to having relatively fewer observations. Lastly, it is straightforward to compute MAPE for all-cancer log-mortality. To compute CRPS, we take advantage of the stochastic feature in GP to simulate the full forecast distribution of all-cancer log-mortality for each Age group in the data.

\subsection{Model Selection}

Next, we compare the predictive performance of multi-level vs single-level MOGP-ICM. Following the same setup as Table~\ref{tbl:mogp-cause}, Table~\ref{tbl:mogp-rank} shows the 3-year median improvement in APE and CRPS for different joint models relative to SOGP models in a multinational context. Joint models produce more accurate mean forecasts with higher credibility but the predictive gains are not uniform across countries. The largest improvements from multi-level ICM are in
Czech Republic; Czech raw data has lower credibility due to its smaller exposures, thus there is more opportunity for data fusion. The results further validate our approach of modelling cause-specific mortality across populations: models that incorporate information from foreign countries (e.g.~Country-Cause GP) have larger predictive improvements compared to Table~\ref{tbl:mogp-cause}. In Poland due to the structural break encountered in mid-2010s (see Fig.~\ref{fig:all_cancer}), the performance of aggregated Country-Cause MOGP is consistently worse than those of all-cause SOGP; this issue is rectified for Country-Cause-Gender MOGPs.

\begin{table}[!ht]
\centering
\caption{Comparison between MOGP models with different ranks $Q$ in terms of  APE and CRPS metrics. The reported values are median relative improvements of MOGP vs SOGP of one-year-out aggregated all-Cancer forecasts for age groups 50--84 based on 3 training periods: 1998--2013 (predict  2014), 1999--2014 (predict 2015), and 2000--2015 (predict 2016). Top half: MOGP models fitted on 5 cancer types and 3 countries. Bottom half: MOGP models fitted on 5 cancer types, 3 countries and 2 genders. \label{tbl:mogp-rank}}
\resizebox{\columnwidth}{!}{%
\begin{tabular}{llcrrrrrr}  \toprule
\multicolumn{2}{l}{\multirow{2}{*}{\large{Country$+$Cause ($L=15$)}}} & \multirow{2}{*}{\begin{tabular}[c]{@{}c@{}}\# Kernel\\ Hyperparameters\end{tabular}}    & \multicolumn{2}{c}{Czech Rep.} & \multicolumn{2}{c}{Germany} & \multicolumn{2}{c}{Poland} \\ \cline{4-9}
\multicolumn{2}{l}{} &   &  APE        & CRPS        & APE         & CRPS         & APE         & CRPS        \\  \midrule
\multirow{3}{*}{\large{ICM}}   & $Q=2$    & 32  & 35.54       & 30.74       & 12.02        & 21.06        & $-$40.44       & $-$59.47      \\
& $Q=3$   &  47   & 45.51       & 50.07       & 8.50         & 20.35        & $-$37.09       & $-$38.48      \\
& $Q=4$   &  62   & 31.64       & 36.56       & 14.08        & 35.66        & $-$13.32       & 9.56        \\  \midrule
% & $Q=5$   &  77   & 37.95       & 42.02       & 27.52        & 40.57        & $-$19.94       & $-$0.38       \\
\multirow{3}{*}{\large{SLFM}}   & $Q=2$   &  34  & 31.51       & 24.57       & 4.44         & 15.87        & $-$47.66       & $-$45.98      \\
& $Q=3$  &   51   & 33.78       & 41.59       & 0.88         & 8.62         & $-$37.25       & $-$24.54      \\
& $Q=4$  &  68   & 44.79       & 43.29       & 8.93         & 20.49        & $-$45.63       & $-$37.66      \\ \midrule
% & $Q=5$  &   85   & 21.05       & 30.70       & 1.32         & 23.22        & $-$1.95        & 16.31       \\
\multirow{3}{*}{Multi-level ICM} & $Q_{ctry}=2,~ Q_{caus} = 2$ & 18 & 26.42       & 24.24       & 4.55         & 6.84         & $-$48.46       & $-$61.53      \\
& $Q_{ctry}=2,~ Q_{caus} = 4$  & 28 & 25.74       & 32.26       & 5.04         & 15.75        & $-$35.94       & $-$44.24      \\
& $Q_{ctry}=3,~ Q_{caus} = 4$  & 36 & 42.03       & 36.80       & 3.07         & 11.12        & $-$44.45       & $-$53.69     \\ \midrule
\multicolumn{2}{l}{\multirow{2}{*}{\large{Country$+$Cause$+$Gender ($L=30$)}}} & \multirow{2}{*}{\begin{tabular}[c]{@{}c@{}}\# Kernel\\ Hyperparameters\end{tabular}}    & \multicolumn{2}{c}{Czech Rep.} & \multicolumn{2}{c}{Germany} & \multicolumn{2}{c}{Poland} \\ \cline{4-9}
\multicolumn{2}{l}{} &   & APE        & CRPS        & APE         & CRPS         & APE         & CRPS        \\  \midrule
\multirow{2}{*}{ICM} & $Q=2$  & 62 & 9.99      & 11.75       & $-$5.40         & $-$13.40        & 12.21       & 6.55     \\
& $Q=3$  & 92 & 0.52      & 3.07       & $-$10.78       & $-$20.35        &    $-$15.31    & 4.09     \\ \midrule
\multirow{2}{*}{Multi-level ICM} & $Q_{ctry} = 2,~Q_{caus} = 4,~Q_{gndr}=2$ & 32 & 14.85      & 24.91       & $-$68.45         & $-$15.22        & $-$25.11       & 7.75     \\
& $Q_{ctry} = 3,~Q_{caus} = 5,~Q_{gndr}=2$ & 40  & 21.34      & 27.22       & $-$40.45         & $-$10.22        & 5.12       & 8.04     \\
 \midrule
\end{tabular}}
\end{table}

\begin{figure}[!ht]
    \centering
    \textbf{STOMACH CANCER} \\
    \vspace{1mm}
    {{\includegraphics[width=15.25cm]{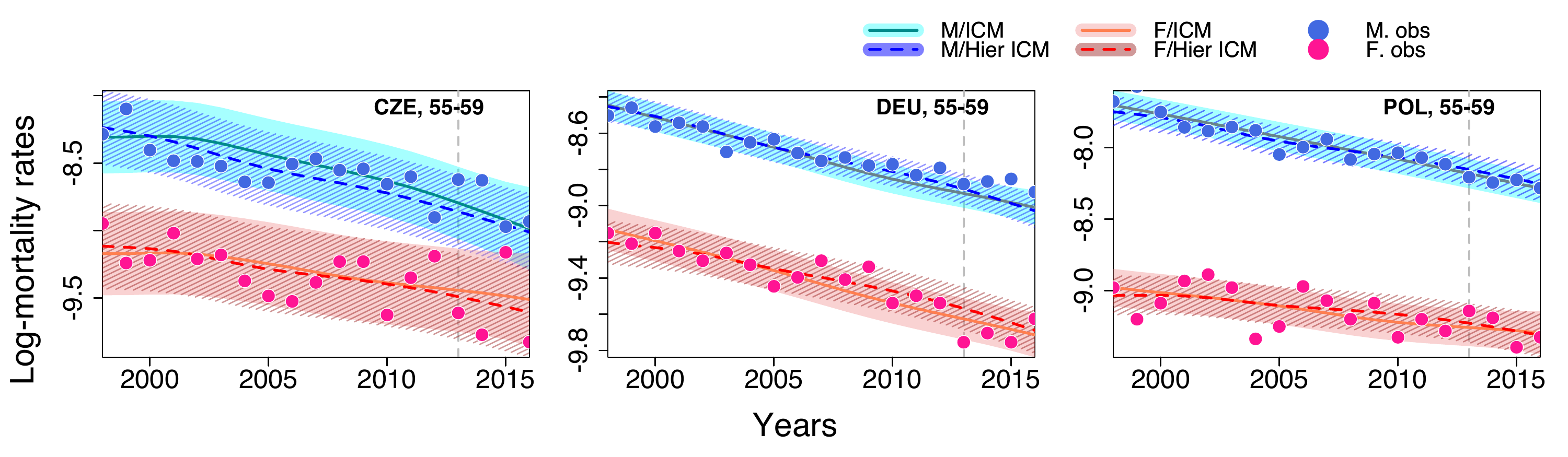}}} \\
    \textbf{COLORECTAL CANCER} \\
    \vspace{1mm}
    {{\includegraphics[width=15.25cm]{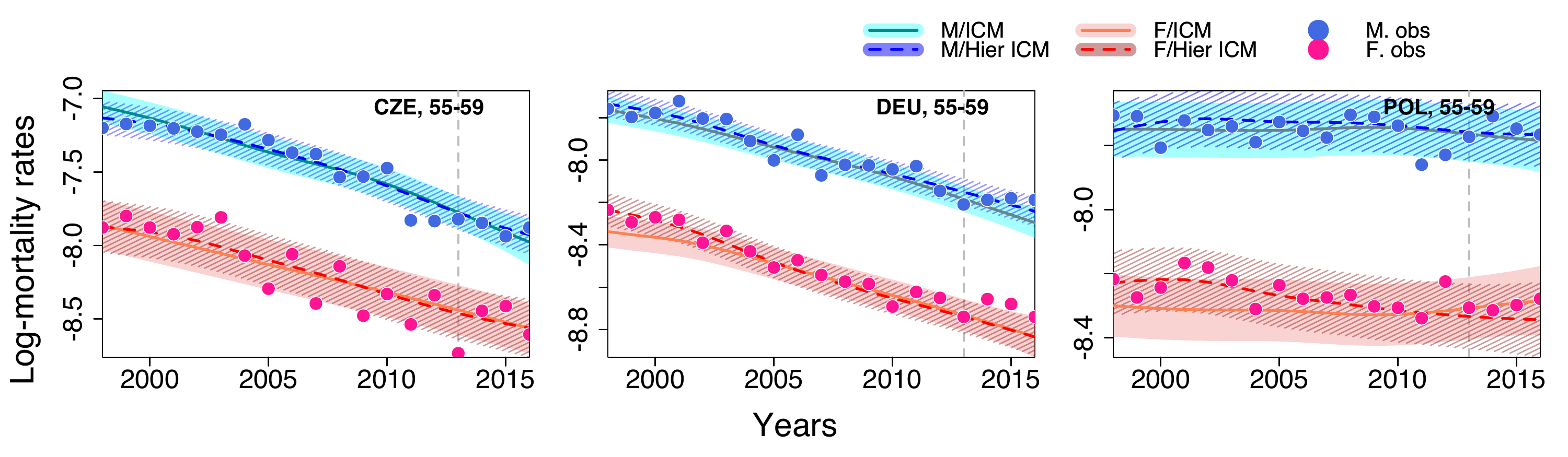}}} \\
    \textbf{{PANCREATIC CANCER}} \\
    \vspace{1mm}
    {{\includegraphics[width=15.25cm]{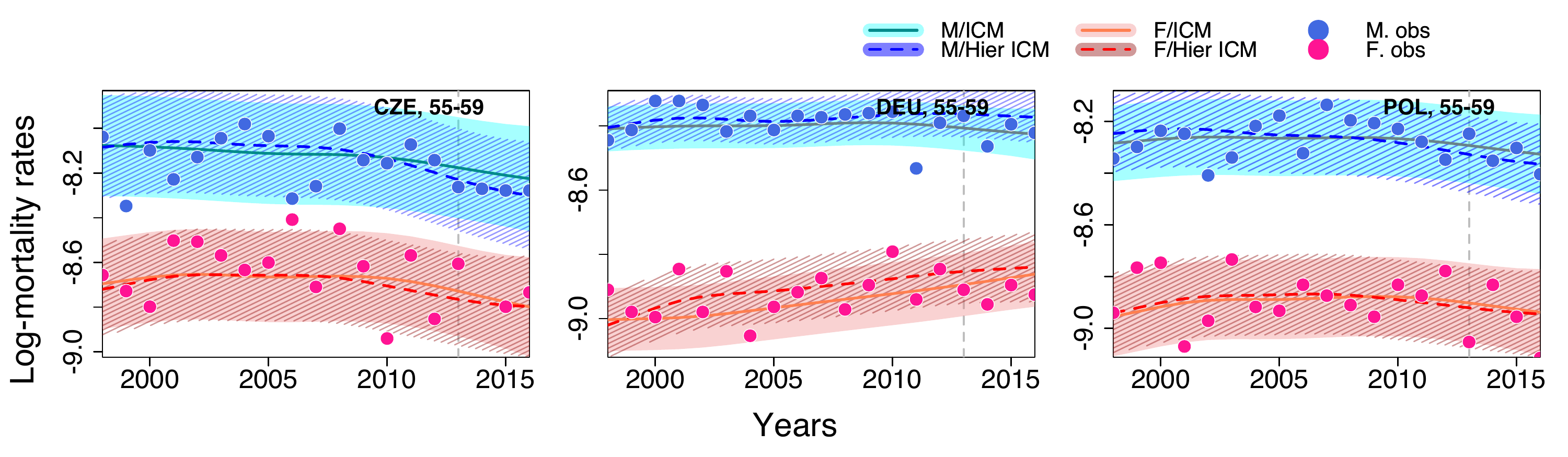}}} \\
    \textbf{LUNG CANCER} \\
    \vspace{1mm}
    {{\includegraphics[width=15.25cm]{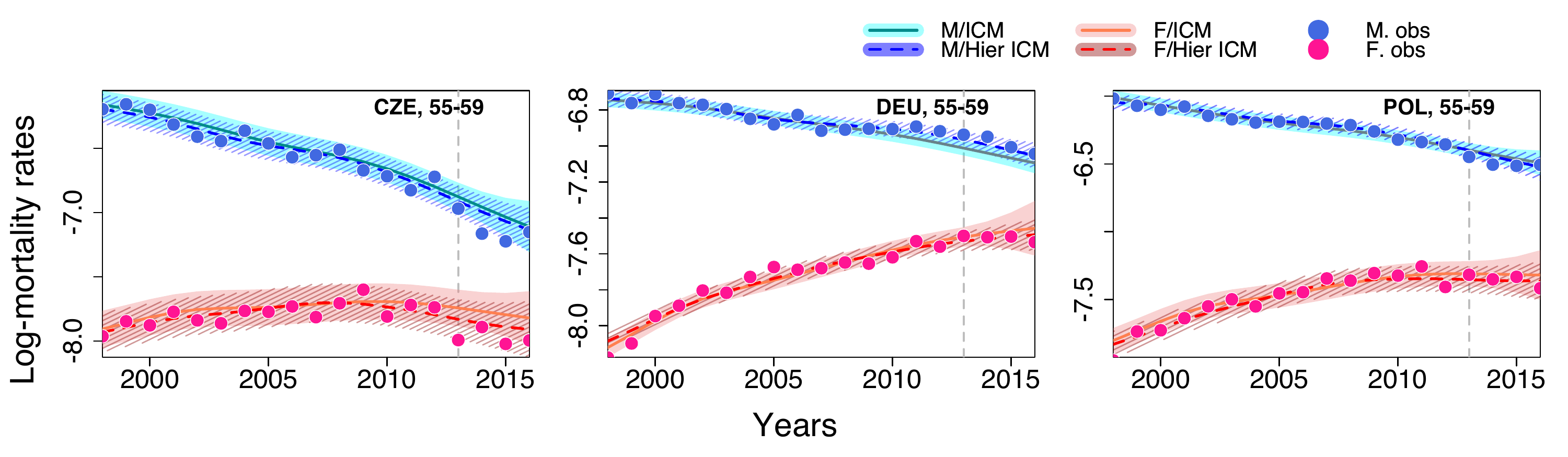}}}
    \caption{Predictive log-mortality distributions from single-level ICM ($Q=2$) and multi-level ICM ($Q_{ctry}=3,~Q_{caus}=5,~Q_{gndr}=2$) models with 3 factor inputs: Country, Cause, and Gender (30 populations). All models are fitted on Age groups 50--84 and Years 1998--2013, and applied for 3-year-out forecasts up to 2016. The shadings indicate 95\% predictive bands; the vertical lines mark the edge of training data. Note different $y$-axes in each panel. \label{fig:hicm-icm-can}}
\end{figure}

In contrast to Section~\ref{sec:results} where the performance of ICM and SLFM was almost identical, with multiple input factors ICM usually outperforms SLFM. Given the strong commonality in mortality trends across Cancer types, a shared Age-Year covariance kernel appears preferable for information fusion. The comparison between single- and multi-level ICM depends upon the number of populations to model. In Country-Cause-Gender setting with $L=30$ populations, the  multi-level ICM ($Q_{ctry}=3,~Q_{caus}=5,~Q_{gndr}=2$) yields better mean APE and CRPS than single-level ICM, and moreover uses fewer hyperparameters. For Country-Cause setting, the performance is comparable.

Table \ref{tbl:mogp-rank} also shows the impact of the latent ranks $Q$ and $Q_p$'s.  For Country-Cause, $Q=4$ tends to yield the best results in single-level models; but for Country-Cause-Gender, ICM with $Q=3$ performs consistently worse than $Q=2$, presumably due to unstable estimates of the more than 90 underlying hyperparameters. For multi-level ICM, we generally find that full-rank $Q_p = L_p$ works best, although low-rank setups $Q_p < L_p$ also yield good predictive performance, indicating the opportunity  to shrink even further the number of kernel hyperparameters.

%We highlight the scalability feature in multi-level ICM by listing the number of kernel hyperparameters in Table \ref{tbl:mogp-rank}.  Even with low rank such as $Q=3$ in ICM,  we already suffer great computational costs with far higher number of hyperparameters to manage.

%Previous section reported a negligible discrepancy in in-sample fitting between MOGP-ICM and SLFM (both with rank $Q=2$) for individual cancers. Despite that, we notice the predictive improvements for all-cancer mortality in ICM models are larger than in SLFM, in both Multi-cause GP and Country-Cause GP cases. Given that there is indeed a commonality in log-mortality trends among cancer types, sharing the length-scales in Age and Year in the spatial covariance kernel strengthens the effect of borrowing information to boost the predictive performance of ICM.  G

%significant improvements over single-population models while having smaller kernel hyperparameter space compared to full-rank approach ($Q_p = L_p, ~1\leq p \leq P$). Similar to Single-level ICM or SLFM, increasing rank values of $B_p$'s in Multi-level ICM should be considered to maximize the predictive performance.

Figure \ref{fig:hicm-icm-can} shows the predicted log-mortality rates for individual Cancers and Age group 55-59 via full-rank multi-level ICM and single-level ICM with $Q=2$. Both models are fitted on Age groups 50--84 and Years 1998--2016 before we perform out-of-sample forecasts for the next 3 years (2014--2016). The single-level ICM produces over-smoothed forecasts $m_*(\cdot)$ for several cancers, especially cancers with large observation noise like Stomach and Pancreas; this problem is mitigated by the shorter length-scale in Year in multi-level ICM ($\theta_{yr} \approx 8.8$ versus $\theta_{yr} \approx 14.8$ in single-level ICM).

\textbf{Coherence in cause-specific trends:} Figure \ref{fig:hicm-icm-can} demonstrates that Males and Females do not always share similar progress in mortality reduction. While the trends in Stomach, Colorectal, and Pancreatic cancers are consistent for both genders, for Lung cancer the Male-Female gap is diminishing rapidly, especially in Germany and Poland. This is driven by a decrease in cigarette consumption among men, while women are more likely to develop Lung cancers that are not associated with smoking. Thus, the concept of forecast coherence (namely extrapolating a stable Male-Female spread, as observed historically) is not always well suited for cause-of-death analysis. %It is still possible to observe the divergence forecasts in the long horizon. Perhaps the most recent divergence is when cancer overtakes heart disease to be the leading cause of deaths worldwide.

\clearpage

\subsection{Borrowing the Latest Datasets from Other Populations}

The period coverage in HCD varies by country as datasets for countries are uploaded asynchronously. This implies that some countries have more up-to-date datasets than others, see Figure \ref{fig:hcd}. This offers opportunities to fuse data from other countries to update domestic forecasts. For Age-Period-Cohort models, such as \cite{LI2005}, this is generally challenging as the model fitting relies on having a rectangular dataset.  Our MOGP framework can straightforwardly handle such `notched’ datasets to take full advantage of additional observations in different countries.

As an illustration, we choose Czech Males as a reference population and examine one-year-out prediction quality for several models. We take Czech observations for 1998--2015 and borrow the more up-to-date dataset from Germany (1998--2016) to implement the notched setup. Table \ref{tbl:notched-setup} in the Appendix describes the setup of the different models we consider in this experiment.

\begin{figure}[H]
    \centering
    %\textbf{ALL-CANCER MORTALITY IN CZECH MALES} \\
    \vspace{0.25cm}
    {\includegraphics[width = 6in]{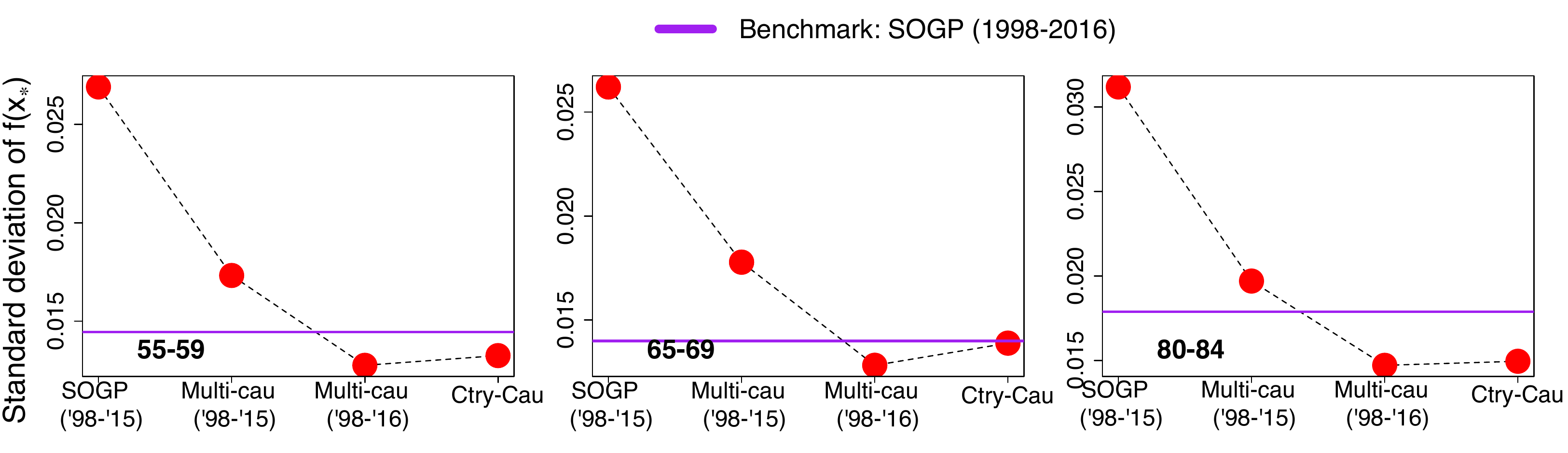}}
    {\includegraphics[width = 6in]{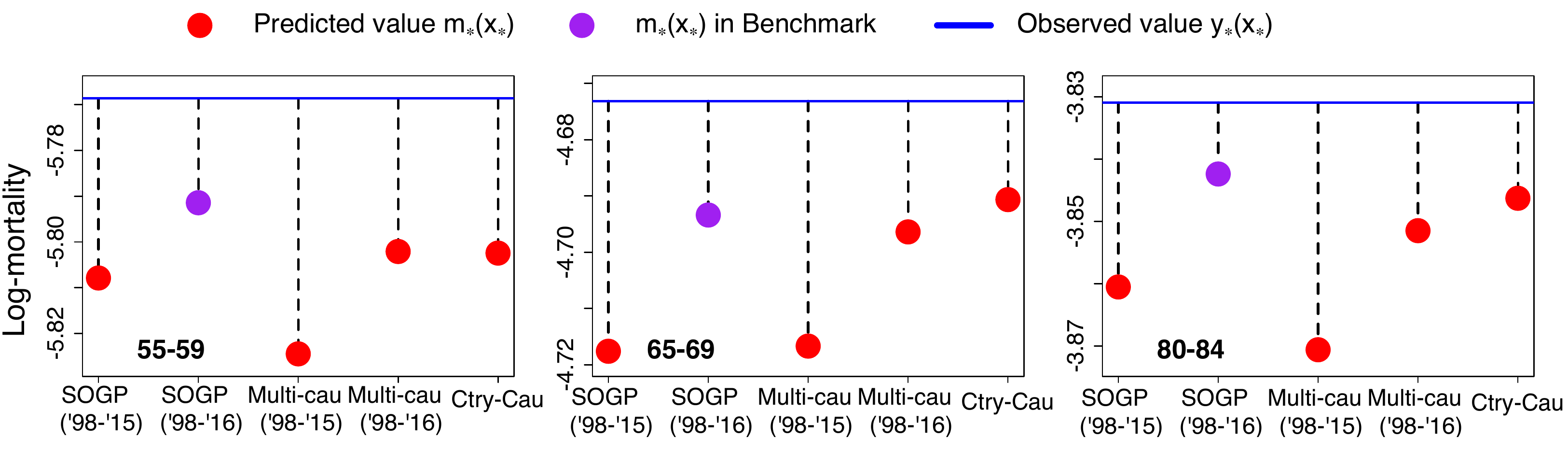}}
    \caption{Comparison of the prediction accuracy for 2016 all-Cancer log-mortality of Czech Males for indicated Age groups between different models with `notched' setup. Top row: Predictive standard deviation $s_*(x_*)$; bottom row: discrepancy between the predictive mean $m_*(x_*)$ and the observed value $y_*(x_*)$.}
    \label{fig:all_cancer_notch}
\end{figure}

Figure \ref{fig:all_cancer_notch} visualizes relative performance by comparing two components: (a) predictive standard deviation of the underlying 2016  $s_*(x_*)$ (top panels), and (b) prediction errors $m_*(x_*)-y_*(x_*)$ relative to the realized 2016 all-Cancer log-mortality rates of Czech Males (bottom). Our benchmark is Czech Males SOGP fitted on all-Cancer log-mortality rates from 1998--2016. We see that a Country-Cause MOGP yields more credible predictions (lower $s_*(x_*)$) than the benchmark model that performs in-sampled smoothing of the 2016 Czech all-Cancer experience. The forecasts from this joint model are nearly as good as the benchmark as both models produce the smallest prediction errors. In fact, the prediction quality of Country-Cause MOGP is as competitive as the Multi-cause MOGP that simultaneously models the log-mortality rates of all 5 cancer types in Czech Males with 2016 observations available. Thus, borrowing the latest information from Germany improves the prediction quality of the recent all-Cancer mortality rates of Czech Males. The results are as good as when we have domestic data available in Czech Republic.

\textit{Remark:} At the moment, the HCD offers cause-of-death data for less than 15 countries. Many countries do not have recent data available yet (e.g.~only up to  2014), leading to limited options in terms of the selection and the number of countries we can incorporate with Czech Males in this experiment. Based on our analysis in \cite{huynh2020}, choosing countries that are highly correlated with Czech Males is essential to maximize the prediction quality.

\section{Conclusion}
\label{sec:summary_cod}

In this article, we develop multi-output GP models for cause-specific mortality modeling within a multi-population context. With the MOGP mechanism, we are able to capture the cross-cause dependencies that allow joint models to gain more predictive power over single-output models that treat each population independently. Among MOGP variants, SLFM offers more flexibility and is recommended for modeling heterogenous causes, such as top-level ICD categories. Multi-level ICM is demonstrated to work well for interpretable modeling across multiple factor inputs. Our case studies show the applicability of MOGPs to understand cause-specific and aggregate mortality trends, both within a country and across nations, whereby our framework is convenient for information fusion and credibility boosting.

% The SLFM bolsters model flexibility by relaxing ICM assumption of the global spatial covariance over Age-Year dimensions. Meanwhile, the multi-level ICM leverages the structured Kronecker covariance kernel to cope with efficiency issues when we want to handle a larger number of populations. Via various examples, we illustrate the applicability of joint models to equip us with more informed insights about the dynamics of the aggregate mortality trends, within a country or in an international context. There is an opportunity to borrow information across national datasets to update the underlying domestic trends, especially for countries with lower data credibility.

The current work focuses on exploiting the structured Kronecker covariance to efficiently learn the joint covariance kernel. This is sufficient for handling a moderately large number of sub-populations (up to 30 in our case studies); additional techniques would be needed to handle larger datasets, for example across more causes-of-death or across all the countries in HCD.  There is an active research area looking at alternative methods (local kernel interpolation, inducing points, block structures, etc., see e.g., \cite{flaxman15}) for massive scalable GP well-suited for gridded mortality datasets. Another methodological extension would be to consider a Linear Coregionalization Model (LCM) for mortality modeling. LCM generalizes ICM and SLFM and allows multiple latent functions from GP priors with different covariance kernels.  A third direction would be to investigate other kernel families, such as composite kernels or kernels that can incorporate structural changes. The latter would be useful to  model (sub-)causes  that exhibit strong disruptions over time in their mortality trends. %Investigating mortality models to identify and capture structural changes is not the area of focus in this study.

%\clearpage

\appendix
\section{Additional Plots for the Case Study on Cancer Sub-Types}
\subsection*{Year ranges available in the HCD by country \label{app:hcd_cover}}
\begin{figure}[H]
    \centering
    \includegraphics[width=6.75cm, height = 6.5cm]{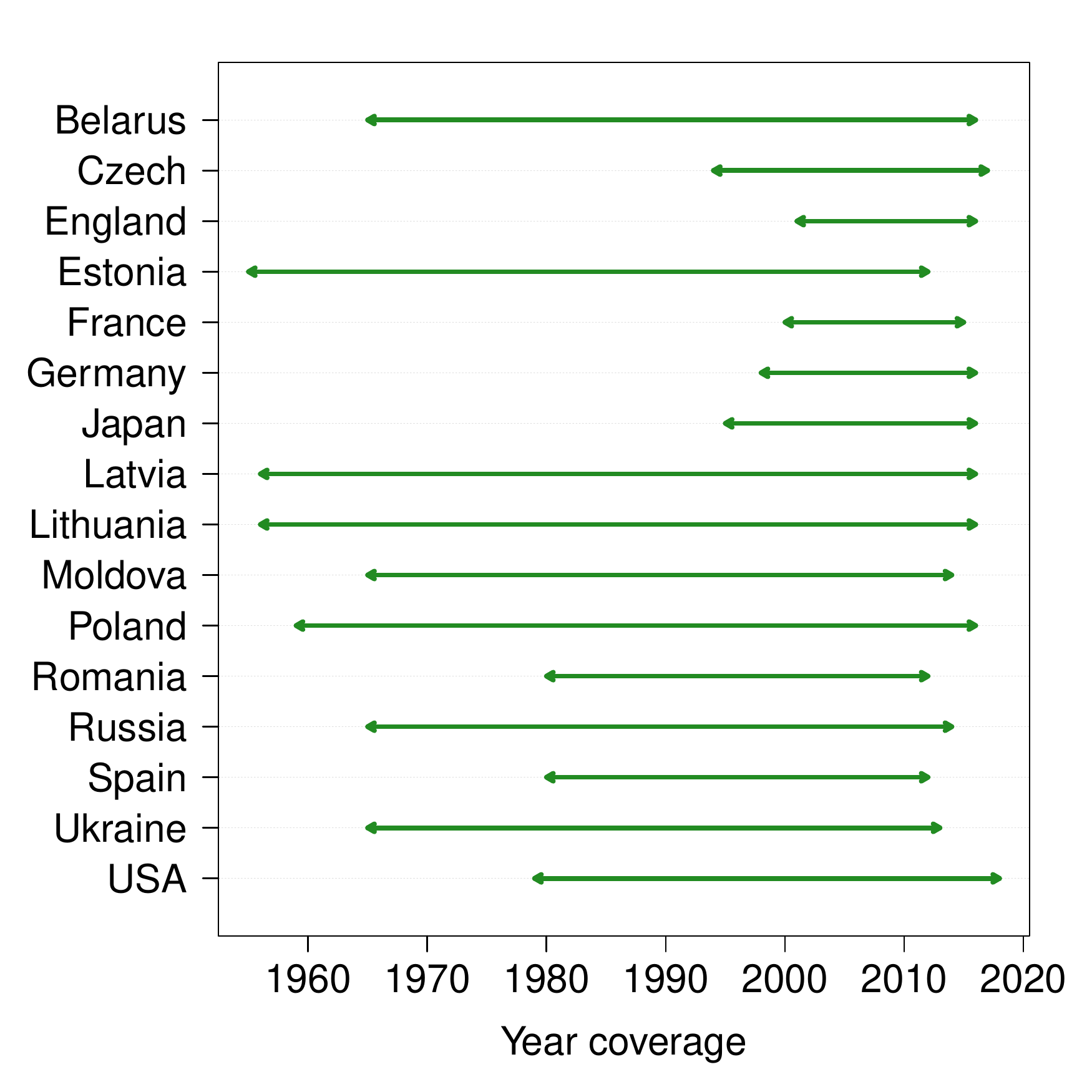}
    \caption{Countries in the HCD and their historical data coverage.}
    \label{fig:hcd}
\end{figure}

\subsection*{Age patterns of log-mortality rates of different cancers \label{app:raw-age}}
\begin{figure}[H]
    \centering
    \includegraphics[trim={0cm 0.25cm 0cm 0cm},clip=true, width=15cm]{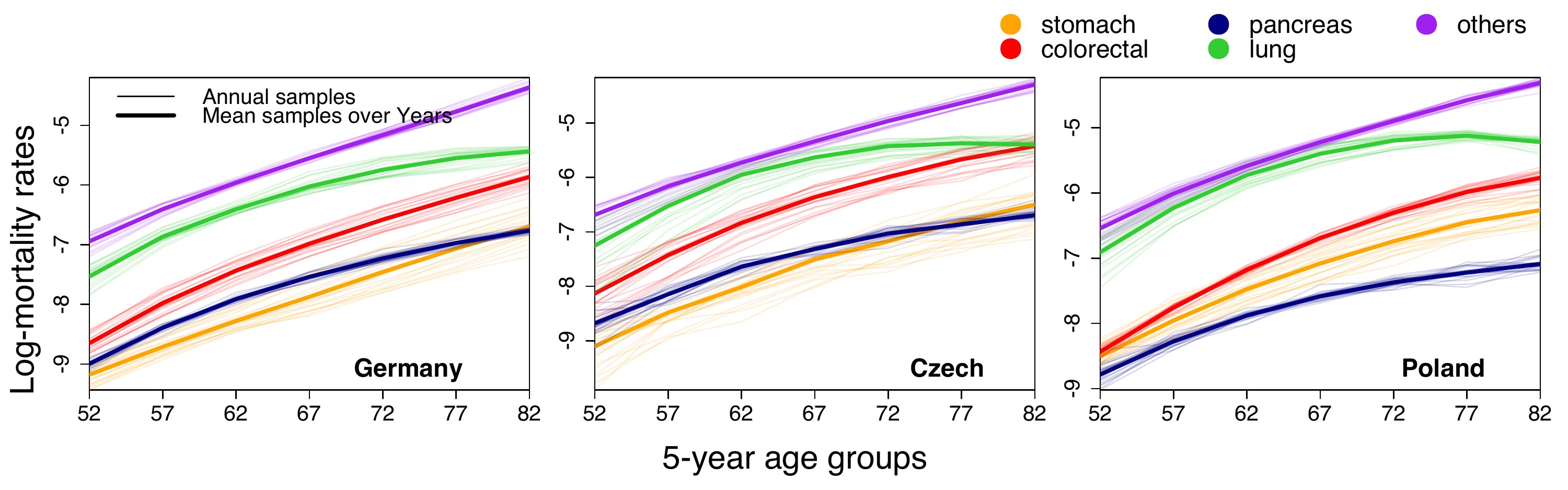} \\
    \caption{Raw log mortality as a function of Age for five Cancer variations in Male populations. We show 19 faint curves for each year in 1998--2016 range used for training, as well as the respective average historical log-mortality rate in bold.\label{fig:age_quad}}
\end{figure}

%\section{ICD Transitions}
%\begin{figure}[!t]
%    \centering
%    \includegraphics[width=15cm, height = 4.75cm]{icd-transition.pdf}
%    \caption{Impact of ICD Revision (moving from ICD-9 to current ICD-10) on mortality trends for several selected causes in the intermediate lisst for Male population in Poland.  \label{fig:transition}}
%\end{figure}
%
%
%
%\todo{I moved this from Intro, cut?} A well-known challenge in cause-of-death modeling is the disruptions in the mortality trends for many causes due to the transition of the International Classification of Disease (ICD).Figure \ref{fig:transition} demonstrates the impact of the ICD transition for several causes in Polish Males, in the period 1980--2016. While the mortality trends for several conditions remained stable across the revision (e.g. lung cancer), we observe discontinuities for sepsis and heart failure.
%
%So far, approaches to minimize this impact have been selecting data period that did not experience any ICD transition, or choosing causes with unsubstantial trend discontinuity. Under multi-population context, these approaches rather limit  the scope of cause-of-death analysis. All the demonstrations in this study use datasets from the Human Cause-of-death Mortality Database that provides ready-to-use cause-specific mortality datasets for 16 countries globally. Unlike other databases, we do not need to perform additional bridge-coding study to adjust the disruptive structure in many causes because of the transition in the ICD.

\subsection*{Kernel Hyperparameter Learning}

Searching for optimal hyperparameters in GP can be challenging if the marginal likelihood features multiple local maxima or flattens around its global maximum \citep{Rasmussen2005}. When the optimizer fails to find the global maximum, unsuitable length-scales in Age and Year sometimes result, leading to a poor fit of the data. Table~\ref{tbl:lengthscales} reports the inferred length-scale in Age ($\theta_{ag}$) and Year ($\theta_{yr}$) of the SOGP models fitted on the 5 Cancer types for  the Male populations in the three considered countries. Many SOGP models have $\theta_{yr}$ being too short (less than 5 or so), resulting in oscillatory fitted $m_*(\cdot)$'s. Such models lack the ability to distinguish between true signal and the inherent randomness in the data. Similarly, when the estimated lengthscales are too large,  the fitted GP surfaces are over-smoothed. Joint models tend to better learn the hyper-parameters by enabling data fusion across multiple populations and utilizing more observations. In Table~\ref{tbl:lengthscales} we show that when we fit multi-cause MOGP (both ICM and SLFM) on all 5 cancer types, the resulting lengthscales are all well calibrated.

\begin{table}[H]
\centering
\caption{GP lengthscales $\theta_{ag},\theta_{yr}$ in Age and Year for different models. All models are fitted on Age groups 50--84 and Years 1998--2016. \label{tbl:lengthscales}}
\resizebox{\columnwidth}{!}{%
\begin{tabular}{crrrrrrrr} \toprule
\multirow{2}{*}{Czech Rep.} & \multicolumn{5}{c}{SOGP on each cancer type}       & Multi-cause ICM & \multicolumn{2}{c}{Multi-cause SLFM} \\ \cline{2-6}
& Stomach & Colorectal & Pancreas & Lung   & Others  & ($Q = 2$)       & \multicolumn{2}{c}{($Q=2$)}          \\ \midrule
$\theta_{ag}$   & 159.92 & 16.32    & 6.21   & 17.69 & 12.52 & 16.90         & 21.26           &  19.67          \\
$\theta_{yr}$  & 38.49 & 9.03    & 7.82   & 13.87 & 3.69 & 10.12         & 13.05           & 11.42  \\ \bottomrule
\end{tabular}} \\
\resizebox{\columnwidth}{!}{%
\begin{tabular}{crrrrrrrr} \toprule
\multirow{2}{*}{Germany} & \multicolumn{5}{c}{SOGP on each cancer type}       & Multi-cause ICM & \multicolumn{2}{c}{Multi-cause SLFM} \\ \cline{2-6}
& Stomach & Colorectal & Pancreas & Lung   & Others  & ($Q = 2$)       & \multicolumn{2}{c}{($Q=2$)}          \\ \midrule
$\theta_{ag}$  & 8.65 & 3.72    & 8.99   & 0.00 & 2.79 & 8.59         &  11.75           & 9.46          \\
$\theta_{yr}$   & 7.35 & 4.90    & 7.69   & 4.82 & 4.44 & 10.26         & 9.92           & 6.77  \\ \bottomrule
\end{tabular}} \\
\resizebox{\columnwidth}{!}{%
\begin{tabular}{crrrrrrrr} \toprule
\multirow{2}{*}{Poland} & \multicolumn{5}{c}{SOGP on each cancer type}       & Multi-cause ICM & \multicolumn{2}{c}{Multi-cause SLFM} \\ \cline{2-6}
& Stomach & Colorectal & Pancreas & Lung   & Others  & ($Q = 2$)       & \multicolumn{2}{c}{($Q=2$)}          \\ \midrule
$\theta_{ag}$   & 23.96 & 16.28    & 5.72   & 15.25 & 22.80 & 16.98         & 16.20           & 21.89         \\
$\theta_{yr}$  & 253.34 & 3.69    & 6.19   & 9.04 & 6.01 &  12.83         & 11.19           & 12.11  \\ \bottomrule
\end{tabular}} \\
\end{table}

\subsection*{Training Designs in Notched Datasets}

\begin{table}[H]
\centering
\caption{Descriptions for models being applied to forecast 2016 all-Cancer log-mortality rates of Czech Males. All the models are fitted on Ages 50--84 (7 age groups) and Year coverage listed below. \label{tbl:notched-setup}}
\resizebox{\columnwidth}{!}{%
\begin{tabular}{llllll} \toprule
GP models  & Outcome variable   & Country       & Year       & Pred. Type & Abbrev.    \\ \midrule
\multirow{2}{*}{SOGP}             & \multirow{2}{*}{All-cancer log-mortality}       & \multirow{2}{*}{Czech R.} & 1998--2016 & In-sample   & SOGP ('98-'16)    \\  \cline{4-6}
&       &       & 1998--2015 & Out-of-sample & SOGP ('98-'15)  \\ \midrule
\multirow{2}{*}{Multi-cause GP}   & \multirow{2}{*}{\begin{tabular}[c]{@{}l@{}}By-cancer log-mortality \\ (5 variations)\end{tabular}} & \multirow{2}{*}{Czech R.} & 1998--2016 & In-sample                                            & Multi-cause ('98-'16)  \\ \cline{4-6}
&       &     & 1998--2015 & Out-of-sample   & Multi-cause ('98-'15)              \\ \midrule
\multirow{2}{*}{Country-Cause GP} & \multirow{2}{*}{\begin{tabular}[c]{@{}l@{}}By-cancer log-mortality\\  (5 variations)\end{tabular}} & Czech  R.         & 1998-2015  & \multirow{2}{*}{Out-of-sample}                       & \multirow{2}{*}{Country-Cause} \\
&       & Germany   & 1998-2016  &      &    \\ \bottomrule
\end{tabular}}
\end{table}

\subsection*{Illustrating Latent Factor Loadings in SLFM \label{app:a_slfm}}

\begin{figure}[H]
    \centering
    \includegraphics[trim=0.2in 0.3in 0.2in 0.3in,height=2.6in]{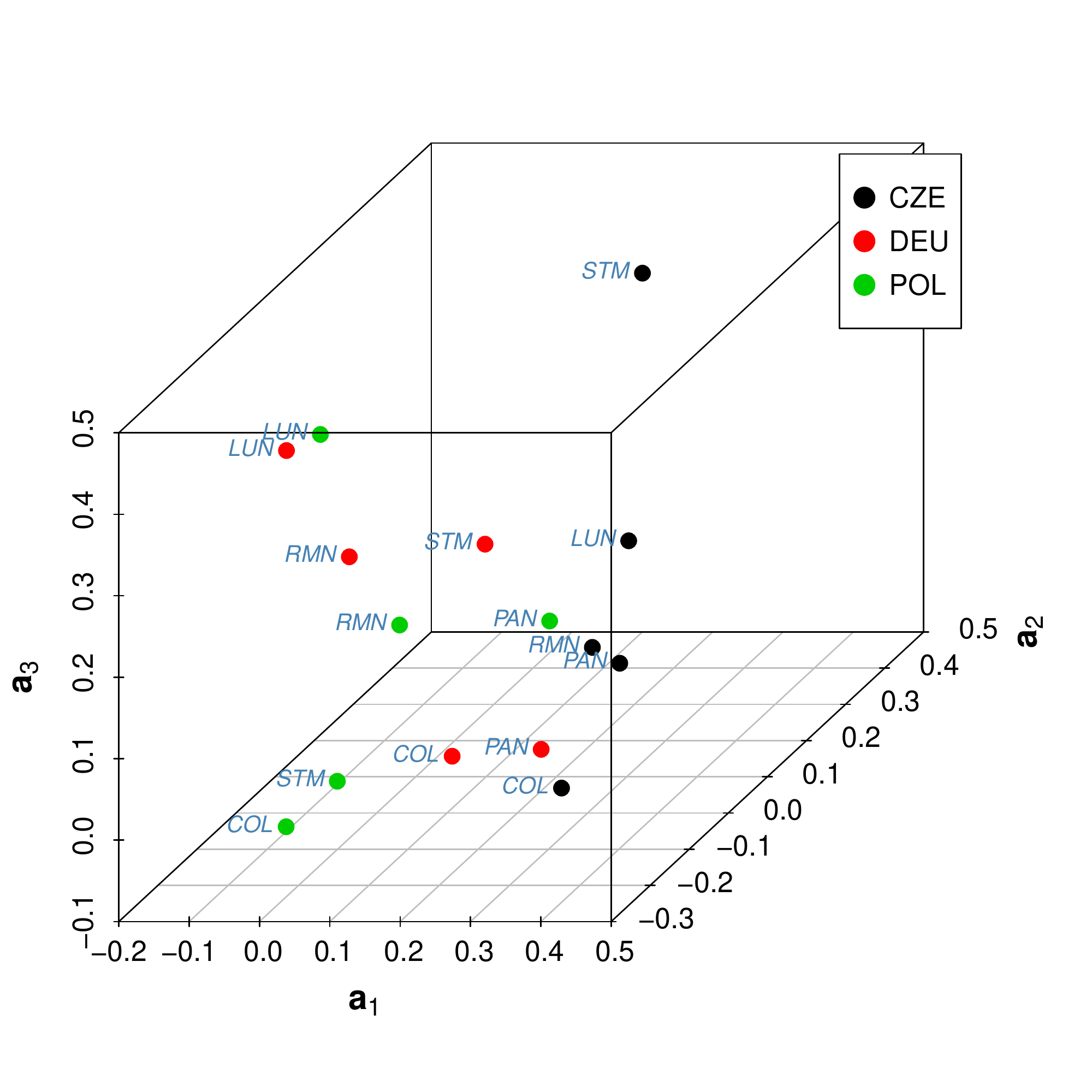}
    \caption{Factor loadings $\mathbf{a}_q=(a_{1,q},\ldots,a_{L,q})^T$ in the Country-Cause SLFM with $Q=3$. The model is fitted on Ages 50--84, Years 1990--2016, over 3 Countries and 5 Cancer types. For each of the 15 populations, we plot $(a_{l,1}, a_{l,2}, a_{l,3})$ as a point in 3-space.
    \label{fig:a-loadings}}
\end{figure}

\section{Additional Plots for the US Top-Level-Cause Study}
\subsection*{Adjusting Drug Overdose Trend} \vspace*{-8pt}

\begin{figure}[H]
    \centering
    %\textbf{AGE 40-44, US MALES} \\
    {{\includegraphics[trim=0.1in 0.1in 0.1in 0.1in,width=2.5in]{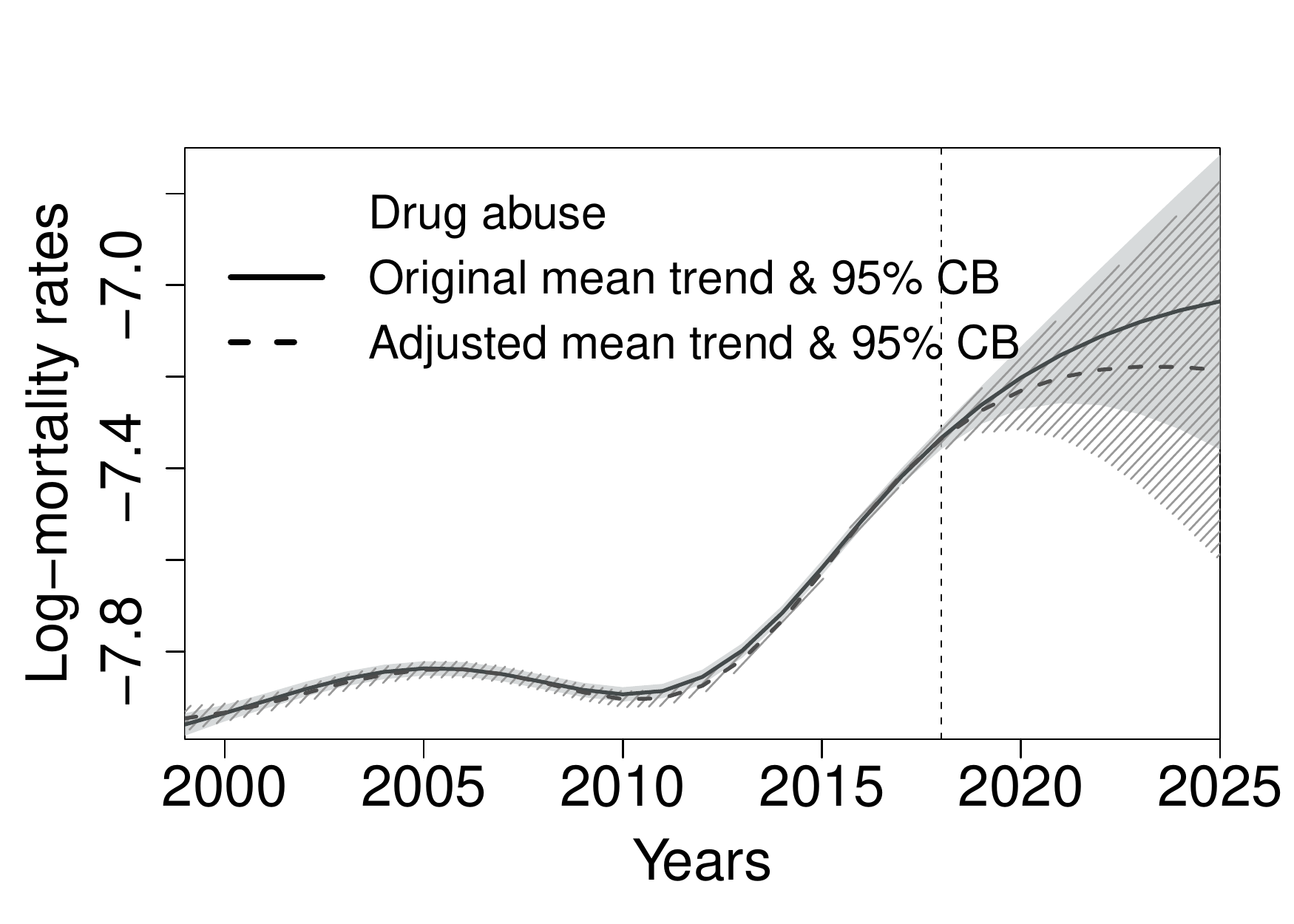}}}
    {{\includegraphics[trim=0.1in 0.1in 0.1in 0.1in,width=2.5in]{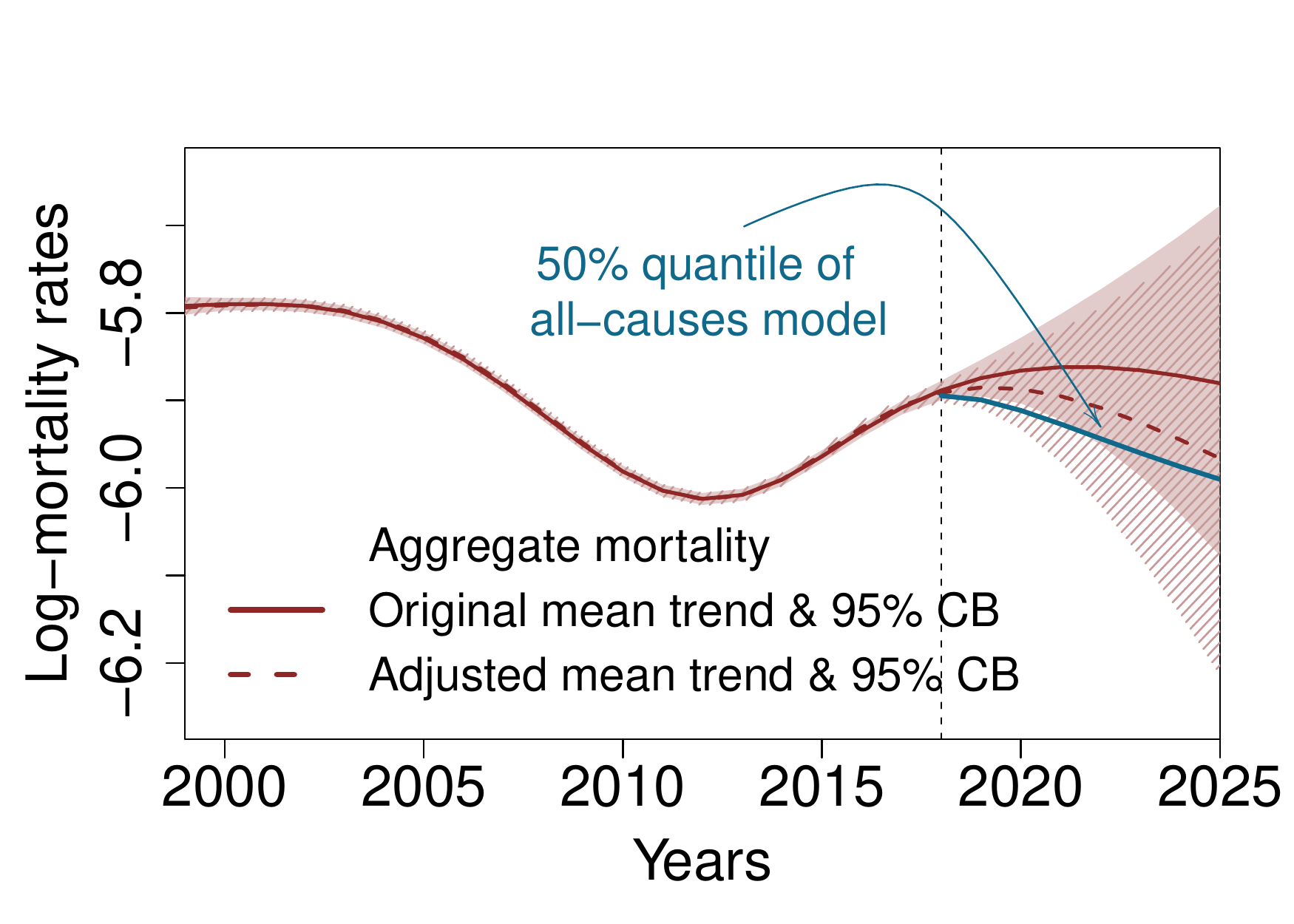}}} \\
     \textbf{Age Group 40--44. Top row:  US Males $\uparrow$. Bottom  row:  US Females $\downarrow$} \\
    {{\includegraphics[trim=0.1in 0.1in 0.1in 0in,width=2.5in]{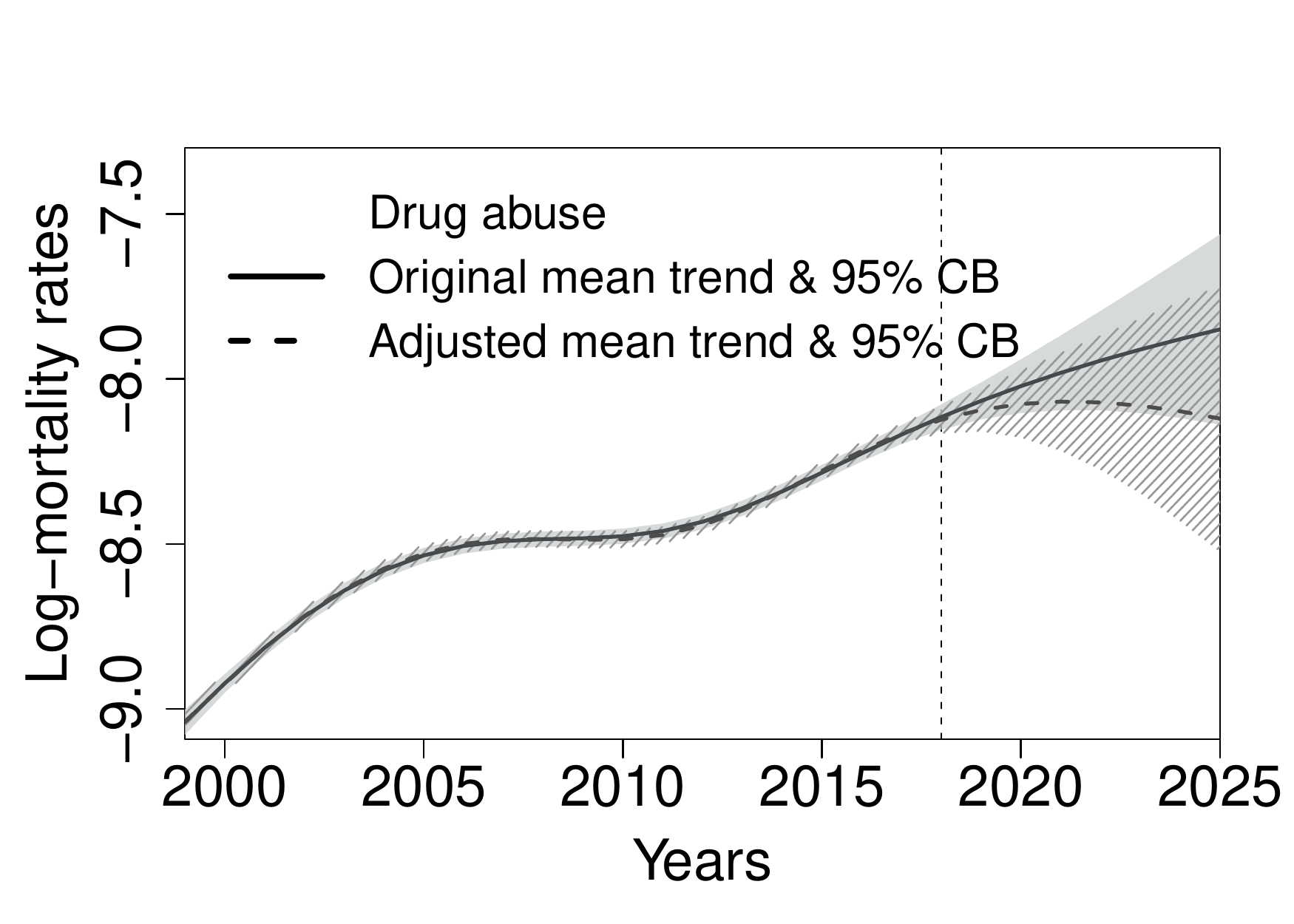}}}
    {{\includegraphics[trim=0.1in 0.1in 0.1in 0in,width=2.5in]{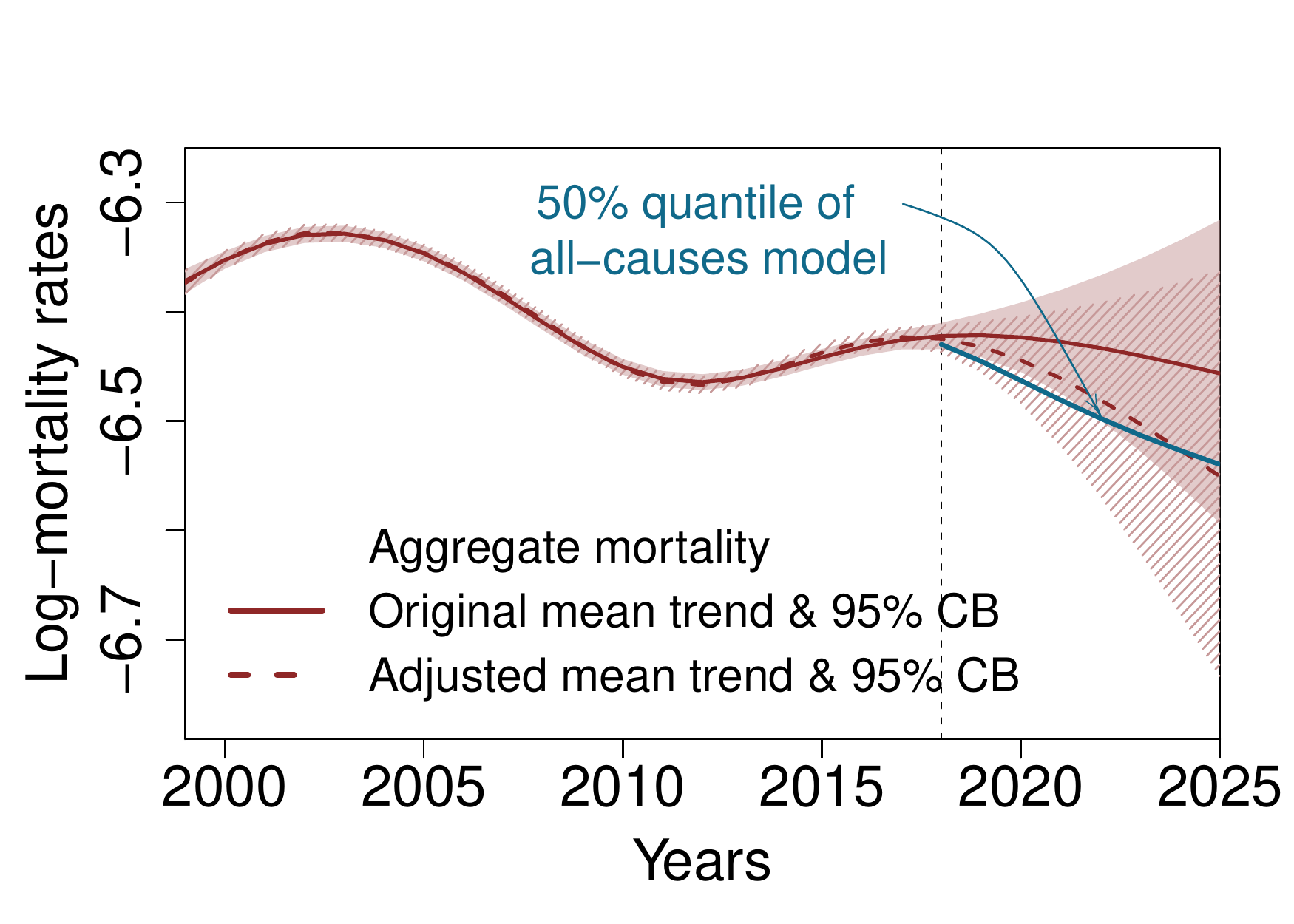}}} \\
    \caption{Comparison of original and adjusted predicted log-mortality for Age group 40--44, US Males and Females. Left: same model after reducing Drug Abuse yearly trend $\beta_{yr}$ by one-third. Right: corresponding aggregated all-cause trends. Vertical lines indicate the edge of training data (1999--2018).}
    \label{fig:expert-based}
\end{figure}

\subsection*{Cross-correlation Matrices for the US All-Cause Analysis}

\begin{figure}[H]
    \centering
    \subfloat[US Males]{{\includegraphics[trim={0cm 3cm 0cm 2.6cm},clip=true, width=8cm]{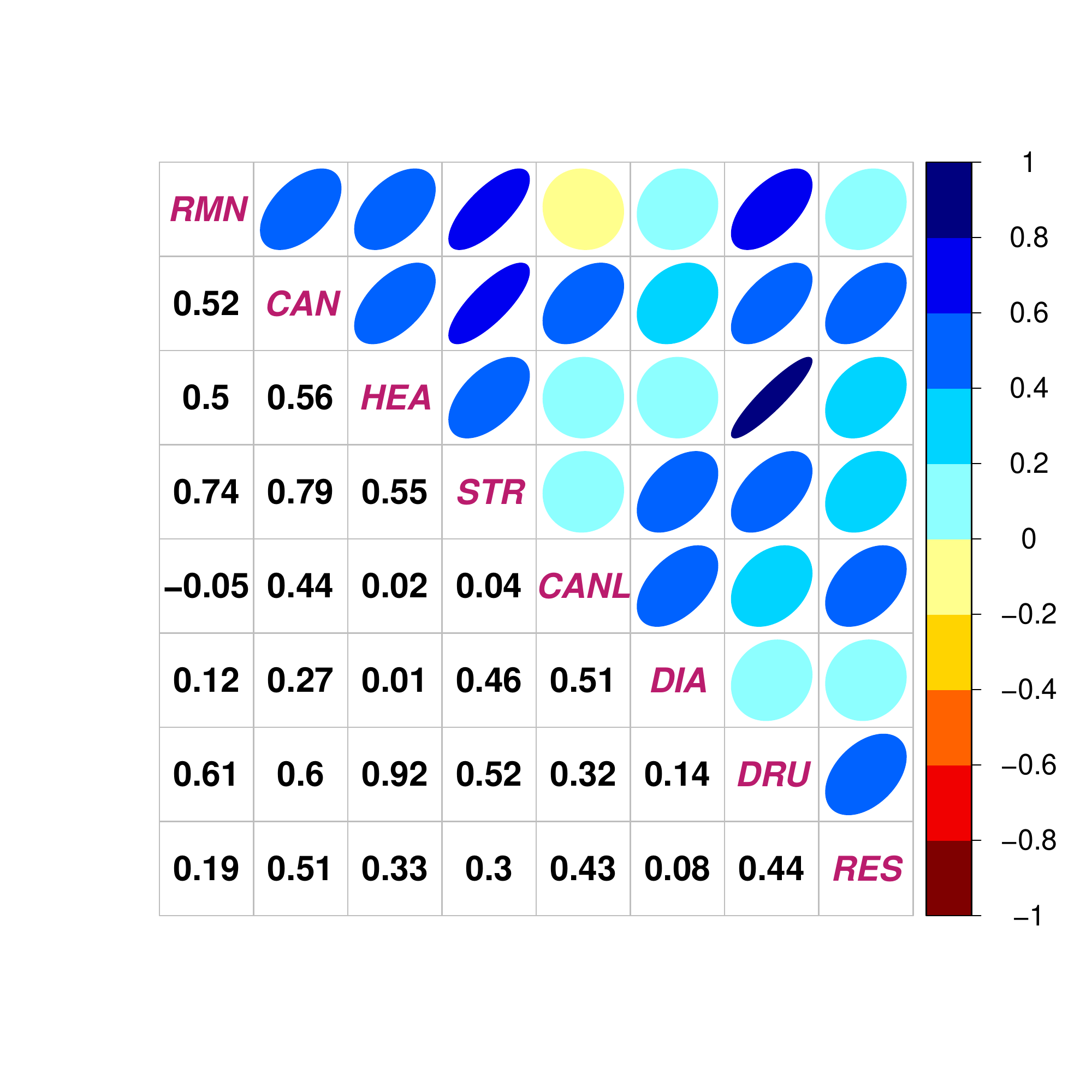}}}
    \subfloat[US Females]{{\includegraphics[trim={0cm 3cm 0cm 2.6cm},clip=true, width=8cm]{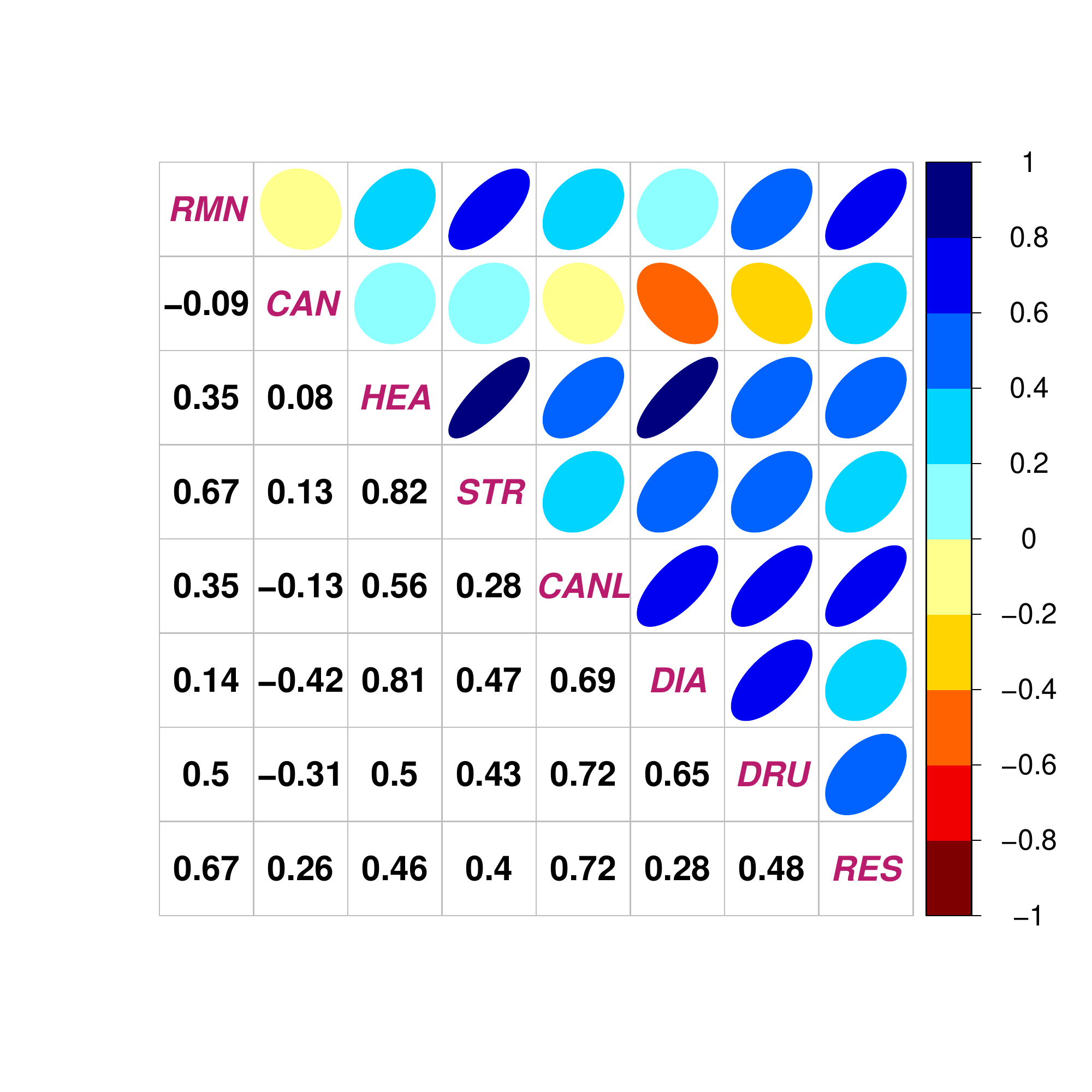}}} \\
    \caption{Cross-cause correlation matrices derived from multi-cause single-level ($Q=6$) ICM MOGP, fitted on Ages 40--69 (6 age groups) and Years 1999--2018, separately for US Males and US Females. \label{fig:corr-us}}
\end{figure}

%\section{Inferred ICM Cross-Correlation Matrix \label{app:cross_corr}}
%We can expand the cross-covariance kernel $B$ in Eq \eqref{eqn:icm-def} such as:
%\begin{align*}
%    B = AA^T & = \sum_{q=1}^Q \mathbf{a}_q\mathbf{a}_q^T \\
%             & = \begin{bmatrix}
%             \sum_{1=q}^Q a^2_{1,q} & \sum_{1=q}^Q a_{1,q}a_{2,q} & \dots & \sum_{1=q}^Q a_{1,q}a_{L,q} \\
%             \sum_{1=q}^Q a_{2,q}a_{1,q} &  \sum_{1=q}^Q a^2_{2,q} & \dots &
%             \sum_{1=q}^Q a_{2,q}a_{L,q} \\
%             \vdots & \vdots & \ddots & \vdots \\
%             \sum_{1=q}^Q a_{L,q}a_{1,q} & \sum_{1=q}^Q a_{L,q}a_{2,q} & \dots &
%             \sum_{1=q}^Q a^2_{L,q} \\
%             \end{bmatrix} \\
%\end{align*}
%The process variance of the l$th$ function output is the l$th$ element of the diagonal in $B$, or  $\eta^2_{l} = \sum_{q=1}^Q a^2_{l,q}$, $1 \leq l \leq L$. The cross-correlation between population $l_1$ and $l_2$ ($1 \leq l_1,l_2 \leq L$) is:
%$$r_{l_1,l_2}:=\frac{\Cov(f_{l_1}(\mathbf{x}),f_{l_2}(\mathbf{x}))}{\sqrt{\Cov(f_{l_1}(\mathbf{x}),f_{l_1}(\mathbf{x}))\times\Cov(f_{l_2}(\mathbf{x}),f_{l_2}(\mathbf{x}))}} = \frac{\sum_{q=1}^Qa_{l_1,q}a_{l_2,q}}{\sqrt
%{\big(\sum_{q=1}^Qa^2_{l_1,q}\big)\big(\sum_{q=1}^Qa^2_{l_2,q}\big)}}.$$
%

\bibliographystyle{plainnat}
\bibliography{mogp-cod-reference}

\end{document}